\documentclass[12pt,a4paper]{article}
\usepackage{amsmath,amssymb,epsfig}
\usepackage{epsfig}
\renewcommand{\theequation}{\thesection.\arabic{equation}}
\newlength{\extraspace}
\newlength{\extraspaces}
\newcommand{\be}{\begin{equation}
\addtolength{\abovedisplayskip}{\extraspaces}
\addtolength{\belowdisplayskip}{\extraspaces}
\addtolength{\abovedisplayshortskip}{\extraspace}
\addtolength{\belowdisplayshortskip}{\extraspace}}
\newcommand{\ee}{\end{equation}}
 
\newcommand{\ba}{\begin{eqnarray}
\addtolength{\abovedisplayskip}{\extraspaces}
\addtolength{\belowdisplayskip}{\extraspaces}
\addtolength{\abovedisplayshortskip}{\extraspace}
\addtolength{\belowdisplayshortskip}{\extraspace}}
\newcommand{\ea}{\end{eqnarray}}

\newcommand{\bas}{\begin{eqnarray*}
\addtolength{\abovedisplayskip}{\extraspaces}
\addtolength{\belowdisplayskip}{\extraspaces}
\addtolength{\abovedisplayshortskip}{\extraspace}
\addtolength{\belowdisplayshortskip}{\extraspace}}
\newcommand{\eas}{\end{eqnarray*}}
 
\newcounter{subequation}[equation]
\makeatletter

\expandafter\let\expandafter
\reset@font\csname reset@font\endcsname

\def\subeqnarray{\arraycolsep1pt
    \def\@eqnnum\stepcounter##1{\stepcounter{subequation}%
        {\reset@font\rm(\theequation\alph{subequation})}}
\jot5mm     \eqnarray}

\def\subarray{\arraycolsep1pt
    \def\@eqnnum\stepcounter##1{\stepcounter{subequation}%
        {\reset@font\rm(\alph{subequation})}}
\jot5mm     \eqnarray}

\makeatother
\newcommand{\newsection}[1]{
\vspace{15mm}
\pagebreak[3]
\addtocounter{section}{1}
\setcounter{equation}{0}
\setcounter{subsection}{0}

\setcounter{footnote}{0}
\addcontentsline{toc}{section}
{\protect\numberline{\arabic{section}}{#1}}
 
\begin{flushleft}
{\large\bf \thesection. #1}
\end{flushleft}
\nopagebreak
\medskip
\nopagebreak}
 
\newcommand{\newsubsection}[1]{
\vspace{1cm}
\pagebreak[3]
\addtocounter{subsection}{1}

\addcontentsline{toc}{subsection}
{\protect\numberline{\thesection.\arabic{subsection}}{#1}}
 
\noindent{ \bf \thesection.\arabic{subsection} #1}
\nopagebreak
\vspace{2mm}
\nopagebreak}

\newcommand{\newappendix}[1]{
\vspace{15mm}
\pagebreak[3]
\addtocounter{section}{1}
\setcounter{equation}{0}
\setcounter{subsection}{0}

\addcontentsline{toc}{section}
{\protect\numberline{\thesection}{#1}}

\renewcommand{\theequation}{\Alph{section}.\arabic{equation}}
\begin{flushleft}
{\large\bf Appendix \Alph{section}: #1}
\end{flushleft}
\nopagebreak
\medskip
\nopagebreak}

\newcommand{\NP}[1]{Nucl.\ Phys.\ {\bf #1}}
\newcommand{\PL}[1]{Phys.\ Lett.\ {\bf #1}}
\newcommand{\CMP}[1]{Comm.\ Math.\ Phys.\ {\bf #1}}
\newcommand{\PR}[1]{Phys.\ Rev.\ {\bf #1}}
\newcommand{\PRL}[1]{Phys.\ Rev.\ Lett.\ {\bf #1}}

\newcommand{\N}{\mathbb{N}}

\newcommand{\Z}{\mathbb{Z}}
\newcommand{\R}{\mathbb{R}}

\newcommand{\T}{\mathbb{T}}

\renewcommand{\H}{\mathbb{H}}

\newcommand{\1}{\mbox{1\hspace{-.8ex}1}}
\newcommand{\bra}{\langle}
\newcommand{\ket}{\rangle}
\newcommand{\ra}{\rightarrow}

\newcommand{\rra}{\ \longrightarrow \ }

\newcommand{\is}{ &\! =\! & }
\newcommand{\nonum}{\nonumber \\[1.5mm]}
\newcommand{\sspace}{\makebox[1cm]{ }}
\newcommand{\bspace}{\makebox[2cm]{ }}
\newcommand{\nspace}{\!\!\!\!\!\!\!\!\!\!}

\newcommand{\inv}{^{-1}}

\renewcommand{\th}{{\theta}}
\newcommand{\eps}{\epsilon}
\newcommand{\lb}{\lambda}
\newcommand{\om}{\omega}
\newcommand{\Om}{\Omega}
\newcommand{\sh}{{\rm sh}}
\newcommand{\ch}{{\rm ch}}
\newcommand{\dd}{{\partial}}

\newcommand{\up}{\uparrow}
 
\newcommand{\cA}{{\cal A}}
\newcommand{\cB}{{\cal B}}
\newcommand{\cC}{{\cal C}}
\newcommand{\cD}{{\cal D}}
\newcommand{\cE}{{\cal E}}

\newcommand{\cH}{{\cal H}}

\newcommand{\cN}{{\cal N}}

\newcommand{\cO}{{\cal O}}
\newcommand{\cP}{{\cal P}}

\newcommand{\cS}{{\cal S}}
\newcommand{\cT}{{\cal T}}

\newcommand{\n}{{\sc n}}

\begin{document}

%%%%%%%%%%%%%%%%%%%%%%%%%%%%%%%%%%%%%%%%%%%%%%%%%%%%%%%%%%%%%%%%%%%%%%%
\begin{titlepage}

%footnotesymbols others than numbers
\renewcommand{\thefootnote}{\fnsymbol{footnote}}
\mbox{}
%\vspace{5mm}

\begin{center}
\mbox{{\Large \bf Non-amenability and spontaneous symmetry breaking}}\\[4mm]
{\Large \bf -- The hyperbolic spin-chain --}
\vspace{1.5cm}

{{\sc M. Niedermaier}%
\footnote{Membre du CNRS.} % e-mail: {\tt max@phys.univ-tours.fr}}}
\\[4mm]
{\small\sl Laboratoire de Mathematiques et Physique Theorique}\\
{\small\sl CNRS/UMR 6083, Universit\'{e} de Tours}\\
{\small\sl Parc de Grandmont, 37200 Tours, France}
\\[4mm]
{\sc E. Seiler}}
\\[4mm]
{\small\sl Max-Planck-Institut f\"{u}r Physik}\\
{\small\sl (Werner-Heisenberg-Institut)}\\
{\small\sl F\"ohringer Ring 6}\\
{\small\sl 80805 M\"unchen, Germany}

\vspace{1.5cm}

{\bf Abstract}
\vspace{-3mm}

\end{center}

\begin{quote}
The hyperbolic spin chain is used to elucidate the notion of 
spontaneous symmetry breaking for a non-amenable internal symmetry group,
here ${\rm SO}(1,2)$. The noncompact symmetry is shown to be spontaneously 
broken -- something which would be forbidden for a compact group  by the 
Mermin-Wagner theorem. Expectation functionals are defined through the $L 
\ra \infty$ limit of a chain of length $L$; the functional measure 
is found to have its weight mostly on configurations boosted by an amount 
increasing at least powerlike with $L$. This entails that despite the 
non-amenability a certain subclass of noninvariant functions is averaged 
to an ${\rm SO}(1,2)$ invariant result. Outside this class symmetry 
breaking is generic. Performing an Osterwalder-Schrader reconstruction 
based on the infinite volume averages one finds that the reconstructed 
quantum theory is different from the original one.    
The reconstructed Hilbert space is nonseparable and contains a separable
subspace of ground states of the reconstructed transfer operator on which
${\rm SO}(1,2)$ acts in a continuous, unitary, and irreducible way. 
\end{quote} 
\vfill

\setcounter{footnote}{0}
\end{titlepage}
%%%%%%%%%%%%%%%%%%%%%%%%%%%%%%%%%%%%%%%%%%%%%%%%%%%%%%%%%%%%%%%%%%%%%%%%%%%%%
\tableofcontents
\newpage
%%%%%%%%%%%%%%%%%%%%%%%%%%%%%%%%%%%%%%%%%%%%%%%%%%%%%%%%%%%%%%%%%%%%%%%%%%%%%

\newsection{Introduction}

Spontaneous symmetry breaking is typically discussed for compact internal
or for abelian translational symmetries, see e.g.~\cite{Ruelle,Sewell,NT}. 
Both share the property of being amenable \cite{Pat}; we recall the definition below but
mention already that all semisimple nonabelian noncompact Lie groups are 
non-amenable. The goal of this note is to elucidate the notion of 
spontaneous symmetry breaking for a non-amenable internal group. This is 
motivated by the ubiquitous appearance of noncompact internal symmetries 
in a gravitational context, specifically in the dimensional reduction of
gravitational theories \cite{qernst}, further in integrable
sectors of QCD \cite{Korch}, or in ghost- or $\theta$-sectors of gauge   
theories, and also in condensed matter physics
\cite{Wegner79,HJKP,Hik,Efetov83, Efetov97}.
The very fact that the group is 
non-amenable turns out to entail a number of surprising new features.
In particular spontaneous symmetry breaking becomes possible in low 
dimensions where it is forbidden by the Mermin-Wagner theorem 
\cite{Ruelle, MW, DS} 
in the case of  compact internal symmetries. In order to have a concrete 
computational framework at hand we consider a definite lattice statistical 
system, the hyperbolic spin chain. This is a spin chain where the 
dynamical variables take values in a hyperbolic (Riemannian) space of 
constant negative curvature and the interaction is through nearest 
neighbors only. The lattice formulation was chosen in order to have 
control over the thermodynamic limit and in preparation to the quantum 
field theoretical case. Indeed we expect that many of the qualitative 
results generalize to generic statistical systems as well as to quantum 
field theories. In an accompanying paper \cite{DNS} we study the 
nonlinear sigma-model with a hyperbolic target space in 2 or more
dimensions.

The systems treated always can be regarded in two different 
ways: either as a system of classical statistical mechanics, or 
as a quantum system in imaginary time. We mostly use the former 
interpretation, but discuss in some detail the reconstruction of the 
associated quantum system.

Following \cite{NT}, in the quantum interpretation we consider dynamical 
systems $(\cC, \tau)$ consisting of a $*$-algebra $\cC$ (``the 
observables'') and a one-parameter group of automorphisms  (``the time 
evolution''), which we take to be discrete here $\tau^x$, $x \in \Z$. 
In addition a group of automorphisms $\rho(g), g \in G$,
(``the symmetry group'') is supposed to act on $\cC$ and to commute with
the time evolution, $\tau \circ \rho = \rho \circ \tau$. A state $\omega$
(positive linear functional over $\cC$) is said to be $\tau$-invariant
if $\omega \circ \tau = \omega$ and extremal $\tau$-invariant if it is not 
a convex combination of different invariant states. The symmetry $\rho$ 
is said to be {\it spontaneously broken} (see e.g.~\cite{Ruelle,Sewell,NT}) 
by an (extremal) $\tau$-invariant state $\omega$ if $\omega \circ \rho \neq 
\omega$.

In the classical statistical mechanics interpretation $\cC$ is a commutative 
$C^\ast$-algebra (though there may be reasons to relax this 
condition) and the `time evolution' really plays the role of space 
translations. A symmetry is again given by a  group of automorphisms 
$\rho(g), g \in G$, acting on $\cC$ and leaving the Hamiltonian (or 
action) invariant, except for possible symmetry violating boundary 
condition (a very precise definition of the notion of symmetry and its 
spontaneous breaking can be found in \cite{Georgii}). The definitions of 
states and their invariance or noninvariance are as in the quantum 
interpretation. Spontaneous symmetry breaking is then said to occur 
if there is an infinite volume Gibbs state (for instance obtained as a 
limit of finite volume Gibbs states) that is noninvariant.

We shall be interested in the above situation when the symmetry group is 
a non-amenable Lie group. A Lie group $G$ is called amenable if there 
exists an (left) invariant state (``a mean'') on the space $\cC_b(G)$ of 
all continuous bounded functions on $G$ equipped with the sup-norm. 
Conversely, $G$ is called {\it non-amenable} if no such invariant mean 
over $\cC_b(G)$ exists. All non-compact semisimple nonabelian Lie groups 
are known to be non-amenable. The notion of amenability has also been
extended from Lie groups to homogeneous spaces (see for instance
\cite{greenleaf,eymard}). Note that {\it if} in the above definition 
$\cC$ was taken to be $\cC_b(G)$, spontaneous symmetry breaking would be 
automatic for all non-amenable symmetries. We shall find however that the 
non-amenability also forces one to consider smaller algebras of 
observables (e.g.~$C^*$-subalgebras of $\cC_b(G)$) so that the issue becomes 
non-trivial again.   

As a guideline it may be helpful to contrast the peculiar features 
we find in the hyperbolic spin chain with those in the corresponding 
compact model.
 
%%%%%%%%%%%%%%%%%%%%%%%%%%%%%%%%%%
\vspace{5mm}

\hspace{6mm}
\begin{tabular}{|c|c|c|}
\hline
quantity      & spherical spin-chain & hyperbolic spin chain   \\[0.5ex] 
\hline \hline
ground state(s)  & unique, normalizable & $\infty$ set, non-normalizable \\
              & SO(3) invariant      &  not ${\rm SO}(1,2)$ invariant  \\ 
\hline 
expectations of        & SO(3) invariant      & SO(1,2) invariant \\
selected SO(2)-invariant        & independent of bc    & depend on bc \\
observables            &                      & bc selects ground state\\
\hline
expectations of        & SO(3) invariant      & SO(1,2) non-invariant \\
generic non-invariant    & independent of bc    & depend on bc \\
observables            &                      &               \\
\hline
reconstructed  & reproduces          & different from \\
quantum theory & original one        & original one \\
\hline
\end{tabular}
%%%%%%%%%%%%%%%%%%%%%%%%%%%%%%%%%%%
\vspace{5mm}

Here the expectations of an observable refer to the thermodynamic limit of 
the chain where the number of sites goes to infinity while the lattice spacing
is still finite. Moreover we require that the expectations are defined through 
a thermodynamic limit that does not involve the selection of `fine-tuned'
subsequences. This defines a subclass of `regular' observables to which we 
mostly limit the discussion. These regular observables (later called 
``asymptotically translation invariant'') presumably include all bounded 
ones, but an explicit formula for their expectations can be derived 
regardless of boundedness. For the hyperbolic chain it turns out that one
has to impose boundary conditions (bc) at the end(s) of the chain which
keep at least one spin fixed. Remarkably, we find that even the
expectations of invariant observables may depend on the choice of bc,
even though in the limit the ends are separated by an infinite number of 
sites from those where the observable is supported!

The bc we are using also single out a preferred subgroup ${\rm SO}(2) 
\subset {\rm SO}(1,2)$ and the expectation functionals turn out to project
any observable onto its ${\rm SO}(2)$ invariant part. Since this averaging
over ${\rm SO}(2)$ does not commute with the action of the full 
${\rm SO}(1,2)$ group generic non-invariant observables will signal 
spontaneous symmetry breaking, i.e.~their expectations are not ${\rm SO}(1,2)$ 
invariant. This is accompanied by an infinite family of nonnormalizable
`ground states' transforming under an irreducible representation
of  ${\rm SO}(1,2)$. This representation becomes unitary under a suitable 
change of the scalar product; such a scalar product will be produced by 
the Osterwalder-Schrader reconstruction described in Section 5.

Somewhat different indications of spontaneous symmetry breaking 
in this context have been obtained in \cite{AmitDav83,vHol87}.  
In a situation of conventional symmetry breaking (say, of a 
compact Lie group symmetry in higher dimensions) one can always switch 
to invariant expectation functionals by performing a group average
over the original noninvariant ones, at the expense of making the 
clustering properties worse. Here, due to the non-amenability of 
${\rm SO}(1,2)$ this cannot be done; the symmetry breaking is more 
severe, and in this respect resembles somewhat the `spontaneous 
collapse' of supersymmetry in a spatially homogeneous state at finite 
temperature \cite{BO}.     
 
It is therefore remarkable that there exists a class of `selected' 
${\rm SO}(2)$ but not ${\rm SO}(1,2)$ invariant observables 
(later called ``${\rm SO}(2)$ and asymptotically invariant'') 
which get averaged to yield a ${\rm SO}(1,2)$ invariant result.
One sees that the impact of the non-amenability is quite subtle:
an invariant mean for all bounded (let alone unbounded) observables cannot
exist, however an invariant  mean on a subalgebra does exist and can be 
constructed explicitly as a thermodynamic limit of probability measures.  
Schematically, the mechanism behind this is that for a finite chain of length
$L$ the functional measure has support mostly at configurations which are 
boosted with a parameter depending on and increasing with $L$. So provided a
limit exists at all it will be ${\rm SO}(1,2)$ invariant as all non-invariant 
contributions die out. This can be paraphrased by saying that the
thermodynamic limit provides a {\it partial invariant mean}, that is a mean 
which is invariant only on the before-mentioned class of `selected' 
noninvariant observables. 

Finally we consider the counterpart of the Osterwalder-Schrader reconstruction
in this context; here it is important not only to consider the regular 
observables but the full algebra $\cC_b$. For the compact chain one 
recovers (a lattice analogue of) the quantum mechanics of a  particle 
moving on a sphere, as expected. In the hyperbolic case, however, the 
reconstructed quantum theory is different from that of a particle moving 
on $\H$: whereas the former has purely continuous spectrum, the latter has 
at least some point spectrum. The reconstructed Hilbert space turns out 
to be nonseparable and the reconstructed quantum theory can be viewed as 
an interacting (though quantum mechanical) version of the ``polymer 
representations'' of the Weyl algebra studied in other contexts 
\cite{LMS,poly1,poly2}. Consistent with these results we find that the 
symmetry breaking disappears in the limit of a flat target space, when 
the symmetry group $\R^2$ becomes amenable again.

For the organization of the rest of the article we refer to the table of
contents.  
%The rest of the article is organized as follows. 
%In the next section we 
%introduce the (iterated) transfer matrix and use its asymptotics in the 
%limit of large separations to identify the ground states. Expectation 
%values for a chain of finite length with various bc are studied in 
%Section 3. The thermodynamic limits for the algebras of observables 
%outlined are constructed in Section 4. Finally these infinite volume 
%expectations are used as the basis for the Osterwalder-Schrader 
%reconstruction.   

%%%%%%%%%%%%%%%%%%%%%%%%%%%%%%%%%%%%%%%%%%%%%%%%%%%%%%%%%%%%%%%%%%%%%%%%%%%
\newpage 
\newsection{The transfer matrix} 

The hyperbolic spin chain can be regarded as a dynamical system in the 
sense outlined above, with the observables being operators on a Hilbert 
space. On the other hand, in the classical statistical interpretation the 
algebra of observables is a suitable algebra of functions over (direct 
products of) $\H$ which we detail in Section 3. We represent $\H$ as 
the hyperboloid
$\H = \{ n \in \R^{1,2}\,|\, n \cdot n =1\,,n^0 > 0\}$, where  
$a\cdot b = a^0 b^0 - a^1 b^1- a^2b^2$ is the bilinear form on 
$\R^{1,2}$. The time evolution of the spin chain is governed by the 
transfer matrix $\T^x$, $x \in \N$, which we study first. The symmetry 
group $G$ is ${\rm SO}_0(1,2)$ which acts unitarily via the (left) 
quasiregular representation $\rho$ on $L^2(\H)$, i.e.~$\rho(A) \psi(n) = 
\psi(A\inv n),\; A \in {\rm SO}_0(1,2)$. Since we use the identity 
component exclusively we write ${\rm SO}(1,2)$ for ${\rm SO}_0(1,2)$. The 
time evolution commutes with the group action 
\begin{equation} 
\T^x \circ \rho = \rho \circ \T^x\,,\quad x \in \N\,,
\label{Tinv} 
\end{equation}  
as required. In the following we analyze the spectrum, the eigenfunctions, 
and the large $x$ limit of $\T^x, \, x \in \N$, in terms of its 
integral kernel $\cT_{\beta}(n\cdot n';x)$. Some results from the harmonic 
analysis on $\H$ are needed which we have collected in appendix A and use 
freely in the following.      

%%%%%%%%%%%%%%%%%%%%%%%%%%%%%%%%%%%%%%%%%%%%%%%%%%%%%%%%%%%%%%%%%%%%%%%%%%%%
\newsubsection{Spectrum and integral kernel of $\T^x$}

The basic (1-step) transfer matrix acts on $L^2(\H)$ and is defined by 
\begin{equation} 
(\T \psi)(n) = \int \! d \Omega(n') \,\mbox{$\frac{\beta}{2\pi}$} 
e^{\beta(1 - n\cdot n')} \psi(n') \,.
\label{Tmatrix} 
\end{equation} 
From (\ref{psibasis}), (\ref{VUsinglet}), one infers that the functions 
$\eps_{\om,k}$ and $\eps_{\om,l}$ defined in (\ref{psis}) and 
(\ref{psils}) are exact generalized eigenfunctions of $\T$ with 
eigenvalues  
\begin{equation} 
\lb_{\beta}(\om) = \sqrt{\frac{2 \beta}{\pi}} \,e^{\beta} 
K_{i\om}(\beta) <1\,.
\label{lb}
\end{equation} 
The eigenvalues are even functions of $\om$ with a unique maximum at 
$\om=0$ (but only $\om\geq$ will appear in the spectral resolution). In 
particular it follows that the operator $\T$ has absolutely continuous 
spectrum given by the generalized eigenvalues $\lambda_\beta(\om)$; the 
spectrum covers an interval
$[-q,\lb_\beta(0)]$ with $0<q<1$ and is infinitely degenerate.
It is interesting to note that, although real and bounded above by $1$, 
the generalized eigenvalues are positive only for $0 < \om < \om_+(\beta)$, 
where $\om_+(\beta)$ increases with $\beta$ like $\om_+(\beta) \sim 
\beta + {\rm const}\, \beta^{1/3}$. For $\om > \om_+(\beta)$ 
the behavior of $\lb_{\beta}(\om)$ is oscillatory with exponentially 
decaying amplitude  
\begin{equation} 
\lb_{\beta}(\om) \sim \sqrt{\frac{\beta}{\om}} 
e^{-\frac{\pi}{2} \om + \beta} \;2 \sin\Big[ \frac{\pi}{4} + 
\om\Big(\ln \frac{2 \om}{\beta} -1 \Big) \Big] 
\sspace \mbox{as} \quad \om \ra \infty \,.
\label{lboscillations} 
\end{equation} 
The fact that some of the spectrum of the transfer operator is negative
means that there is no reflection positivity under reflections between 
the lattice points. However positivity of the eigenvalues is 
restored in the continuum limit: introducing momentarily the lattice 
spacing $a$, physical distances $x_{\rm phys} = x a$,
as well as a coupling $g^2= 1/(\beta a)$ one has 
\begin{equation} 
\lim_{a \ra 0} \; [\lb_{\frac{1}{g^2 a}}(\om)]^{\frac{x_{\rm phys}}{a}} = 
\exp\Big\{ - x_{\rm phys} \frac{g^2}{2}
\Big(\frac{1}{4} + \om^2\Big) \Big\}\,.
\label{lb_cont}
\end{equation} 
These `eigenvalues' are readily recognized as those of the heat kernel
$\exp(-\frac{g^2}{2}{\bf C} x_{\rm phys})$, see~(\ref{psis}); $1/g$ 
could be removed by rescaling the $n$-fields; $g$ then parameterizes the 
curvature of the hyperboloid.

Besides (\ref{lboscillations}) another feature distinguishing the 
non-compact spin chain from the compact ones is that the iterated 
transfer matrix is bounded but, having continuous spectrum, is not trace 
class. Heuristically this is because due to the invariance (\ref{Tinv}) 
the infinite volume of ${\rm SO}(1,2)$ gets ``overcounted'' in any trace 
operation. 

More precisely we have the following:

{\bf Lemma 2.1.} {\it Let $K$ be a self-adjoint operator on $L^2(\H)$ 
commuting with the unitary representation $\rho$. Then $K$ has only essential 
spectrum, implying that K cannot be compact. In particular $K$ cannot be 
trace class.}

{\it Proof.} Assume that $K$ has an eigenvalue $\lambda$. The 
corresponding 
eigenspace $\cH_\lb\subset L^2(\H)$ then is invariant under the action 
of $\rho$ and therefore the representation $\rho$ can be restricted to a 
unitary subrepresentation $\rho_\lb$. Since ${\rm SO}(1,2)$ is noncompact,  
$\rho_\lb$ is either infinite dimensional or it is a direct sum of 
copies of the trivial representation. But the trivial representation 
cannot be a subrepresentation of $\rho$ since the only functions carrying 
the trivial representation are constants, and thus are not square 
integrable.
\hfill$\blacksquare$

{\it Remark 1.} There is a stronger version of the last statement in the 
proof: the trivial representation also is not even weakly contained in 
the direct integral decomposition of $L^2(\H)$ because ${\rm SO}(1,2)$ is 
not amenable \cite{Pat}. 
 
{\it Remark 2.} As noted above, $\T^x$ has only absolutely continuous 
spectrum.

Since $\T^x$ is not trace class, correlators cannot be defined by  
the usual expressions involving traces. The obvious remedy is
gauge-fixing. This could be done by introducing a damping factor at one site 
and by adopting twisted boundary conditions. Then analytic computations 
are still feasible but are not much different from those in the simpler
gauge fixing approach in which one completely freezes one spin. This 
is the procedure we use in section 3. 

Also the iterated transfer matrix acts as an integral operator on $L^2(\H)$ 
with kernel 
\ba
(\T^x \psi)(n) \is \int \! d \Omega(n') 
\cT_{\beta}(n\cdot n';x) \psi(n') \,,\quad x =1,2,3,\ldots\,,
\nonum
\cT_{\beta}(n\cdot n';x) \is 
\int_0^{\infty} \! \frac{d\om}{2\pi} \,\om \tanh \pi \om \,
\cP_{-1/2 + i\om}(n\cdot n') \,
[\lb_{\beta}(\om)]^x\;,
\label{Txmatrix} 
\end{eqnarray}
where the kernels have the semigroup property
\begin{equation} 
\int d \Omega(n') \cT_{\beta}(n\cdot n';x)  \cT_{\beta}(n'\cdot n'';y)
= \cT_{\beta}(n\cdot n'';x+y)\,.
\label{Txconv}
\end{equation} 
Manifestly the naive expression for the trace, i.e.~the $d \Omega (n)$  
integral over $\cT_{\beta}(1;x)$ does not exist due to the infinite volume 
of $\H$. In passing we note that in terms of the Legendre functions (\ref{Txconv}) 
amounts to the following identity (``projection property'')
\begin{equation} 
\int \! d\Omega(n') \,\cP_{-1/2 + i \om}(n\cdot n') 
\cP_{-1/2 + i \om''}(n'\cdot n'') = 
\frac{ 2\pi \delta(\om - \om'')}{\om \tanh \pi \om}
\;\cP_{-1/2 + i \om}(n\cdot n'')\,,
\label{Pconv}
\end{equation} 
which can also be verified directly from (\ref{Pprop}).
Integral kernels of spectral projections in the proper sense are easily 
obtained by integrating over intervals $I\ni\om$:
\begin{equation}
P_I(n\cdot n') := \int_{\om\in I} \frac{d\om}{2\pi}\,\om\, 
\tanh \pi\om  \,\cP_{-1/2+i\om}(n\cdot n')\,.
\end{equation}
Using Eq. (\ref{Pconv}) one easily verifies for two intervals $I\,,J$ 
\begin{equation}
\int d\Omega(\n') P_I(n\cdot n') P_J(n'\cdot n'')=P_{I\cap J}(n\cdot n'')\,,
\end{equation}
showing that the operators $P_I$ are spectral projections for an interval 
in $\om$ and hence for a corresponding spectral interval for $\T$. 
Absolute continuity of the spectrum follows from the completeness 
relation of the generalized eigenfunctions given in appendix A.

Before proceeding let us note the continuum limit of the iterated transfer matrix.
Using the notation of (\ref{lb_cont}) one has 
\ba
&& \cT_c(\xi; g^2 x_{\rm phys}) := \lim_{a \ra 0} \cT_{\frac{1}{g^2 a}}
\Big(\xi; \frac{x_{\rm phys}}{a} \Big) 
\nonum
&& \quad =  \int_0^{\infty} \! \frac{d \om}{2\pi} \,\om \,
\tanh \pi \om \, \cP_{-1/2 + i \om}(\xi) 
\exp\Big\{ - x_{\rm phys} \frac{g^2}{2}
\Big(\frac{1}{4} + \om^2\Big) \Big\}\, , 
\label{Tcont}
\end{eqnarray} 
where the limit is understood in the strong sense.
With $t = -i x_{\rm phys}$ this is the correct result for the Feynman kernel 
evolving a wave function for time $t$, see 
e.g.~\cite{GroschSt88,Schaefer} and \cite{Campo} for the propagators on 
other homogeneous spaces.. 
Most of the discussion on the large $x$ limit of $\cT_{\beta}(\xi;x)$ below 
transfers directly to the large $x_{\rm phys}$ limit of 
$\cT_c(\xi, g^2 x_{\rm phys})$. 

Clearly for the further analysis the properties of the
transfer matrix (\ref{Txmatrix}) will be crucial. By
(\ref{Tmatrix}) and by iteration of the convolution property
$\xi \ra \cT_{\beta}(\xi;x)$ is a positive function for
all $x \in \N$ and $\beta>0$. For small $x$ it can be evaluated explicitly
\ba
\cT_{\beta}(\xi; 0) \is \frac{1}{2\pi} \delta(\xi -1) \,,
\nonum
\cT_{\beta}(\xi; 1) \is  \frac{\beta}{2\pi}e^{\beta}\,
e^{-\beta\xi} \,,
\nonum
\cT_{\beta}(\xi; 2) \is \frac{\beta}{2\pi} e^{2 \beta}\,
\frac{e^{-\beta \sqrt{2(1+ \xi)}}}{\sqrt{2(1 + \xi)}}\,,
\label{T012}
\end{eqnarray}
with $\xi = n \cdot n' \geq 1$. The fact that $\cT_{\beta}(\xi;2)$ can
be given in closed form could be used to define a coarse grained action
corresponding to decimation of half of the spins. Note also the strictly
monotonic decay in $\xi$, stronger than any power, which is masked by the
rapidly oscillating integrand in (\ref{Txmatrix}). Numerical evaluation
of some $x \geq 3$  transfer matrices suggests that these are generic
features, see Fig.~1.

%%%%%%%%%%%%%%%%%%%%%%%%%%%%%%%%%%%
\begin{figure}[htb]
\leavevmode
\vskip 10mm
\epsfxsize=15cm
\epsfysize=10cm
\epsfbox{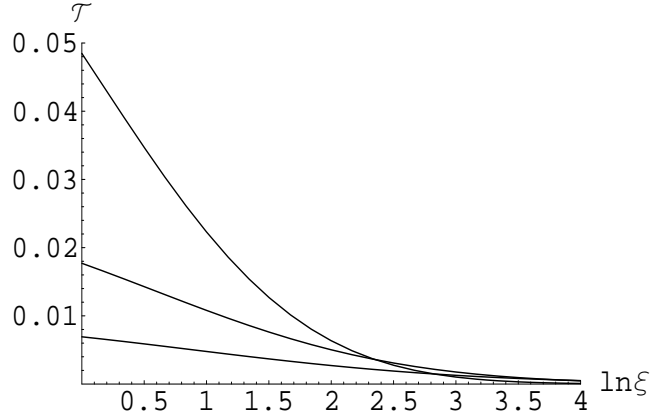}
\vspace{-5cm}

\caption{\small
x-step transfer matrix $\cT_{\beta=1}(\xi;x)$ for $x=3,6,10$,
in order of decreasing slope. Note the non-uniformity:
$\cT_{\beta}(\xi;x+1)$ is smaller/larger than  $\cT_{\beta}(\xi;x)$
for $\xi$ smaller/larger than an intersection point $\xi_x$.
By (\ref{Tmoments}b) the enclosed area is always the same; the value
at $\xi =1$ is the x-site partition function.
}
\label{Txfig}
\end{figure}
%%%%%%%%%%%%%%%%%%%%%%%%%%%%%%%%%%%

We proceed to prove these and some further properties of the kernels of $\T^x$:

{\bf Lemma 2.2.} {\it For fixed $x$ the kernel $\cT_\beta(\xi;x)$ has the 
following properties:
\vspace{-3mm}

\begin{itemize}
\itemsep 3pt
\item[(i)] For any integer $p\ge 0$
\be 
\frac{1}{2\pi} t_{\beta}(p;x):= \int_1^{\infty}\!d\xi\,\xi^p \,
\cT_\beta(\xi;x)<\infty\,.
\label{tmom1}
\end{equation}
\item[(ii)] $\cT_\beta(\xi;x)$ is strictly decreasing in $\xi$ and vanishes for 
$\xi\to\infty$.
\item[(iii)]  $\cT_{\beta}(\xi;x) \leq \cT_{\beta}(1;x) \,\cP_{-1/2}(\xi)$ for
all $\xi \geq 1$. 
\item[(iv)] Let $f: [1,\infty) \to \R^+$ be a strictly positive locally 
integrable function 
satisfying  
\be
\sup_{n\cdot n^\up >K}\frac {f(n\cdot n')} {f(n\cdot n^\up)}\le C 
(n'\cdot n^\up)^p\,,
\label{cond}
\end{equation}
for some constants $p\ge 0$ and $C,K>0$. Then 
\be
\left| \frac {1} {f(n\cdot n^\up)} \int\! d\Om(n') \cT(n'\cdot n^\up;x) 
f(n\cdot n')\right| \le C'\,, 
\end{equation}
with some constant $C'$.
\end{itemize}}

{\it Remark.} Condition (\ref{cond}) holds for any function $f$ with 
power-like growth or decay at $\infty$. This follows from the fact that 
the geodesic distance between two points $n,n'$ behaves aymptotically like 
$\ln (n\cdot n')$ and the globally valid triangle inequality for the 
geodesic distance on $\H$. 

{\it Proof.} (i) The proof proceeds by induction in $x$. Note that 
naively exchanging the order of integrations in (\ref{Txmatrix}) would 
suggest a divergent answer already for the zero-th moment. 

The point to observe is that the convolution property   
(\ref{Txconv}) implies the recursion relation
\ba
\cT_{\beta}(\xi;x+1) \is
\int_1^{\infty} \! du \,j_{\beta}(\xi,u)\,\cT_{\beta}(u;x)\,,
\nonum
j_{\beta}(\xi,u) &:=& \beta e^{\beta(1- \xi \,u)}
I_0(\beta \sqrt{u^2 -1} \sqrt{\xi^2 -1})\,,
\label{Txrecursion}
\end{eqnarray}
where $I_0(u)$ is a modified Bessel function. The kernel
$j_{\beta}(\xi,u)$ has the following properties: its integral wrt to
either variable equals 1; for fixed (not too small) $\xi$ it
is a bell-shaped function of $u$ decaying like
$\exp\{ -\beta (\xi - \sqrt{\xi^2 -1}) u \}/\sqrt{u}$ for
large $u$, and with a single maximum whose position grows linearly
in $\xi$ and whose value decays like $1/\xi$, for large $\xi$.
In particular the troublesome rapidly oscillating
integrand of (\ref{Txmatrix}) is gone. So in the expression defining the 
$\xi$-moments the interchange of the $\xi$ and $u$ integrations is 
legitimate. The $\xi$-integral can be done by repeated differentiation
with respect to $\alpha$ of the formula (\cite{Grad}, p.722)
\be
\int_1^\infty d\xi e^{-\alpha\xi} j_\beta(\xi,u)=\beta e^\beta
\frac 
{e^{-\sqrt{\beta^2+2u\alpha\beta+\alpha^2}}}
{\sqrt{\beta^2+2u\alpha\beta+\alpha^2}}\ =:\ F_\beta(\alpha,u)\,.
\label{besselint}
\end{equation}   
This will be used below to obtain explicit expressions for the low 
moments. Note that both sides of the equation are holomorphic functions of 
$\alpha$ for $|\alpha|<r_0=\beta(u-\sqrt{u^2-1})$, so that we may 
freely differentiate at the origin. By Cauchy's estimate
\be
\left| \left(-\frac{\partial}{\partial\alpha}\right)^p
F_\beta(\alpha,u) \right|_{\alpha=0} \le p!\ r^{-p} M(r,u)\ , 
\label{cauchy}
\end{equation}
where $M(r,u)$ is the maximum of $|F|$ on the cirle $\alpha=r$, 
$r<r_0$. With the choice $r_1=\beta(u-\sqrt{u^2-1/2})$ it is not hard to 
see that the maximum is attained for $\alpha=-r$ -- this follows from 
the fact that the zeros of the quadratic form 
$Q(\alpha):=\beta+2u\alpha\beta+\alpha^2$ are both real and negative, so 
both the real part and the modulus of $Q(r e^{i\phi})$ take on their 
minimal value $\beta^2/2$ for $\phi=\pi$. One concludes from (\ref{cauchy}) 
\be
\left| \left(-\frac{\partial}{\partial\alpha}\right)^p F_\beta(\alpha,u)
\right|_{\alpha=0}
\le p!\ [\beta (u-\sqrt{u^2-1/2}]^{-p} \sqrt{2}\, e^{\beta(1-1/\sqrt{2})}\ .
\label{cauchy2}
\end{equation} 
If we finally use the fact that $u-\sqrt{u^2-1/2}\ge {\rm const} u^{-1}$, 
and insert into the integral defining $t_\beta(p;x+1)$ the convolution 
formula (\ref{Txrecursion}) we obtain 
\be
t_\beta(p;x+1)\le p!\ {\rm const}^p\  t_\beta(p;x)\, 
e^{\beta(1-1/\sqrt{2})}\,.
\end{equation}

Since for $x=1$ all moments exist trivially, this inequality shows the 
existence of all moments for all $x$ and (i) is proven.

(ii) For $x=1$ this is manifest from (\ref{T012}).
For $x>1$ we again proceed by induction. Assuming that $\cT_{\beta}(\xi;x)$ 
is already known to be strictly decreasing, we want to show
\begin{equation} 
\dd_{\xi} \cT_{\beta}(\xi;x + 1) = \int_1^{\infty} \! du \,
\dd_{\xi} j_{\beta}(\xi,u) \, \cT_{\beta}(u;x) 
\,\stackrel{\displaystyle{!}}{<} \,0 \,.
\label{Tdecrease1}
\end{equation}    
This follows from the properties of the kernel $j_{\beta}$, namely 
\begin{equation} 
\dd_{\xi} j_{\beta}(\xi,u) \;\;\left\{ 
\begin{array}{ll} <0 & \mbox{for} \;\; u<u_0(\xi) \,,\\
                  >0 & \mbox{for} \;\; u>u_0(\xi) \,,
\end{array}
\right.
\quad \mbox{and} \quad 
\int_1^{\infty} \! du \, \dd_{\xi} j_{\beta}(\xi,u) =0\;.
\label{Tdecrease2}
\end{equation} 
Using (\ref{Tdecrease2}) one gets for the rhs of (\ref{Tdecrease1})
\ba
&& \int_1^{u_0(\xi)} du \, \dd_{\xi} j_{\beta}(\xi,u) \cT_{\beta}(u;x) +
\int_{u_0(\xi)}^{\infty} du \, \dd_{\xi} j_{\beta}(\xi,u) \cT_{\beta}(u;x)
\nonum
&& \sspace \quad <
\int_1^{\infty}\! du  \,\dd_{\xi} j_{\beta}(\xi,u) 
\;\cT_{\beta}(u_0(\xi);x) =0\,,
\label{jkernel2}
\end{eqnarray}  
where in the first integral $\cT_{\beta}(u;x)$ was replaced by its 
minimum and in the second one by its maximum over the range of integration, 
using the induction hypothesis. Thus $\xi \mapsto \cT_{\beta}(\xi;x)$
is strictly decreasing for all $x$. The vanishing for $\xi \ra \infty$ 
follows from (iii). 

(iii) This is obtained from (\ref{Txmatrix}) and the estimate 
$|\cP_{-1/2 + i \omega}(\xi)| \leq \cP_{-1/2}(\xi)$, which is
manifest from (\ref{Pdef}).

(iv) The proof is an elementary consequence of (i).
\hfill $\blacksquare$

{\it Remark 1.} Iteration of Eq.~(\ref{Txrecursion}) provides an efficient 
way to compute numerically $\cT_\beta(\xi;x)$ for moderately large $x$. 
This was used to produce Figs.~1 and 2. 

{\it Remark 2.} Explicit expressions for the low 
moments are obtained by differentiating (\ref{besselint}) and inserting 
the result in the recursion (\ref{Txrecursion}). This gives
\ba
\label{momrecursion}
p=0: && t_{\beta}(0;x+1) =  t_{\beta}(0;x) \,,
\nonum
p=1: && t_{\beta}(1;x+1) = \Big(1 + \frac{1}{\beta} \Big) 
\,t_{\beta}(1;x)\,,
\\
p=2: && t_{\beta}(2;x+1) = - \frac{1}{\beta^2}(1+ \beta^2)
 \,t_{\beta}(0;x) + \frac{1}{\beta^2} (3 + 3 \beta + \beta^2)
t_{\beta}(1;x)\,,
\nonumber 
\end{eqnarray}
etc. Since for $x=1$ all moments are known
\begin{equation}
t_{\beta}(p;1) = \beta e^{\beta} (-\dd_{\beta})^p (\beta 
e^{\beta})^{-1}\,,
\end{equation}
(which is basically a Laguerre polynomial in $\beta$) the $x$-recursions
can be solved successively for $p=0,1,2,\ldots$. The solution of the
first two is trivial and gives 
\begin{equation}
t_{\beta}(0;x) = 1\,,\sspace t_{\beta}(1;x) =
\Big(1 + \frac{1}{\beta} \Big)^x \;,\quad \;\;\forall \, x \in \N \,.
\label{Tmoments}
\end{equation}
The higher ones won't be needed explicitly.

In summary, the qualitative properties of all the $\cT_{\beta}(\xi;x)$,
$x \in \N$, are very much like the ones exemplified in Fig.~1.  
The rate of decrease becomes softer with increasing $x$ but remains 
faster than any power. The overall scale is set by the maximum 
$\cT_{\beta}(1;x)$, which turns out to 
decrease like $x^{-3/2}$ for large $x$. (This is to be contrasted with the 
flat case of the Euclidean plane $\R^2$, where the decay is only like 
$x^{-1}$). 

%%%%%%%%%%%%%%%%%%%%%%%%%%%%%%%%%%%%%%%%%%%%%%%%%%%%%%%%%%%%%%%%%%%%%%%%%%%%%%

\newsubsection{Large $x$ asymptotics of $\cT_{\beta}(\xi;x)$} 

We next determine the large $x$ asymptotics of 
$\cT_\beta(\xi,x)$. This is of interest because in this limit the 
iterated transfer matrix normally becomes a (generalized) 
projector onto the ground state(s), which can in particular be 
used to identify the latter. In a compact spin chain the kernel of the 
iterated transfer matrix (normalized such that the largest eigenvalue is $1$) 
tends to a constant for $x \ra \infty$, which is indeed the ground state 
(unique eigenstate to the highest eigenvalue) of the transfer matrix. This is a 
reflection of the Mermin-Wagner theorem, i.e.~of the absence
of spontaneous symmetry breaking. The decay in $x$ is exponential
due to the gap in the spectrum. In our noncompact model, since the spectrum 
is gapless, one expects the limit of large separations $x$ to be approached 
power-like rather than exponentially. This is correct, but concerning the 
structure of the limit we are in for a surprise: the large $x$ behavior 
is  {\it not} invariant under the symmetry group ${\rm SO}(1,2)$. Instead 
one finds 
\begin{equation}
\lim_{x\to\infty} {\cT_\beta(\xi;x)\over 
\cT_\beta(1;x)}=\cP_{-1/2}(\xi)\,,
\label{limit} 
\end{equation}
as will be shown below.

So in some sense $\cP_{-1/2}(\xi)$ plays the role of a ground state, but
unlike the compact case, where there is a unique, invariant and
normalizable ground state, in our case we have a whole family of
generalized non-normalizable ground states
$\psi_{n_0}(n)=\cP_{-1/2}(n_0\cdot n)$, spanning a representation
space of ${\rm SO}(1,2)$. We shall explore the consequences of 
(\ref{limit})
in more detail below. However already at this point it is clear that in
this 1D noncompact model the Mermin-Wagner theorem cannot hold in the 
usual
sense.

For later use we also introduce the ${\rm SO}^{\up}\!(2)$ averaged
versions of the iterated transfer matrix  and the corresponding bounds. 
The former is given by
\be
\overline{\cT}_{\!\beta}(\xi,\xi';x) = \frac{1}{2\pi} \int_{-\pi}^{\pi} d
\varphi \,
\cT_{\beta}\Big(\xi \,\xi' - (\xi^2 -1)^{1/2} \,({\xi'}^2 -1)^{1/2}
\cos(\varphi - \varphi');x\Big) \,,
\label{TavgO2}
\end{equation}
where $n = (\xi, \sqrt{\xi^2 -1} \sin \varphi, \sqrt{\xi^2-1} \cos 
\varphi)$, etc.. Note that $\overline{\cT}_{\beta}(\xi,\xi';1) = 
j_{\beta}(\xi,\xi')$ is the convolution kernel in (\ref{Txrecursion}). 

We collect our results on the asymptotics of the kernels 
$\cT_\beta(\xi;x)$ 
and $\overline{\cT}_{\!\beta}(\xi,\xi';x)$, which contain (\ref{limit}) as 
a special case, in the following 

{\bf Proposition 2.3.} {\it The large $x$ asymptotics of $\cT_{\beta}(\xi;x)$ 
is 
governed by the relations: 
\vspace{-4mm}

\begin{itemize}
\itemsep 3pt
\item[(i)]
\begin{equation}
\lim_{x \ra \infty} \frac{\cT_{\beta}(\xi;x-y)}{\cT_{\beta}(\xi';x)}
=\frac{\cP_{-1/2}(\xi)}{\cP_{-1/2}(\xi')}\,\lb_{\beta}(0)^{-y}\,.
\label{TDlimit1}
\end{equation}
\item[(ii)]
\be
\bigg| \frac{\cT_{\beta}(\xi;x)}{\cT_{\beta}(1;x)} - \cP_{-1/2}(\xi) 
\bigg|
\leq {\rm Const} \,(\ln\xi)^2 \, \cP_{-1/2}(\xi)\, 
\frac{c(\beta)}{x}\;,
\end{equation}
where $c(\beta)$ is given below in (\ref{TDlemma})
\item[(iii)]
\be
\lim_{x \ra \infty} \frac{\overline{\cT}_{\!\beta}
(\xi, \xi';x)}{\cT_{\beta}(1;x)} = \cP_{-1/2}(\xi) \cP_{-1/2}(\xi')\,,
\label{TlimitO2}
\end{equation}
\item[(iv)]
\be
 \bigg| 1-
\frac{\overline{\cT}_{\!\beta}(\xi,\xi';x)}%
{\cT_{\beta}(1;x)\cP_{-1/2}(\xi')\cP_{-1/2}(\xi)} \bigg|
\leq \,[\ln^2\xi + \ln^2 {\xi'}]\, O(x^{-1})\;.
\label{limitav}
\end{equation}
\end{itemize}}

The main ingredient in the proof of this proposition is contained in 

{\bf Lemma 2.4.} {\it Let $f$ be an even function of $\om\in\R$, which is 
at 
least twice differentiable at 0 and grows at most polynomially as 
$\om\to\infty$. Then
\ba
&\nspace & \int_0^{\infty} \!d\om\,\om \,\sh \pi \om  f(\om)
\,[\tilde\lb_{\beta}(\om)]^x \sim
\frac{\pi}{[c(\beta)x]^{3/2}} \,
\left[ \sqrt{\frac{\pi}{2}} f(0) + 2 f'(0) [c(\beta) x]^{-1/2}
+ O(x^{-1}) \right]
\nonumber \\[4mm]
&& \bspace \quad \;\;
\mbox{with}\quad  \tilde\lb_{\beta}(\om)=
\frac{\lb_{\beta}(\om)}{\lb_{\beta}(0)} \quad
\mbox{and} \quad c(\beta) =
\dfrac{\int_0^{\infty} dt \,t^2 \exp(- \beta \ch t)}%
{\int_0^{\infty} dt \exp(- \beta \ch t)}\,.
\label{TDlemma}
\end{eqnarray}}

{\it Proof of Lemma 2.4:} The idea of the proof is that the 
contributions of all $|\tilde\lambda_\beta(\om)|$ for $\om>0$ get
exponentially suppressed, because they are less than 1, so only the 
$\om=0$ contribution survives for $x\to\infty$. The leading power
$x^{-3/2}$ arises from the double zero of the integrand at $\om =0$ and
the structure of $\tilde{\lb}_{\beta}(\om)$, which has a unique maximum at
$\om=0$. In more detail (\ref{TDlemma}) one applies the
Laplace expansion (see e.g.~\cite{Olver}) to the kernel
$\exp(-x h_{\beta}(\om))$, where $h_{\beta}(\om) = -
\ln\tilde{\lb}_{\beta}(\om)$ is strictly increasing in
$0<\om<\om_+(\beta)$ with $h_{\beta}(0)
= h_{\beta}'(0)=0$ and $h_{\beta}''(0) = c(\beta) >0$.
Here $\om_+(\beta)$ is the position of the first zero of
$\lb_{\beta}(\om)$ described after Eq.~(\ref{Tmatrix}).
The fact that $\lb_{\beta}(\om)$ changes sign at $\om_+(\beta)$ is
inconsequential because $|\tilde\lb_\beta(\om)|<1$ also for
$\om\ge\om_+(\beta)$ and the contribution of this region to the integral
is exponentially suppressed. 
\hfill $\blacksquare$

{\it Proof of Proposition 2.3:}
We first prove (ii). To this end set
\ba
D_p &:=& \int_0^\infty \!\frac{d\om}{2\pi} \,\om^{1+p} \tanh\pi\om\,
[\tilde\lambda_\beta(\om)]^x\,,\quad p=0,1,\ldots \,,
\nonum
N &:=& \int_0^\infty \! \frac{d\om}{2\pi} \,\om \tanh\pi\om \,  
[\tilde\lambda_\beta(\om)]^x\,
[\cP_{-1/2+i\om}(\xi)-\cP_{-1/2}(\xi)]\,.
\nonumber
\end{eqnarray}
This is chosen such that $\cT_{\beta}(\xi;x)/\cT_{\beta}(1;x) - 
\cP_{-1/2}(\xi)
= N/D_0$, as is manifest from (\ref{Txmatrix}) and $\cP_{-1/2+i\om}(1)=1$.
On the other hand, using the integral representation (\ref{Pdef}) given in
Appendix A one obtains the bound
\be   
|\cP_{-1/2+i\om}(\xi)-\cP_{-1/2}(\xi)|
\le {\rm Const}\, \om^2 \,\cP_{-1/2}(\xi) \,(\ln\xi)^2\,.
\label{pxi}
\end{equation}
Thus
\be
\bigg| \frac{\cT_{\beta}(\xi;x)}{\cT_{\beta}(1;x)} - \cP_{-1/2}(\xi) 
\bigg|
\leq {\rm Const} \,(\ln\xi)^2 \, \cP_{-1/2}(\xi)\, \frac{D_2}{D_0}\;.
\label{limitder}
\end{equation}
Using again Laplace's theorem, one finds $D_2/D_0 = O((c(\beta)x)^{-1})$,
which establishes (ii) (and therefore also (\ref{limit})).

(i): We apply Lemma 2.4 to $D_0=\cT_\beta(1;x)$ to obtain 
\be
\cT_{\beta}(1;x) \sim
\frac{\sqrt{\pi}}{[2 c(\beta)\, x]^{3/2}} \,\lb_{\beta}(0)^x + \ldots \,.
\label{ZTDlimit}
\end{equation}
Combining (\ref{limit}) with 
\be  
\frac{\cT_{\beta}(\xi;x-y)}{\cT_{\beta}(\xi';x)}
= \frac{\cT_{\beta}(\xi;x-y)}{\cT_\beta(1;x-y)}\,   
\frac{\cT_\beta(1;x-y)}{\cT_\beta(1;x)} \,
\frac{\cT_\beta(1;x)}{\cT_\beta(\xi';x)} \,,
\end{equation}
and (\ref{ZTDlimit}) gives (i).

(iii): This follows from averaging (\ref{TDlimit1}) over 
${\rm  SO}^{\uparrow}\!(2)$ and using (\ref{Pprop}c).

(iv): This is proven similarly as (\ref{limitder}) starting from the 
spectral representation for the kernels $\overline{\cT}_{\!\beta}$, which 
is obtained from (\ref{Txmatrix}) by averaging over the angles using 
(\ref{Pprop}c).

This concludes the proof of Proposition 2.3. 
\hfill $\blacksquare$

We want to mention a stronger bound for which we do not have a 
complete proof.

{\bf Conjecture 2.5.} {\it The following global bound holds for 
all $x \in \N$ and for all $\xi \geq 1$: 
\be
\cT_{\beta}(\xi;x) \leq \cT_{\beta}(1;x) \,\cP_{-1/2}(\xi)\,
E\Big(\frac{\ln \xi}{\sqrt{x}}\Big)\,, 
\label{Tbound}
\end{equation}
for some function $E:\R_+ \ra \R_+$ having finite moments of all orders.} 

{\it Remark.} The asymptotics of (\ref{limitder}) for large $x$ 
suggests that $E(t) = 1 - \frac{c_1}{c(\beta)}t^2 + O(t^3)$, with a 
constant $c_1$  of order unity and $c(\beta)$ as in (\ref{TDlemma}). The 
proposal $E(t) = \exp(-\frac{c_1}{c(\beta)}t^2)$ is thus plausible. 
In the continuum limit (\ref{Tbound}) then reduces 
to a known global bound on the heat kernel (see e.g.~\cite{Anker}), noting that 
the geodesic distance from the origin is ${\rm arccosh}\xi \sim \ln \xi$ 
(for large $\xi$) and $c(\beta) \sim 1/\beta$ for large $\beta$. 
Given (\ref{Tbound}) a similar global bound on 
$\overline{\cT}_{\!\beta}(\xi,\xi';x)$ can be obtained 
from (\ref{Tbound}) by using 
$ \xi \xi'-(\xi^2-1)^{1/2}({\xi'}^2-1)^{1/2}\cos(\varphi-\varphi')\geq 
\xi [\xi'-({\xi'}^2-1)^{1/2}]$, and then averaging over ${\rm SO}^\up(2)$.

Let us now explore the consequences of (\ref{limit}) in more detail.
Consider the map $\psi \mapsto P \psi$
\be
(P\psi)(n) := \int \! d\Omega(n') \,\cP_{-1/2}(n \cdot n') \psi(n')\,.
\label{Pop}
\end{equation}  
As map from $L^2(\H)$ to itself this would have only the null
vector in its domain, because it maps even strongly decreasing functions
$\psi$ into functions with a decrease so
slow that they are not square integrable. But it may be regarded for
instance as a map from the test function space $\cS$ into its dual space 
$\cS'$ (see appendix A). However, the range of $P$ does not intersect its
domain of definition, so the map (\ref{Pop}) cannot even be iterated.
This is in strong contrast to the situation in a compact model, where the
corresponding operator is a well defined projection onto the
1-dimensional subspace of constant functions. Here, on the other hand,
the image $(P\psi)(n)$ is in general not even invariant under
some ${\rm SO}(2)$ subgroup. The Fourier transform of $P\psi$ can be
defined nevertheless in a distributional sense. Using
Eq.~(\ref{psilbasis}) and one finds $\om \tanh \pi \om \, 
\widehat{P\psi}(\om,l)
= 2 \pi \delta(\om) \widehat{\psi}(0,l)$, consistent with the
picture that  the limit (\ref{limit}) lowers the `energy' as much as
possible.

There are two further important properties that encode the  
ground state property of $\cP_{-1/2}(n\cdot n')$. The first one is

{\bf Lemma 2.6.} {\it Let $K$ be an ${\rm SO}(1,2)$ invariant integral 
operator with kernel $\kappa(n\cdot n')$, $\kappa\in L^1(\xi^{-1/2}\ln\xi 
d\xi)$. Then $P\psi$ is a generalized eigenfunction of $K$ with eigenvalue 
$\widehat \kappa(0)$; explicitly
\be
\label{eigenf}
\int \!d\Omega(n') \kappa(n\cdot n') \cP_{-1/2}(n'\cdot n'') =
 \widehat{\kappa}(0) \cP_{-1/2}(n\cdot n'')\,.
\end{equation}}

{\it Proof.} Applying the Mehler-Fock transformation (\ref{MehlerFock}) 
for $\kappa$ and the convolution formula for the Legendre functions the 
left hand side becomes
\ba
&&\int d\Omega(n')\int_0^\infty \! \frac{d\om}{2\pi} \, \om \tanh(\pi\om)
\widehat{\kappa}(\om)\cP_{-1/2+i\om}(n\cdot n')\cP_{-1/2}(n'\cdot
n'')\nonum
&&= \int_0^\infty \!d\om \delta(\om) \,\hat\kappa(\om)
\cP_{-1/2+i\om}(n\cdot n'')
= \widehat{\kappa}(0)\,\cP_{-1/2}(n\cdot n'')\,,
\end{eqnarray}
(where the integral and the $\delta$ function have to be interpreted   
suitably to include $\om=0$).  
The $L^1$ condition ensures that $\kappa \cP_{-1/2}$ is integrable from 
$1$ to $\infty$; this follows from the global bound $\cP_{-1/2}(\xi) 
\leq (1+\ln\xi)/\sqrt{\xi}$, valid for all $\xi \geq 1$.   
\hfill $\blacksquare$

Taking for $K$ the iterated transfer matrix one has in particular
$\T^x P \psi = \lb_{\beta}(0)^x P \psi$. For $x=1$ this gives explicitly 
\begin{equation}
\int_1^{\infty} \! du \,j_{\beta}(\xi,u)\,\cP_{-1/2}(u)
= \lb_{\beta}(0)\, \cP_{-1/2}(\xi) \;,
\label{TDlimit2}
\end{equation}
using (\ref{lbP}) and the fact that $j_{\beta}$ is the ${\rm
SO}^{\up}\!(2)$
average of $\cT_{\beta}(n''\cdot n;1)$. Thus $\cP_{-1/2}$ is also an  
eigenfunction of the recursion relation (\ref{Txrecursion}) with the 
correct eigenvalue.

The second property is the `cyclicity' of the function      
$\psi_{\up}(n):=\cP_{-1/2}(n \cdot n^\up)$ for the ${\rm
SO}^{\up}\!(2)$
invariant subspace of $L^2(\H)$ under the action of ${\rm
SO}^{\up}\!(2)$ invariant
operators. See appendix A for an explicit description of this subspace
and the ${\rm SO}^{\up}\!(2)$ invariant operators acting on it. We
repeat that
the ${\rm SO}^{\up}\!(2)$ denotes the stability subgroup of
$n^\up =(1,0,0)$.
The cyclicity of $\psi_{\up}$ then follows trivially   
from the fact that $\cP_{-1/2}$ does not vanish anywhere, so any  
${\rm SO}^{\up}\!(2)$
invariant element $\psi\in L^2(\H)$ can be obtained by acting on it with a
multiplication operator. On the other hand $\psi_{\up}$ has no nice
properties
with respect to operators that are not ${\rm SO}^{\up}\!(2)$ 
invariant.
Defining $\overline{P}$ as the integral operator with kernel
$\cP_{-1/2}(n^{\up}\cdot n) \cP_{-1/2}(n^{\up} \cdot n')$ one 
has
$(\overline{P}\psi)(n) = \cP_{-1/2}(n^{\up} \!\cdot \!n) \,C_{\psi}$ 
for
some constant $C_{\psi}$. As with $P$ one needs sufficiently
strong falloff of $\psi$ for this to be well defined and the image is 
again
not an element of $L^2(\H)$.  As expected, $\overline{P}$ automatically
projects out the part of a wave function lying in the orthogonal 
complement of
the ${\rm SO}^{\up}\!(2)$ invariant subspace.

%%%%%%%%%%%%%%%%%%%%%%%%%%%%%%%%%%%%%%%%%%%%%%%%%%%%%%%%%%%%%%%%%%%%%%%%%%%%%%%%%%%
\newpage 
\newsection{Expectation functionals for finite length}

Since the transfer operator is not trace class the overall ${\rm SO}(1,2)$ 
invariance has to be (`gauge-') fixed already for a chain of finite length.
We do this by keeping the spin at one end of the chain fixed and impose 
various boundary conditions at the other end. Expectation functionals 
(mapping observables, i.e.~functions of the spins into complex numbers) 
then are always well defined. However in the limit of infinite length interesting 
statements can only be made about certain subalgebras of observables which we 
also introduce here. As before, ${\rm SO}^{\up}\!(2)
\subset {\rm SO}(1,2)$ denotes the stability group of the vector $n^\up$.

\newsubsection{Boundary conditions and algebras of observables}

We consider chains of length $2L+1$, with sites
$x=-L,L+1, \ldots L-1, L$, and spins $n_x$ on them, in order to make the 
boundary go to infinity as $L\to\infty$, so as to obtain an infinite 
volume Gibbs state for the chosen action. As discussed earlier, 
some `gauge fixing' is needed, which is accomplished conveniently by fixing
the spin at the left boundary of our chain: $n_{-L}=n^{\up} = 
(1,0,0)\in \H$. At the other end we consider the following choices: fixed  
(Dirichlet) bc $n_L=An_{-L}$, with $A\in  {\rm SO}(1,2)$, or free bc 
(integrating with the invariant measure of $\H$ over $n_L$). We refer to 
Dirichlet bc with $A=\1$ as `periodic bc' and with $A\ne \1$ as 
`twisted bc'. The fixing of the spin $n_{-L}$ avoids the overcounting  
of the infinite volume of  $\H$ induced by the invariance (\ref{Tinv}); 
since the associated Faddeev-Popov determinant is just $1$, it is
justified to refer to fixed bc with $A=\1$ as `periodic bc'. 
In some cases a nontrivial twist matrix would explicitly break the otherwise
manifest ${\rm SO}^{\up}\!(2)$ invariance. In those cases we shall average 
$n_L$ over an ${\rm SO}^{\up}\!(2)$ orbit, thereby maintaining the 
${\rm SO}^{\up}\!(2)$ invariance of the bc.

We consider several classes of observables, all of which consist of 
functions of finitely many spins. They form algebras with addition and 
multiplication defined pointwise. Of particular interest is the algebra 
$\cC_b$ of bounded continuous functions on direct products of $\H$. 
Equipped with the sup-norm and completed with respect to this norm, 
this is a commutative $C^*$-algebra, and the expectation functionals 
constructed later fit the usual concept of a `state' $\om$ as a normalized 
positive (and therefore bounded) functional on the observable algebra, see 
e.g.~\cite{Ruelle}. More generally we consider the $*$-algebra $\cC_p$ of 
polynomially bounded functions. For the construction of expectation 
functionals we introduce a system of subsets of $\cC_p$, closed under a 
suitable norm and designed such that explicit results for thermodynamic 
limit can be obtained. 

It turns out that the expectations of a multilocal observables 
$\cO \in \cC_p$ can always be expressed in terms of a kernel $K^{\cO}$ 
associated with $\cO$ as follows:

{\bf Definition 3.1.} For $\cO\in\cC_p$ and $\ell\ge 2$ set 
\be
K^{\cO}(n, n') := \int\prod_{i=2}^{\ell-1} d\Omega(n_i)
\cO(n_1,\dots,n_\ell)\prod_{i=2}^\ell\cT_\beta(n_{i-1}\cdot
n_i;x_i-x_{i-1})\,, 
\label{kappaO}
\end{equation}
where $n_1=n$ and $n_\ell=n'$. For observables $\cO$ depending only on one 
spin set 
\be
K^{\cO}(n,n'): = \cO(n) \,\delta(n,n'),
\end{equation}
where $\delta(n,n')$ is the delta-distribution (point measure) 
concentrated at $n = n'$, defined with respect to the measure $d\Om$.

{\bf Lemma 3.2.} {\it The assignment $\cO \mapsto K^{\cO}$ mapping 
observables $\cO \in \cC_p$ into integral operators $K^{\cO}$ on 
$L^2(\H)$ with kernel (\ref{kappaO}) has the following properties: 
\begin{itemize}
\itemsep 3pt
\item[(i)] let $\cA,\cB\in\cC_p$ be two observables of ordered non-overlapping 
`support', i.e. $\cA$ depends on  $n_{x_1}, \ldots n_{x_k}$ and $\cB$ 
on $n_{x_{k+1}}, \ldots n_{x_\ell}$ with  $x_{k+1}\geq x_k$; then 
\ba
\label{kappaconv}
&& K^{\cA \cB} = K^{\cA} \,\T^{x_{k+1} - x_k} \,K^{\cB}\;,
\\
&& K^{\cA \cB}(n_1,n_{\ell}) = 
\int \! d \Omega(n)d\Omega(n')\, 
K^{\cA}(n_1,n) \, \cT_{\beta}(n\cdot n';x_{k+1}-x_k) 
K^{\cB}(n',n_{\ell}) \,,
\nonumber
\end{eqnarray}
where $(\cA \cB)(n_{x_1}, \ldots, n_{x_{\ell}}) = 
\cA(n_{x_1},\ldots,n_{x_k}) \, \cB(n_{x_{k+1}},\ldots,n_{x_{\ell}})$, 
$k,\ell-k \geq 2$. If $x_{k+1} = x_k$, the transfer matrix 
$\cT_{\beta}(n\cdot n';0)$ is interpreted as $\delta(n,n')$.  
\item[(ii)] The action of ${\rm SO}(1,2)$ on $\cC_p$, 
i.e.~$\rho(A)\cO(n_{x_1},\ldots,n_{x_l}) 
= \cO(A \inv n_{x_1}, \ldots , A \inv n_{x_{\ell}})$ induces an action on 
the kernels 
\be 
K^{\rho(A)\cO}(n_1,n_{\ell}) = 
K^{\cO}(A\inv n_1, A \inv n_{\ell}) \,.
\label{kapparho}
\end{equation}
\item[(iii)] For the unit $\1 \in \cC_p$ one has: 
$K^{{\scriptsize \1}}(n_1,n_{\ell}) = 
\cT_{\beta}(n_1\cdot n_{\ell} ; x_{\ell} - x_1)$.
\end{itemize}}

{\it Proof.} This is a straightforward computation.

{\it Remark 1.} The last property also implies that the correspondence 
$\cO \mapsto K^{\cO}$ is unique only for the equivalence classes 
obtained by inserting into a given $K^{\cO}$ extra powers of $\T$. 
For example taking in (i)  
for $\cA=\1$  one obtains $K^{{\scriptsize \1}\cB} 
= \T^{x_{k+1} - x_1} K^{\cB}$. In the multipoint functions 
this just means that not all of the `unobstructed' integrations 
have been performed. We shall therefore usually work with a 
reduced representative, i.e.~one which cannot written in 
the form $\T^{y_1} K^{\cA_1} \T^{y_2} K^{\cA_2} \ldots $ with some 
smaller $y_1,y_2, \ldots \geq 0$.    

{\it Remark 2.} For observables depending only on one spin neither 
(\ref{kappaO}) nor (\ref{kappaconv}) are directly applicable. 
However the assignment
$K^{\cO}(n_1,n_2) = \cO(n_1) \,\delta(n_1,n_2)$ is compatible 
with the formulas for the 1-point functions (\ref{1pt_per}), 
(\ref{1pt_free}) and the convolution (\ref{kappaconv}), provided we 
associate $n_1$ and $n_2$  with the same lattice point.   

We now introduce various classes of observables, where the ${\rm SO}^\up(2)$ 
average of an observable $\cO$ is denoted by $\overline \cO$.

{\bf Definition 3.3.}
\vspace{-1mm}

\begin{itemize}
\item[(i)]  An observable $\cO = \overline{\cO} \in\cC_p$ is called 
{\it invariant} if
\be
[K^{\cO}, \rho] =0\,,
\end{equation}
i.e.~$\overline{\cO}(A n_1, \ldots , A n_{\ell}) = 
\overline{\cO}(n_1, \ldots , n_{\ell})$ for all $A \in {\rm SO}(1,2)$.\\ 
The set of these observables is denoted by $\cC_{\rm inv}$.
\item[(ii)] An observable $\cO\in\cC_p$ is called {\it asymptotically 
invariant} if 
\be
\lim_{A\to\infty} \rho(A) [K^{\overline\cO}, \rho] =0\ . 
\label{ainv}
\end{equation}
The set of these observables is denoted by $\cC_{\rm ainv}$.
\item[(iii)] An observable $\cO\in\cC_p$ is called {\it translation invariant} if 
\be
[K^{\overline\cO}, \T] =0\ .
\end{equation}
The set of these observables is denoted by $\cC_{\rm \T\, inv}$.
\item[(iv)] An observable $\cO\in\cC_p$ is called {\it asymptotically translation 
invariant} if 
\be
\lim_{A \ra \infty} \rho(A)[ K^{\overline\cO}, P] =0\ . 
\label{tainv}
\end{equation}
The set of these observables is denoted by $\cC_{\rm  \T \,ainv}$.
\end{itemize}
In (\ref{tainv}) $P$ is the integral operator (\ref{Pop}). Both in 
(\ref{ainv}) and (\ref{tainv}) $A \ra \infty$ refers to a sequence of 
${\rm SO}(1,2)$ transformations such that $\Vert A \Vert \ra \infty$, and 
the commutator has to obey some decay condition detailed in the next 
section (Definitions 4.2 and 4.5).

These subsets of $\cC_p$ are related as follows:
\be 
\begin{array}{ccc}
\cC_{\rm \T \, inv} & \subset&  \cC_{\rm \T \, ainv} \\[1mm]
\cup                &        & \cup                 \\
\cC_{\rm inv} & \subset&  \cC_{\rm ainv} \\
\end{array}
\label{subalgebras}
\end{equation} 
where all inclusions are proper. 

{\bf Definition 3.4.} The sets $\cC^\up_{\rm ainv}$, $\cC^\up_{\rm \T 
\,inv}$ and 
$\cC^\up_{\rm \T \,ainv}$ are defined as the ${\rm SO}^{\up}\!(2)$ 
invariant subsets of $\cC_{\rm ainv}$, $\cC_{\rm \T \,inv}$ and 
$\cC_{\rm \T \,ainv}$, respectively. 

Of course the inclusion relations are preserved and the counterpart of the 
diagram (\ref{subalgebras}) remains valid for the  ${\rm SO}^{\up}\!(2)$ 
invariant subsets. 

%%%%%%%%%%%%%%%%%%%%%%%%%%%%%%%%%%%%%%%%%%%%%%%%%%%%%%%%%%%%%%%%%%%%%%%%%%%%%%%%%%%%

\newsubsection{Expectation functionals} 

The expectation functionals for finite $L$ are defined by explicitly given 
measures and for the largest class of observables $\cC_p$. For states over 
$\cC_b$ it follows from the general though not very constructive
Banach-Alaoglu theorem \cite{ReedS} that thermodynamic limits always exist. 
The system of algebras (\ref{subalgebras}) is designed to make useful and
explicit statements about the limit, even for unbounded observables. 
Sometimes we refer to the expectation values as `correlators' by a common 
abuse of language.  
 
With twisted bc the finite volume average of an observable $\cO(\{n\})$ 
is then defined as 
\ba
\bra \cO \ket_{L,\beta,\alpha} = \frac{1}{Z_{\beta,\alpha}(2L)} 
\int \prod_{x=-L}^{L-1} d \Omega(n_x) \,\mbox{$\frac{\beta}{2\pi}$} 
e^{\beta(1 - n_x\cdot n_{x+1})} \,\cO(\{n\}) \, \delta(n_{-L}, 
n^{\up})\,,
\label{Oavg} 
\end{eqnarray}
Here we anticipate 
that in the cases of interest the dependence on the twist matrix $A$ is only 
through the scalar product $n^{\up}\cdot n_L$ or equivalently the 
``twist parameter'' $\alpha := {\rm arcosh} \,n^{\up} \!\cdot \!n_L \geq
0$. $Z_{\beta,\alpha}(2L)$ is the  partition function normalizing the 
averages, 
$\bra \1 \ket_{L,\beta,\alpha} =1$. The technique to evaluate 
expressions like (\ref{Oavg}) is well known from the compact models: one
uses the semigroup property (\ref{Txconv}) to perform all integrations not 
`obstructed' by the variables in $\cO(\{n\})$. 
For the partition function there are no obstructions and one readily finds
\begin{equation} 
Z_{\beta,\alpha}(2L) = \cT_{\beta}(\ch \alpha ;2L)\;.
\label{Z}
\end{equation}       
For the expectation value of some multilocal observable $\cO$ one has 

{\bf Proposition 3.5.} (twisted bc): {\it For $\ell\ge 2$ 
\begin{eqnarray}
\label{lpoint_twist}
&& \langle \cO(n_{x_1},\ldots n_{x_{\ell}}) \rangle_{L,\beta,\alpha}
\\[1mm]
&& = \frac{1}{Z_{\beta,\alpha}(2L)}
\int \! d\Omega(n_1) d\Omega(n_{\ell})\, \cT_{\beta}(n^{\up}\cdot 
n_1; L+
x_1) K^{\cO} (n_1,n_{\ell})\,
\cT_{\beta}(n_{\ell} \cdot n_L;L-x_{\ell})\,,
\nonumber
\end{eqnarray}   
where $x_1 < \ldots < x_{\ell}$. For $\ell=1$ we have
\be
\bra \cO(n_x) \ket_{L,\beta,\alpha} =  \frac{1}{Z_{\beta,\alpha}(2L)}
\int \!d \Omega(n) \,\cO(n) \cT_{\beta}(n^{\up} \cdot n ; L+ x)
\cT_{\beta}(n\cdot n_L; L-x)\,.
\label{1pt_per}
\end{equation}}
{\it Proof.} This is a simple consequence of (\ref{Oavg}) and the 
definition of $K^\cO$.                         \hfill$\blacksquare$
 
{\it Remark.} As will become clear later for a ${\rm SO}^{\up}\!(2)$ 
noninvariant field $\cO$ one should average $n_L$ over ${\rm 
SO}^{\up}\!(2)$, which amounts to 
replacing $\cT_{\beta}(n_{\ell}\cdot n_L; L- x_{\ell})$ by 
$\overline{\cT}_{\!\beta}(n^{\up}\!\cdot \!n_{\ell}, n^{\up}
\!\cdot \!n_L; L- x_{\ell})$ defined in (\ref{TavgO2}). 
For a field $\cO$ which is ${\rm SO}^{\up}\!(2)$ invariant the replacement is 
an identity. Since the expectation value is taken with a positive  
probability measure, for observables $\cO\in \cC_b$ we have $|\bra \cO 
\ket|\le ||\cO||$ where  $\|\cO\|$ is the supremum norm, and for 
nonnegative $\cO$ the expectation value is nonnegative. Observe also that 
due to the gauge fixing the functions (\ref{lpoint_twist}) are in general 
not translation invariant; we shall later find a simple supplementary condition
which restores translation invariance even at finite $L$. 

For free boundary conditions at $x=L$ the situation is similar:
First note that the partition function with free bc~at $L$ is
\be
Z_{\beta,{\rm free}}(2L)=1.
\label{Zfree}
\end{equation}
This follows from the normalization $\int \!d\Omega(n')\, \cT_\beta(n \cdot 
\n';2 L)=1$ and the semigroup property of 
$\cT_\beta(n\cdot n';x)$, see Eqs.~(\ref{T012}) and (\ref{Txconv}).
Thus the expectation of an observable 
$\cO(\{n\})$ with free bc~at $L$ is simply
\ba
\bra \cO \ket_{L,\beta,{\rm free}} = 
\int \prod_{x=-L}^{L} d \Omega(n_x) 
\,\prod_{x=-L}^{L-1}\mbox{$\frac{\beta}{2\pi}$} 
e^{\beta(1 - n_x\cdot n_{x+1})} \,\cO(\{n\}) \, \delta(n_{-L}, n^{\up})\,.
\label{Oavg_free} 
\end{eqnarray}
Again these expectation values can be rewritten similarly as in Proposition 
3.5:

{\bf Proposition 3.6.} (free bc): 
{\it For $\ell\ge 2$
\be
\label{lpoint_free}
\langle \cO(n_{x_1},\ldots n_{x_{\ell}}) \rangle_{L,\beta,{\rm free}}
= \int \! d\Omega(n_1) d\Omega(n_{\ell})\, \cT_{\beta}(n^{\up} \cdot 
n_1;L+ x_1)
\,K^{\cO}(n_1,n_{\ell}) \,,
\end{equation}
where again $x_1 < \ldots < x_{\ell}$ and $K^{\cO}$ is as in
(\ref{kappaO}). For $\ell =1$ 
\be
\bra \cO(n_x) \ket_{L,\beta,{\rm free}} =
\int \! d \Omega(n) \,\cO(n) \cT_{\beta}(n^{\up} \cdot n ; L+ x)\,.
\label{1pt_free}   
\end{equation}}
{\it Proof.} Again a simple consequence of (\ref{Oavg_free}) and the 
definition of $K^\cO$.                     \hfill$\blacksquare$

{\it Remark 1.} By comparing Eqs (\ref{lpoint_free}) and 
(\ref{lpoint_twist}) 
one sees that the expectation values of observables with free and twisted 
bc are related by 
\be 
\int \! d \Omega(n_L) \, \cT_{\beta}(n^{\up}\cdot n_L;2L) \, 
\bra \cO \ket_{L,\beta,\alpha} = \bra \cO\ket_{L, \beta,{\rm free}}\,.
\label{pervsfree}
\end{equation}
In other words for finite $L$ the free expectation is some kind of 
weighted average over the twisted expectations. In the thermodynamic 
limit this is no longer true, as we will find below.

{\it Remark 2.} Due to (\ref{kapparho}) a ${\rm SO}(1,2)$ transformation
on the observable can always be compensated by a change in the bc
\be
\bra \rho(A) \cO \ket_{L,\beta,{\rm bc}} = 
\bra \cO \ket_{L,\beta,{\rm bc}}\Big|_{n^{\up} \ra A\inv n^{\up},\,
n_L \ra A\inv n_L}\,.
\label{rhobc}
\end{equation} 
Of course our interest will be in the invariance or noninvariance of the 
expectations when the bc are kept fixed as $L\to\infty$.

For translation invariant observables the expectation values can be 
simplified. Recall that for $\cO \in \cC_{\rm \T \,inv}$  
\ba 
\label{Otrans}
&& [ K^{\cO},\,\T]=0 \quad \Longleftrightarrow \quad \forall n,n' \in \H
\\[1mm]
&& \int \! d\Omega(n') \, K^{\cO}(n,n') \, \cT_{\beta}(n'\cdot n'';1) 
= \int \! d\Omega(n') \, \cT_{\beta}(n\cdot n';1)\,K^{\cO}(n',n'')\,. 
\nonumber
\end{eqnarray}
For these expressions to make sense, one has to impose some technical 
conditions; it suffices to demand that $K^\cO$ is a bounded operator.
Using the convolution property (\ref{Txconv}) 
it is then easy to show that for translation invariant observables the 
expressions (\ref{lpoint_twist}) and (\ref{lpoint_free}) simplify to 

{\bf Proposition 3.7.} (translation invariant observables):
{\it For $\cO\in \cC_{\rm \T \,inv}$
\ba
\label{lpoint_inv}
\langle \cO(n_{x_1},\ldots n_{x_{\ell}}) \rangle_{L,\beta,\alpha}
\is \frac{1}{Z_{\beta,\alpha}(2L)} 
\int \! d\Omega(n) \, K^{\cO}(n^{\up}, n) 
\cT_{\beta}(n \cdot n_L;2L+x_1 -x_{\ell})\,,
\nonum
\langle \cO(n_{x_1},\ldots n_{x_{\ell}}) \rangle_{\beta,{\rm free}}
\is  \int \! d\Omega(n) \,K^{\cO}(n^{\up}, n) \,.
\end{eqnarray}}

{\it Remark 1.} For twisted bc also the equivalent form of the integrand
$K^{\cO}(n, n_L) \cT_{\beta}(n \cdot n^{\up} ;2L+x_1 -x_{\ell})$
could be used. Observe that these expectations are translation invariant 
already for finite $L$. Moreover for free bc they are $L$ independent
altogether, so that taking the thermodynamic limit becomes trivial.

{\it Remark 2.} For observables whose kernels admit a Fourier expansion 
(\ref{KFT}) a necessary and sufficient 
condition for (\ref{Otrans}) to hold is that expansion takes the form 
\be
K^{\cO}(n,n') = \sum_{l,l' \in \Z} (-)^{l + l'} 
\int_0^{\infty} \! \frac{d\om}{2\pi}\, 
\om \tanh \pi \om \,\widehat{\kappa}_{l,l'}(\om) \,
\eps_{\om,-l}(n)\eps_{\om,-l'}(n')\,.
\label{kappaOtrans}   
\end{equation}
It differs from the most general one in (\ref{KFT}) only by the fact that 
it is diagonal in the energy parameter $\om$, as expected. An important 
special case is when the spectral weight is (up to a sign factor) 
independent of $l_1,l_2$. Due to the addition theorem (\ref{Pprop}c) the 
kernel becomes a function of $n_1 \cdot n_{\ell}$ only. In this case the 
corresponding observables $\overline{\cO}$ can be characterized directly 
as being ${\rm SO}(1,2)$ invariant. 

{\it Remark 3.} As already seen in section 2 the `vacuum structure' can 
be explored by taking the thermodynamic limit of the discrete system. 
Equivalently one 
can first take the continuum limit and then consider its behavior for 
large Euclidean times. The continuum limit of the correlators 
in Propositions 3.5, 3.6, and 3.7 is obtained  by substituting 
\be 
(x_i)_{\rm phys} = a x_i\,,\quad L_{\rm phys} = a L\,,\quad 
\beta = \frac{1}{a g^2}\,,
\label{cont1}
\end{equation}
and taking the limit $a \ra 0$. In view of (\ref{Tcont}) this basically 
amounts to 
replacing $\cT_{\beta}$ by $\cT_c$ everywhere, with the rescaled arguments. 
This procedure yields the additional bonus of restoring reflection 
positivity. 

%%%%%%%%%%%%%%%%%%%%%%%%%%%%%%%%%%%%%%%%%%%%%%%%%%%%%%%%%%%%%%%%%%%%%%%%%%%%%%%%

\newsubsection{Projection onto  ${\rm SO}^{\up}\!(2)$ invariant observables} 

For ${\rm SO}(1,2)$ non-invariant observables we did not assume special 
symmetry properties. It turns out, however, that one needs to consider 
only ${\rm SO}^{\up}\!(2)$ invariant observables (bounded or unbounded) 
since with our gauge fixing ${\rm SO}^{\up}\!(2)$ noninvariant ones are 
effectively projected onto ${\rm SO}^{\up}\!(2)$ invariant ones. In order 
to see this let us 
apply an ${\rm SO}^{\up}\!(2)$ rotation $A(\varphi)$, $A(\varphi) n^{\up} =
n^{\up}$ (with $\varphi$ the rotation angle) to an ${\rm SO}^{\up}\!(2)$ 
noninvariant observable $\cO$. Using (\ref{kapparho}) one finds for $\ell 
\geq 2$
\begin{subeqnarray}
\label{O2lpoint}
&& \langle \cO(A(\varphi) n_{x_1},\ldots, A(\varphi)n_{x_{\ell}}) 
\rangle_{L,\beta,\alpha}
\\[1mm]
&& = \frac{1}{Z_{\beta,\alpha}(2L)} 
\int \! d\Omega(n_1) d\Omega(n_{\ell})\, \cT_{\beta}(n^{\up}\cdot n_1; L+
x_1) K^{\cO} (n_1,n_{\ell})\,
\cT_{\beta}(n_{\ell} \cdot A(\varphi) n_L;L-x_{\ell})\,,
\nonumber
\\[1mm]
&& \langle \cO(A(\varphi) n_{x_1},\ldots, A(\varphi)n_{x_{\ell}}) 
\rangle_{L,\beta,{\rm free}}
\\[1mm]
&& = \int \! d\Omega(n_1) d\Omega(n_{\ell})\, \cT_{\beta}(n^{\up} \cdot n_1;L+ x_1)
\,K^{\cO}(n_1,n_{\ell}) \,,
\nonumber 
\end{subeqnarray} 
and similarly for $\ell =1$. 
For free bc one sees that the dependence on the rotation angle drops out, so
that the expectations with these bc are ${\rm SO}^{\up}\!(2)$ invariant even if the 
observable is not. Equivalently ${\rm SO}^{\up}\!(2)$ noninvariant observables 
have the same expectations as their ${\rm SO}^{\up}\!(2)$ averages. 
For twisted periodic 
bc this is not quite true. However the ${\rm SO}^{\up}\!(2)$ noninvariance of 
(\ref{O2lpoint}) is evidently caused by the noninvariance of the bc. To 
retain the ${\rm SO}^{\up}\!(2)$ invariance of the bc one can average 
$n_L$ over an ${\rm SO}^{\up}\!(2)$ orbit. Then $\cT_{\beta}$ is replaced with  
$\overline{\cT}_{\!\beta}$ in Eq.~(\ref{TavgO2}) and the situation is the 
same as with free bc. In summary, the expectations (\ref{Oavg}) (when $n_L$ is
averaged over an ${\rm SO}^{\up}\!(2)$ orbit) and (\ref{Oavg_free}) for 
finite $L$ are already ${\rm SO}^{\up}\!(2)$ invariant and hence we need not 
distinguish between ${\rm SO}^{\up}\!(2)$ noninvariant and 
${\rm SO}^{\up}\!(2)$ invariant observables. In terms of the algebras 
introduced in section 3.1 a projection $\cC_p \ra \cC_p^{\up}$, takes 
place upon insertion into the expectation functionals. In terms of the 
kernels $K^{\cO}$ the projection amounts to the replacement
\be 
K^{\cO}(n_1,n_{\ell}) \rra \frac{1}{2\pi}
\int_{-\pi}^{\pi} d \varphi \, K^{\cO}(A(\varphi)n_1,A(\varphi)n_{\ell}) 
=: \overline{K}^{\cO}(n^{\up} \cdot n_1, n^{\up} \cdot n_{\ell})\,.
\label{SOproj}
\end{equation}
For later reference let us issue the warning 
\be 
\overline{K}^{\rho(A)\cO}(n^{\up}\!\cdot \!n_1,
n^{\up}\!\cdot \!n_{\ell})
\neq 
\overline{K}^{\cO}(n^{\up}\!\cdot \!A\inv n_1,
n^{\up}\!\cdot \!A\inv n_{\ell})\,,
\label{kappaO2}
\end{equation}
that is, ${\rm SO}^{\up}\!(2)$ averaging does not commute with the 
${\rm SO}(1,2)$ action.

In compact sigma models, where there is no need for gauge fixing, one
can choose invariant bc, so that the expectation of any noninvariant 
observable is equal to that of its group average. By the Mermin-Wagner 
theorem in dimensions 1 and 2 this remains also true in the thermodynamic 
limit, irrespective of the bc used. Here we find an analogous situation 
only with respect to the maximal compact subgroup ${\rm SO}^{\up}\!(2)$, 
singled out by the gauge fixing.    

In contrast, for the full ${\rm SO}(1,2)$ group the expectations of 
noninvariant and of invariant observables {\it cannot} be related by 
group averaging. This is because -- due to the nonamenablity of 
${\rm SO}(1,2)$, such averages (invariant means) do not exist \cite{Pat}. 
Heuristically this can be understood by viewing the group averaging as a 
projector onto the trivial subrepresentation in the direct integral 
decomposition of tensor products of $L^2(\H)$ functions. By the 
nonamenability the trivial representation does not occur, though. This 
lack of amenability is the source of many peculiarities in the vacuum 
structure of the noncompact model.

%%%%%%%%%%%%%%%%%%%%%%%%%%%%%%%%%%%%%%%%%%%%%%%%%%%%%%%%%%%%%%%%%%%%%%%%%%%%%%%%%%%%
\newpage 
\newsection{The thermodynamic limit as a partial invariant mean} 

By the non-amenability of SO(1,2) an invariant mean on $\cC_b$ cannot 
exist; a fortiori this holds for the unbounded functions $\cC_p$. 
It is known, however, that there are subspaces of the 
space of bounded continuous functions on any group, such as the spaces of 
almost periodic or weakly almost periodic functions on which a unique 
invariant mean exists \cite{Pat}. These spaces are defined rather abstractly 
by relative compactness resp.~weak compactness of their orbits under the group action. 
In the following we will introduce concretely a subspace $\cC^{\up}_{\rm ainv} 
\subset \cC_p$ for which there is a unique, {\it invariant}, and explicitly 
{\it computable} thermodynamic limit. The infinite volume averages therefore 
define a `partial invariant mean'. We presume that the bounded subalgebra 
$\cC^{\up}_{\rm ainv} \cap \cC_b$ of our class $\cC^{\up}_{\rm 
ainv}$ (viewed as functions on SO(1,2)) consists of weakly almost 
periodic functions, but not of almost periodic functions (the latter set 
contains only the constant functions \cite{dixmier}).  For the 
construction of the thermodynamic limit we proceed in several steps, 
where we first construct the thermodynamic limit for the algebras 
in the top row of the diagram (\ref{subalgebras}). The limit is shown 
to be explicitly computable and unique (but different) for free and for 
twisted bc. The construction does not require the selection of 
subsequences, i.e.~works without recourse to the Banach-Alaoglu theorem. 
In each case we then proceed to show that this limit is ${\rm SO}(1,2)$ 
invariant for the described subalgebras in the bottom row of the diagram, 
trivially for $\cC_{\rm inv}$ and nontrivially for $\cC^{\up}_{\rm 
ainv}$.

%%%%%%%%%%%%%%%%%%%%%%%%%%%%%%%%%%%%%%%%%%%%%%%%%%%%%%%%%%%%%%%%%%%%%%%%%%%%%%%%%%%%
\newsubsection{Thermodynamic limit for translation invariant observables} 

We begin by studying the thermodynamic limit of translation invariant
observables. The distinction between the bounded observables
and the polynomially bounded observables turns out to be inessential
and we assume $\cO \in \cC_{\rm \T\,inv}$ throughout.
With free bc, as seen in Eq.~(\ref{lpoint_inv}), there is no $L$
dependence left -- so no limit has to be taken. For $\cC_{\rm \T\,inv}$  
expectations defined with twisted bc the existence of an
$L \ra \infty$ limit needs to be established.

First there is a slight complication that needs to be taken care of: 
twisted bc
$n_{L} = A n_{-L}$, $n_{-L} = n^{\up}$, with $A \neq \1$ explicitly 
break
${\rm SO}^{\up}\!(2)$ invariance. 
Since in this study we are interested in the spontaneous symmetry
breaking for the nonamenable ${\rm SO}(1,2)$, we restore the 
${\rm SO}^{\up}\!(2)$ invariance of the bc by performing an average 
of $n_L=An^\up$ over ${\rm SO}^{\up}\!(2)$.
For finite length $L$ the
expectations will then still depend on the `height' $n_L^0 = \ch \alpha$. 
In a slight abuse of terminology we shall keep referring to these bc as
`twisted' ones and also keep the original notation 
$\bra\,.\,\ket_{L,\beta, \alpha}$. Only when a confusion is possible we 
emphasize the additional averaging by denoting the corresponding 
expectations by $\bra\,.\,\ket_{L,\beta, \alpha,{\rm av}}$.

{\bf Proposition 4.1.} {\it For $\cO \in \cC_{\rm \T\,inv}$ and twisted bc 
the thermodynamic limit is given by the equivalent expressions:  
\be
\label{TDtwinv}
\langle \cO(n_{x_1},\ldots n_{x_{\ell}})
\rangle_{\infty,\beta,\alpha}
= \lb_\beta(0)^{x_1-x_\ell} \, 2 \pi \,\int_1^{\infty} \!d\xi \,
\overline{K}^{O}(\xi,1) \, \cP_{-1/2}(\xi)\,.
\end{equation}
\ba
\label{TDtwinvab}
\langle \cO(n_{x_1},\ldots n_{x_{\ell}})
\rangle_{\infty,\beta,\alpha}
&=& \lb_\beta(0)^{x_1-x_\ell} \, 2 \pi \,\int_1^{\infty} \!d\xi \,
\overline{K}^{O}(1, \xi) \, \cP_{-1/2}(\xi)
\\[2mm]
&=& \lb_\beta(0)^{x_1-x_\ell} \, 2 \pi \,\int_1^{\infty} \!d\xi \,
\frac{\overline{K}^{O}(\xi, n^{\up}\!\cdot \!n_L)}%
{\cP_{-1/2}(n^{\up}\!\cdot \!n_L)} \, \cP_{-1/2}(\xi)\,. 
\nonumber
\end{eqnarray}}
{\it Proof.} For Eq.~(\ref{TDtwinv}) we use the ${\rm SO}^{\uparrow}\!(2)$ 
invariance of the bc to replace the transfer matrix $\cT_{\beta}$ by
$\overline{\cT}_{\!\!\beta}$ (see (\ref{TavgO2})) and then $K^{\cO}$ 
by its  ${\rm SO}^{\uparrow}\!(2)$ average (see (\ref{SOproj})). Since by 
assumption the integral  $\int \!d\Omega(n) |K^{\cO}(n,n')|$ exists
one can in the first equation of (\ref{lpoint_inv}) take the $L \ra
\infty$ limit inside the integral. To obtain Eq.~(\ref{TDtwinvab}) one 
uses the fact that for translation invariant observables $\cO$ the 
integral operators $K^{\cO}$ commute with $P$ in (\ref{Pop}).
\hfill $\blacksquare$ 

{\it Remark 1.} There are no elements of $\cC_{\rm \T\,{\rm inv}}$ 
depending 
only on one spin, except constants. For free bc no thermodynamic limit 
has to be taken, see Proposition 3.6.

{\it Remark 2.} In particular Proposition 4.1 is valid for ${\rm SO}(1,2)$ 
invariant observables $\overline{\cO} \in \cC_{\rm inv} \subset 
\cC_{\T\,{\rm inv}}$ where the kernel 
$K^{\overline{\cO}}(n_1,n_{\ell})$ depends only on the invariant 
distance $n_1\! \cdot\! n_{\ell}$. The thermodynamic limit (\ref{TDtwinv}) 
is then independent of the twist  $n_{-L}\cdot 
n_L=n^\up\cdot n_L=\cosh\alpha$, i.e.
\be
\label{indeptw}
\langle \overline{\cO}(n_{x_1},\ldots n_{x_{\ell}})
\rangle_{\infty,\beta,\alpha}=
\langle \overline{\cO}(n_{x_1},\ldots n_{x_{\ell}})
\rangle_{\infty,\beta,0}\,,\quad \overline{\cO} \in \cC_{\rm inv}\,.
\end{equation}
This can be verified directly using the ground state property (\ref{eigenf}).
Indeed, if one does not take the thermodynamic limit in (\ref{lpoint_inv}) with the 
${\rm SO}^{\up}\!(2)$ averaged transfer matrix one obtains initially 
an alternative version of the second Eq.~in (\ref{TDtwinvab})
\be
\langle \cO(n_{x_1},\ldots n_{x_{\ell}})
\rangle_{\infty,\beta,\alpha}
= \frac{\lb_\beta(0)^{x_1-x_\ell}}{\cP_{-1/2}(\ch \alpha)} \, \int 
\!d\Omega(n) \,
\overline{K}^{\overline{O}}(n^{\up} \cdot n) \, 
\cP_{-1/2}(n\cdot n_L)\,,\quad \overline{\cO} \in \cC_{\rm inv}\,. 
\label{indeptw2}
\end{equation}
Averaging over $n_L$ and use of the addition theorem (\ref{Pprop}c) shows
that the dependence on $\alpha$ drops out. Alternatively one can use 
(\ref{eigenf}) to verify (\ref{indeptw}). 

{\it Remark 3.} For generic translation invariant observables the 
infinite volume expectations
are in general not ${\rm SO}(1,2)$ invariant. Rather one finds from 
(\ref{rhobc}) the following induced action on the kernels by $\cO \ra 
\rho(A\inv) \cO$: 
\begin{subeqnarray}
\label{rhokappa}
\overline{K}^{\cO}(1, \xi) & \rra & 
\overline{K}^{\cO}(n^{\up} \!\cdot \!A n^{\up},
n^{\up} \!\cdot \!A n)\,, 
\\[1mm]
\overline{K}^{\cO}(1, \xi) & \rra & 
\frac{\cP_{-1/2}(n^{\up} \!\cdot \!A n_L)}{\cP_{-1/2}(n^{\up} \!\cdot \!n_L)}\,
\overline{K}^{\cO}(n^{\up} \!\cdot \!A n^{\up}, \xi) 
\\[1mm]
\overline{K}^{\cO}(\xi, n^{\up}\!\cdot \!n_L) & \rra & 
\cP_{-1/2}(n^{\up} \!\cdot \!A n^{\up}) \,
\overline{K}^{\cO}(\xi, n^{\up} \!\cdot \!A n_L)\,,
\end{subeqnarray}
where for free bc only (\ref{rhokappa}a) applies while for twisted bc all 
three (equivalent) expressions are applicable. We shall return to 
these formulae later but note already here that observables in 
$\cC_{\rm \T\,inv}\setminus \cC_{\rm inv}$ will in general show 
spontaneous symmetry breaking: 
$\bra \rho(A) \cO \ket_{\infty, \beta, {\rm bc}} \neq   
\bra \cO \ket_{\infty, \beta, {\rm bc}} $. 
  
For the rest of this subsection we now focus on the special case of 
${\rm SO}(1,2)$ invariant observables. Then symmetry breaking is not an issue,
nevertheless the result (\ref{TDtwinv}) is surprising. Besides the mere 
existence of a thermodynamic limit one would of course expect 
that the effect of the different bc is washed out. While we have found 
that the dependence on the twist $\ch\alpha$ actually does disappear, free 
bc~in general give a different thermodynamic limit. In other words, even 
{\it invariant} observables show a {\it dependence on the boundary conditions}, 
even after the boundary is removed to infinity!

To illustrate this consider specifically the usual `spin-spin' two-point
functions with the various bc. For twisted bc the thermodynamic limit is 
obtained from 
(\ref{TDtwinv}) and (\ref{kappaO}) (for $\ell =2$ with
$\overline{\cO}(n_1,n_2) = n_1\cdot n_2$) as  
\begin{equation} 
\lim_{L\ra \infty} \langle n_0 \cdot n_x \rangle_{L,\beta,\alpha} =
\lim_{L\ra \infty} \langle n_0 \cdot n_x \rangle_{L,\beta,0} 
= \frac{2\pi}{\lb_{\beta}(0)^{x}} \int_1^{\infty} \! 
d\xi \, \xi\, \cT_{\beta}(\xi;x) \cP_{-1/2}(\xi)\,. 
\label{tpt_TDtwist}
\end{equation} 
The independence of the twist angle has been seen before to be a general 
feature. However the same expectation with 
free bc at the right end of the chain gives a different result. One finds 
\begin{equation} 
\langle n_0 \cdot n_x \rangle_{\beta,{\rm free}} =   
2\pi \int_1^{\infty} \! d\xi \, \xi\,  
\cT_{\beta}(\xi;x) = \Big( 1 + \frac{1}{\beta} \Big)^x\,,
\label{tpt_TDfree} 
\end{equation} 
using Eq.~(\ref{Tmoments}b) in the second step. As seen generally in
Eq.~(\ref{lpoint_inv}) the correlator is $L$-independent and thus coincides with its 
thermodynamic limit. But this thermodynamic limit is now different from 
the previous one. To make sure that the analytical expressions
(\ref{tpt_TDtwist}) and (\ref{tpt_TDfree}) really define different functions
we evaluated them numerically; the results are shown in Figure 2 below.   
For periodic bc also the approach to the thermodynamic limit is shown, 
which turns out to be nonuniform and extremely slow.

For the `internal energy' $E_{\beta, {\rm bc}} :=\lim_{L \ra \infty} 
\bra \n_0\cdot n_1\ket_{L,\beta,{\rm bc}}$ the discrepancy can be 
seen immediately:
\begin{eqnarray}
\label{Elimit}
E_{\beta, {\rm bc}} \is 1+\frac{1}{\beta}- 
\lim_{L\to\infty}\frac{1}{2L}\frac{\partial}{\partial\beta}
\ln Z_{\beta,{\rm bc}}(2L) 
\nonum
\is \left\{ \begin{array}{ll} 
{\displaystyle 1+\frac{1}{\beta}-\frac{\partial}{\partial\beta}\ln \lb_\beta(0)} & 
\mbox{for twisted bc}, \\[3mm]
{\displaystyle 1 + \frac{1}{\beta}} & \mbox{for free bc}, 
\end{array} \right.
\end{eqnarray}
using Eq.~(\ref{ZTDlimit}) and (\ref{Zfree}), respectively.  

Technically the discrepancy can be traced back to the fact that in
Eq.~(\ref{pervsfree}) the operations `averaging' and `taking the thermodynamic
limit' do not commute, schematically: 
$\lim_{L \ra \infty} \int \!d \Omega(n_L) (\ldots) \neq  
\int \!d \Omega(n_L) \lim_{L \ra \infty}(\ldots) $.
Indeed, the lhs is $L$-independent and equals 
$\bra \overline{\cO} \ket_{\beta,{\rm free}}$ while the integrand and 
hence the integral on the rhs vanishes pointwise. In fact the integrand on 
the right hand side behaves very nonuniformly for $L\to\infty$: for instance 
the two-point function with twisted bc is unbounded as a function of $\alpha$ 
and the convergence as $L\to\infty$ takes place more and more slowly as 
$\alpha$ increases. 
%%%%%%%%%%%%%%%%%%%%%%%%%%%%%%%%%%%
\begin{figure}[htb]
\leavevmode
\vskip 15mm
\epsfxsize=16cm
\epsfysize=10cm
\epsfbox{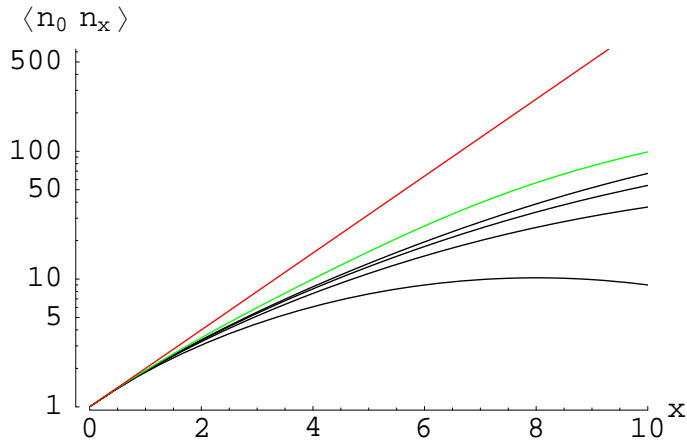}
\vspace{-5.5cm}

\caption{\small
Spin two-point function for $\beta =1$: for periodic bc and $L =8,16,32,64,
\infty$, and for free bc, in order of increasing values at fixed $x$. 
}
\end{figure}
%%%%%%%%%%%%%%%%%%%%%%%%%%%%%%%%%%%

These features  are in sharp contrast to those of the compact ${\rm 
O}(N)$ spin chains where it is well known that all boundary conditions 
yield the {\it same} thermodynamic limit for the correlators of invariant 
as well as noninvariant quantities; see for instance \cite{SeilerY03}. In 
the compact models no gauge fixing is required, but one could fix a spin 
at the boundary just as we did here, and the thermodynamic limit would be 
insensitive to it. This is a consequence of the Mermin-Wagner theorem, 
which holds in this case.

One might suspect that this `long range order' in the non-compact model
reflects the poor choice of observables, i.e.~that the kernel 
$\overline{\cO}(n,n') = n \cdot n'$ does not define an 
operator on $L^2(\H)$ (as explained after Eq.~(\ref{Pasym})).
However the situation is the same for invariant kernels
$\overline{\cO}(n, n') = \kappa(n \cdot n')$ which obey (\ref{MehlerFock2}) 
and which therefore do define integral operators on $L^2(\H)$. 
The thermodynamic limit of the corresponding two-point functions is 
obtained simply by replacing $\xi \cT_{\beta}(\xi;x)$ with 
$\kappa(\xi)  \cT_{\beta}(\xi;x)$ in Eqs.~(\ref{tpt_TDtwist}) and (\ref{tpt_TDfree}). 
These two-point functions will be conventional, decreasing functions of $x$. 
Nevertheless they will in general be different for free and for 
periodic bc.

Another potential problem could be the lack of clustering. However for 
${\rm SO}(1,2)$ invariant observables the situation turns out
to be peculiar -- there is perfect clustering even at finite distance.
Consider two invariant observables, $\cA(n_{x_1},\ldots n_{x_{\ell}})$
and $\cB(n_{y_1},\ldots n_{y_{k}})$ such that
$x_1<\ldots<x_\ell \leq y_1<\ldots<y_k$. We claim that for all bc 
\be
\label{clus}
\langle \cA \cB\rangle_{\infty,\beta,{\rm bc}}=
\langle \cA \rangle_{\infty,\beta,{\rm bc}}
\langle \cB \rangle_{\infty,\beta,{\rm bc}}\,.
\end{equation}
For twisted periodic bc the derivation proceeds along the lines leading to  
Eq.~(\ref{indeptw}) via (\ref{indeptw2}): we define kernels 
$K^\cA$ and $K^\cB$ as above and use the 
ground state property Eq.~(\ref{eigenf}) of $\cP_{-1/2}$. It turns out 
that both sides of Eq.~({\ref{clus}) are equal to the same multiple of 
$\widehat\kappa_\cA(0)\widehat\kappa_\cB(0)$ (and thus in particular are independent of
the twist parameter). For free bc the expectations of invariant observables
are already $L$-independent; the asserted factorization can be seen in a way similar
to the step from (\ref{lpoint_free}) to (\ref{lpoint_inv}).

This `hyperclustering' property is unpleasant, because it means that from 
the correlators of invariant fields one can only reconstruct a one-dimensional 
Hilbert space. The latter is suggested by the fact that all vectors obtained by 
applying invariant kernels to the ground state will by (\ref{eigenf}) be 
proportional to it. Technically it follows from the Osterwalder-Schrader 
reconstruction of the Hilbert space, as detailed in section 5. On the other 
hand this feature is a peculiarity present likewise for other 
one-dimensional spin models, like the compact ${\rm O}(N)$ chains or the 
harmonic chains. In these models, since they are based on amenable 
symmetries, there exists a unique thermodynamic limit also for 
noninvariant correlators and therefore one obtains by the reconstruction 
a nontrivial infinite dimensional Hilbert space. We now show that in the 
noncompact models the situation encountered for invariant observables 
persists for a class of noninvariant ones: for all observables in $ 
\cC_{\rm \T \, ainv}$ for fixed bc a unique thermodynamic limit exists but 
is in general different for periodic and for free bc. The hyperclustering, 
however, does not carry over to those observables, as we will see.

%%%%%%%%%%%%%%%%%%%%%%%%%%%%%%%%%%%%%%%%%%%%%%%%%%%%%%%%%%%%%%%%%%%%%%%%%%%%%%%%%%%% 

\newsubsection{TD limit for asymptotically translation invariant observables} 

We now relax the condition of translation invariance to ``asymptotic 
translation invariance''. It suffices to consider ${\rm SO}^{\up}\!(2)$ 
invariant bc (such as free, periodic, or ${\rm SO}^{\up}\!(2)$ averaged
twisted bc). As explained in section 3.3 this allows one to restrict attention to 
${\rm SO}^{\up}\!(2)$ invariant observables. As before we denote by 
$K^{\cO}$ and $P$ the integral operators with kernels $K^{\cO}(n,n')$ in 
(\ref{kappaO}) and $\cP_{-1/2}(n\cdot n')$, respectively. Similarly 
$[K^{\cO},P](n,n')$ is the kernel of the commutator of $K^{\cO}$ with $P$. 

We give now the precise version of Definition 3.3 (iv):

{\bf Definition 4.2.}~$\cO \in \cC_p$ is called {\it asymptotically 
translation invariant} iff its ${\rm SO}^{\up}\!(2)$ average satisfies
\be 
\label{Tainv_def}
\Big| [K^{\overline\cO}, P](n,n')\Big| \leq p(n^{\up}\cdot n) \,
 p(n^{\up}\cdot n') \,, \sspace p(\xi) \sim \xi^{-1/2} (\ln\xi)^{-3}\,,
\end{equation}
for some fixed $n$  and all $n' \in \H$ or vice versa.  For observables 
$\cO(n)$ depending on a single spin only we define asymptotic translation 
invariance by the condition that their ${\rm SO}^{\up}\!(2)$ average 
$\overline{\cO}(n)$ has a limit as $n \ra \infty$. 

The function $p(\xi)$ needs to be bounded but it is mainly the 
large $\xi$ asymptotics that matters; for definiteness we 
take $p(\xi) = p_1\,\xi^{-1/2}(1+ \ln \xi)^{-3}$, for some $p_1 =p(1) >0$. 

To motivate the terminology ``asymptotically translation invariant'' recall
from section 2 that $P$ can be viewed as a weak limit of transfer operators 
$\T^L/\cT_{\beta}(1;L)$ for $L \ra \infty$. Further, for 
$\cO \in \cC_{\rm \T \,ainv}$ one has 
\be 
\label{Tainv}
\lim_{A \ra \infty} \rho(A) [K^{\cO}, P] = 0\;. 
\end{equation}
Here we assumed that the commutator acts on $L^1$ wave functions so that 
$\rho(A)([K^{\cO},\T]\psi)(n)$ can be bounded by $\frac{\beta}{2\pi}p_1 p(A n^{\up} \cdot n) 
\Vert \psi\Vert_1$. 

The thermodynamic limit for asymptotically translation invariant multi-spin
observables and twisted bc is given by the {\it same} expressions as for 
translation invariant observables: 

{\bf Proposition 4.3.}{\it 
\vspace{-3mm}

\begin{itemize}
\item[(i)] Let $\cO\in \cC_{\rm \T \, ainv}$ be a 1-point observable, 
i.e.~any function of one spin such that its ${\rm SO}^{\uparrow}\!(2)$ 
average has a limit $\overline{\cO}(\infty)$. Then: 
\be
\lim_{L \ra \infty} \bra \cO(n_x) \ket_{L,\beta,{\rm bc}} = 
\overline{\cO}(\infty)\,.
\label{1pt_TD}
\end{equation}
\item[(ii)] Let  $\cO\in \cC_{\rm \T \, ainv}$ be a multi-point 
observable, $\ell \geq 2$. If Conjecture 2.5 holds then:  
\be
\label{TDTainv_twist}
\langle \cO(n_{x_1},\ldots, n_{x_{\ell}})
\rangle_{\infty,\beta,\alpha}
= \lb_\beta(0)^{x_1-x_\ell} \, 2 \pi \,\int_1^{\infty} \!d\xi \,
\overline{K}^{O}(\xi,1) \, \cP_{-1/2}(\xi)\,. 
\end{equation}
\end{itemize}}
{\it Proof.} 
(i)  By definition the ${\rm SO}^\up(2)$ average of $\cO(n)$ has a limit 
$\overline\cO(\infty)$ for $n\to\infty$ (and therefore is a bounded 
function). We write 
\be
\bra \cO(n_x) \ket_{L,\beta, {\rm bc}} = \overline\cO(\infty) +
\int \! d\mu_{L,\beta, {\rm bc}}(n;x) [\cO(n) - \overline\cO(\infty)] \,.
\label{1pt_TDder2}
\end{equation} 
Decomposing the second term into an integral over $n^{\uparrow} \cdot n 
\in
[1, \Lambda]$ and  $n^{\uparrow} \cdot n \in [\Lambda, \infty[$, given
$\epsilon$ choose $\Lambda$ so large that
${\rm sup}|\overline\cO(\overline\infty) - \cO(n)|<\epsilon$,
with the supremum over $n^{\uparrow}\! \cdot \!n \in [\Lambda, \infty[$.
In Lemma 4.7 below is shown that the 1-spin measure of any bounded set 
in $\H$ goes to 0 as $L\to\infty$, so sending $L\to\infty$ the first 
integral vanishes. This shows that the total integral goes to 0 for 
$L\to\infty$ and one obtains
(\ref{1pt_TD}).
 
(ii) The proof is based on a reduction to the case (i) of a one-spin 
observable. It is convenient to write $(A B)(n,n')$ for the kernel of 
$AB$, for any pair of integral operators $A,B$. With this notation one 
starts from 
\be
\bra \cO\ket_{L,\beta,\alpha} = \int \! d\Omega(n_{\ell}) \,
(\T^{L+ x_1} K^{\cO})(n^{\uparrow}, n_{\ell}) \,
\frac{\overline{\cT}_{\beta}(n^{\uparrow} \! \cdot \! n_{\ell},
n^{\uparrow} \! \cdot \! n_L; L-x_{\ell} )}%
{\cT_{\beta}(n^{\uparrow}\!\cdot \!n_L; 2 L)}\,.
\label{der1}
\end{equation}
These multipoint averages can be written as one-point averages as follows
\ba 
&& \bra \cO\ket_{L,\beta,\alpha} = 
\bra \cO_{0,L,\alpha} \ket_{L,\beta,\alpha}\,, \quad \mbox{with} 
\nonum
&& \cO_{0,L,\alpha}(n_0) := \int \! d\Omega(n) \, 
(\T^{x_1} K^{\cO})(n_0,n) \,
\frac{\overline{\cT}_{\beta}(n^{\up} \! \cdot \! n, 
n^{\up} \!\cdot \! n_L;L-x_{\ell})}%
{\overline{\cT}_{\beta}(n^{\up} \! \cdot \! n_0, n^{\up}\! \cdot \! n_L;L)}\,.
\label{der2}
\end{eqnarray}
For the time being (\ref{der2}) is just an identity (Fubini's 
theorem); later we shall 
put it in the context of the Osterwalder-Schrader reconstruction. 
Next we observe that $\cO_{0,L,\alpha}(n_0)$ has a $L\ra \infty$  
limit, pointwise for all $n_0 \in \H$, which is independent of 
the twist parameter $\alpha$ defining $n_L$ modulo 
${\rm SO}^{\up}\!(2)$ rotations:
\ba
&& \lim_{L \ra \infty}  \cO_{0,L,\alpha}(n_0) = \cO_{0,\infty}(n_0)\,,
\quad \mbox{with} 
\nonum
&&  \cO_{0,\infty}(n_0) := \int \! d\Omega(n)\, (\T^{x_1} K^{\cO})(n_0,n) 
\frac{\cP_{-1/2}(n^\up \cdot n)}{\cP_{-1/2}(n^\up \cdot n_0)}\,
\lb_{\beta}(0)^{-x_{\ell}}\,.
\label{der3}
\end{eqnarray}
Here we used Eqs.~(\ref{TlimitO2}). The crucial identity now is 

{\bf Lemma 4.4.} {\it Assume that Conjecture 2.5 holds. Then
\be 
\lim_{L \ra \infty} \bra \cO_{0,L,\alpha}  \ket_{L,\beta,\alpha} = 
\lim_{L \ra \infty} \bra \cO_{0,\infty}  \ket_{L,\beta,\alpha}\,,
\quad \mbox{for all} \;\; \cO \in \cC_b\,.
\label{der4}
\end{equation}}
{\it Proof of Lemma 4.4:} 
We start with the bound
\ba
\label{der5}
&& \Big| \bra (\cO_{0,L,\alpha}-\cO_{0,\infty}) \ket_{L,\beta,\alpha} 
\Big|
\\[2mm]
&& \leq \int \!
d\Omega(n_0)\,\Big|\cO_{0,L,\alpha}(n_0)- \cO_{0,\infty}(n_0)\Big| \,
\frac{
\cT_\beta(n_0\cdot n^\uparrow;L)
\overline{\cT}_{\beta}(n^{\uparrow}\!\cdot
\! n_0,n^{\uparrow}\! \cdot
\!n_L;L)}%
{\cT_{\beta}(n^{\uparrow}\! \cdot \! n_L ;2L)} \,,
\nonumber
\end{eqnarray}

To examine the difference $\vert \cO_{0,L,\alpha}(n_0)- \cO_{0,\infty}(n_0) \vert$ 
we write
\be
\cO_{0,L,\alpha}(n_0)-\cO_{0,\infty}(n_0)= \cD_1(n_0) + \cD_2(n_0)
\end{equation}
with
\ba
\cD_1(n_0)&:=&\cO_{0,L,\alpha}(n_0)-\cE(n_0)\nonum
\cD_2(n_0)&:=&\cE(n_0)-\cO_{0,\infty}(n_0)
\end{eqnarray}
and
\be
\cE(n_0):=\int \!d\Omega(n)(\T^{x_1}K^\cO)(n_0,n)\, 
\frac{\overline{\cT}_{\!\beta}(n^\up \!\cdot \! n;
n^{\up}\!\cdot \! n_L; L-x_\ell)}{\cP_{-1/2}(n^\up\cdot n_0)
\cP_{-1/2}(n^{\up}\!\cdot \!n_L) \cT_{\beta}(1;L)}.
\end{equation}
Using the bound
\be
\Big|(\T^{x_1}K^\cO)(n_0,n)\Big|\leq \Vert\cO\Vert \,\cT(n_0\cdot n;x_\ell)
\end{equation}
and the convolution property of the transfer matrices, a bound for $\cD_1$ is 
\be
|\cD_1(n_0)|\leq \Vert \cO\Vert 
 \bigg| 1- \frac{\overline{\cT}_{\!\beta}(n^{\up}\!\cdot \!n_0,
   n^{\up}\!\cdot \! n_L;L)}%
{\cP_{-1/2}(n^{\up} \cdot n_0) \cP_{-1/2}(n^{\up}\!\cdot \!n_L) 
\cT_{\beta}(1;L)}\bigg|\,,
\label{D1bound}
\end{equation}
while for $\cD_2$ one obtains simply
\ba
|\cD_2(n_0)| & \leq & \Vert \cO\Vert \int\!d\Omega(n) \,
\frac{\cT_{\beta}(n_0\cdot n;x_{\ell})}{\cP_{-1/2}(n^{\up}\cdot n_0)}
\nonum
& & \times  
\bigg| \cP_{-1/2}(n^{\up} \cdot n) \lb_{\beta}(0)^{-x_{\ell}} -
\frac{\overline{\cT}_{\beta}(n^{\up}\!\cdot \!n,n^{\up}\!\cdot \!
  n_L;L-x_{\ell})}{\cP_{-1/2}(n^{\up}\!\cdot \!n_L)\cT_{\beta}(1;L)}
\bigg|\,.
\label{D2bound}
\end{eqnarray}
According to (\ref{der5}) we have to estimate
\be
d_{1,2}:= \int 
\!d\Omega(n_0)\,|\cD_{1,2}(n_0)|\,
\frac{
\cT_\beta(n_0\cdot n^\up;L) 
\overline{\cT}_{\beta}(n^{\up}\!\cdot \! n_0,n^{\up}\! \cdot \!n_L;L)}%
{\cT_{\beta}(n^{\up}\! \cdot \! n_L ;2L)} \,, 
\end{equation}
In a first step we use $\overline{\cT}_{\!\beta}(\xi_0,\xi_L;x) 
\leq \cT_{\beta}(1;x) \cP_{-1/2}(\xi_0) \cP_{-1/2}(\xi_L)$ and the 
fact that $\cD_{1,2}(n_0)$ depend on $\xi_0 = n^{\uparrow} 
\! \cdot \! n_0$ only to write 
\be 
d_{1,2} \leq 2\pi 
\frac{\cP_{-1/2}(\xi_L)\cT_{\beta}(1;L)}{\cT_{\beta}(\xi_L;2L)} 
\int_1^{\infty} d\xi_0 |\cD_{1,2}(\xi_0)| \cT_{\beta}(\xi_0,L) 
\cP_{-1/2}(\xi_0)\,. 
\label{d12bound1}
\end{equation}
Next we claim
\be 
|\cD_{1,2}(\xi_0)| \leq [\ln^2 \xi_0 + \ln^2 \xi_L] O(1/L)\,.
\label{D12bound}
\end{equation} 
For $\cD_1(\xi_0)$ this follows directly from 
(\ref{D1bound}) and Proposition 2.3(iv). For $\cD_2(\xi_0)$
we likewise use Proposition 2.3(iv) and then apply 
Lemma 2.2(iv) with the function $f(\xi) = \cP_{-1/2}(\xi) 
[\ln^2\xi + \ln^2 \xi_L]$ (see the remark after Lemma 2.2). This gives    
\ba
\label{d2bound2}
|\cD_2(n_0)| &\leq & O(1/L) \int_1^{\infty} \!d\xi \, 
\overline{\cT}_{\!\beta}(\xi_0,\xi;x_{\ell}) 
\frac{\cP_{-1/2}(\xi)}{\cP_{-1/2}(\xi_0)} [\ln^2 \xi + \ln^2 \xi_L]
\nonum 
& \leq & O(1/L) [\ln^2 \xi_0 + \ln^2 \xi_L]\,,
\end{eqnarray} 
as asserted. 

On account of (\ref{D12bound}) the integrand $I(\xi_0)$ in 
(\ref{d12bound1}) vanishes pointwise for $L \ra \infty$. Using 
(\ref{D12bound}), assuming Conjecture 2.5 and recalling that 
$\cP_{-1/2}(\xi_0) \leq (1 + \ln \xi_0)/\sqrt{\xi_0}$ we can bound 
$I(\xi_0)$ by
$O(1/L)\cT_\beta(1;L) \xi_0^{-1}(1+\ln \xi_0)^2 E(\ln\xi_0/\sqrt{L}) 
[\ln^2 \xi_0 + \ln^2 \xi_L]$. Changing now the integration variable to 
$t:= (\ln\xi_0)/\sqrt{L}$ the new integrand is bounded by
\be
F(t):={\rm const}\ (t+\frac{1}{\sqrt{L}})^2\ (t^2+\frac{\ln^2 
\xi_L}{L})E(t)
\end{equation}
and the right hand side is bounded uniformly in $L$ by an integrable 
function. By the dominated convergence theorem we can interchange the 
limit with the integration and conclude that $d_{1,2} \ra 0$ for $L \ra 
\infty$, completing  the proof of Lemma 4.4. 
\hfill$\blacksquare$

This lemma, combined with (\ref{der2}), reduces the computation 
of the thermodynamic limit for multipoint functions to that of one-point 
functions: from Eqs.~(\ref{1pt_TD}), (\ref{der2}) it follows that the 
thermodynamic limit of a multipoint observable can be computed 
as 
\be 
\lim_{L \ra \infty} \bra \cO\ket_{L,\beta,\alpha} = 
\lim_{n_0 \ra \infty} \overline{\cO}_{0,\infty}(n_0) \,,
\label{der6}
\end{equation}
whenever the limit exists. We claim that for all $\cO \in \cC_{\rm \T
\,ainv}$ the limit does exist and is given by the rhs of Eq.~(\ref{TDTainv_twist}).   
To see this we return to (\ref{der3}) and swap the order of $K^{\cO}$ and $P$:
\ba
\label{der7}
\cO_{0,\infty}(n_0) \is \frac{\lb_{\beta}(0)^{-x_{\ell}}}%
{\cP_{-1/2}(n^{\up}\cdot n_0)}\,(\T^{x_1} K^{\cO} P)(n_0,n^{\up})
\nonum
\is \frac{\lb_{\beta}(0)^{x_1-x_{\ell}}}%
{\cP_{-1/2}(n^{\up}\cdot n_0)}
\int \!d\Omega(n) \cP_{-1/2}(n_0 \cdot n) K^{\cO}(n,n^{\up}) 
\nonum
&+& \frac{\lb_{\beta}(0)^{-x_{\ell}}}{\cP_{-1/2}(n^{\up}\cdot n_0)}
\int \!d\Omega(n) \,
\cT_{\beta}(n_0\cdot n;x_1) [K^{\cO},P](n,n^{\up})\,.
\end{eqnarray}
We now take the ${\rm SO}^{\up}\!(2)$ average wrt $n_0$. In the 
first term the $n_0$ dependence then drops out by (\ref{Pprop}c) and 
produces the announced result. For the second term we use the defining 
bound (\ref{Tainv_def}) and distinguish between $x_1 =0$ and $x_1 \neq 0$. 
In the first case a bound on the second term is $\lb_{\beta}(0)^{-x_{\ell}}
p_1 p(\xi_0)/\cP_{-1/2}(\xi_0)$, which vanishes for $\xi_0 \ra \infty$. 
For $x_1 \neq 0$ we bound the integral by $2\pi p_1 \int\!d\xi \,
\overline{\cT}_{\!\beta}(\xi_0,\xi;x) p(\xi)$, using the definition 
(\ref{Tainv_def}). To this integral we apply Lemma 2.2(iv) to 
get ${\rm const} p(\xi_0)$, which again vanishes for $\xi_0 \ra \infty$.    
This completes the proof of Proposition 4.3(ii).
\hfill $\blacksquare$

Let us add a number of comments on (\ref{TDTainv_twist}), (\ref{1pt_TD}). 
First one should note that the thermodynamic limit can be computed explicitly 
for all of $\cC_{\rm \T\,ainv}$ without having to select `fine-tuned' 
subsequences, i.e.~without recourse to the Banach-Alaoglu theorem. 
Second one observes that translation invariance is 
restored in the thermodynamic limit even though for $\cO \in \cC_{\rm
\T\,ainv}$ the finite volume expectations are not translation invariant.
Third, just as for translation invariant observables the expectations 
(\ref{TDTainv_twist}), (\ref{1pt_TD}) will in general not be 
${\rm SO}(1,2)$ invariant. An exception are observables in 
a subclass $\cC_{\rm ainv}^{\up} \subset \cC_{\rm \T\, ainv}$ 
to be discussed below.

Just as $\cC_{\rm \T\, inv}$ contained the ${\rm SO}(1,2)$ invariant
observables $\cC_{\rm inv}$ as special cases, here there is a subspace 
$\cC_{\rm ainv}$ of observables which decay sufficiently fast to an 
${\rm  SO}(1,2)$ invariant one {\it after} averaging over 
${\rm SO}^{\up}\!(2)$. We denote the limiting observable by 
\be
\cO_{\infty}(n_1,\ldots n_{\ell}) :=\lim_{A\to\infty}
\cO(An_1,\ldots An_{\ell})\,,
\label{ainvO}
\end{equation} 
and specify the rate of approach to the limit below. Provided the limit 
exists it will automatically be ${\rm SO}(1,2)$ invariant. For example one
can build a large class of $\cC_p^{\up}$ observables satisfying 
(\ref{ainvO}) by replacing in a function of $n_i \cdot n_j$ each $n_i \cdot
n_j$ with $n_i \cdot n_j \, f(n_i, n_j)$ or with 
$n_i \cdot n_j + f(n_i, n_j)$, for some  
${\rm SO}^{\up}\!(2)$ -- but not ${\rm SO}(1,2)$ -- invariant function
$f$ that goes to a constant in the limit. Note that the dependence on the 
invariant part may correspond to an unbounded function. In addition any 
dependence on the
$n_i^0$ is allowed, constrained only by the requirement that the limit 
(\ref{ainvO}) exists. Of course $\cC_{\rm ainv}$ contains 
the ${\rm SO}(1,2)$ invariant observables $\cC_{\rm inv}$  as a proper 
subset. For observables depending only on one spin ($\ell =1$)  
asymptotic translation invariance just reduces to the existence of the limit 
in (\ref{ainvO}), as it did for $\cC_{\rm \T\,ainv}$ observables with 
$\ell =1$.      

For $\ell>1$ we specify the rate of approach in which the limit in 
(\ref{ainvO}) is reached in terms of the kernels $K^{\cO}$ as 
follows, thereby giving a technically precise version of Definition 
3.3(ii): 

{\bf Definition 4.5.}
$\cO \in \cC_p$ is called {\it asymptotically invariant},
$\cO \in \cC_{\rm ainv}$, iff after ${\rm SO}^{\rm \up}\!(2)$ averaging
the associated kernel obeys   
\ba
\label{ainv_def}
&& \Big| K^{\cO}(n,n') - K^{\cO_{\infty}}(n\cdot n') \Big| 
\leq p(n^{\up}\!\cdot \!n)\,p(n^{\up}\!\cdot \!n')\,, 
\quad \mbox{with}
\nonum
&& K^{\cO_{\infty}}(n_1\cdot n_\ell)
:= \lim_{A\to\infty}
K^{\rho(A)\cO}(n_1, n_\ell)
=\lim_{A\to\infty}
K^\cO(A\inv n_1, A\inv n_\ell)\,,
\nonumber
\end{eqnarray}
where $p(\xi)$ is as in (\ref{Tainv_def}). 

Note that in analogy with (\ref{Tainv}) this implies 
$\lim_{A \ra \infty} \rho(A) [K^{\cO}, \rho ] =0$, for 
$\cO \in \cC_{\rm ainv}$,  which was used as the defining property 
in 3.3(ii). Further
\be 
\cC_{\rm ainv} \subset \cC_{\rm \T\,ainv}\,.
\label{ainv_Tainv}
\end{equation}
To see this one writes $[K^{\cO}, \T^L] = [(K^{\cO} - K^{\cO_{\infty}}), \T^L]
+[K^{\cO_{\infty}}, \T^L]$. The second commutator vanishes because 
${\rm SO}(1,2)$ invariant observables are translation 
invariant. The kernel of the first commutator is bounded in modulus by 
$\cP_{-1/2}(n^{\up}\cdot n)\,p(n^{\up} \cdot n') 
\cT_{\beta}(1;L)$. It follows that $[K^{\cO},P]$ satisfies 
(\ref{Tainv_def}), which verifies (\ref{ainv_Tainv}).

It follows that the formulae (\ref{TDTainv_twist}), (\ref{1pt_TD})
are valid also for observables in $\cC_{\rm ainv}$. Moreover 
the kernel $\overline{K}^{\cO}$ can in fact be replaced with the invariant 
limiting kernel $K^{\cO_{\infty}}$. 

{\bf Proposition 4.6.}
\vspace{-3mm}

{\it
\begin{itemize}
\item[(i)] For a multi-point observable $\cO \in \cC_{\rm ainv}$, 
$\ell \geq 2$:
\be
\label{TDainv}
\langle \cO(n_{x_1},\ldots, n_{x_{\ell}})
\rangle_{\infty,\beta,\alpha}
= \lb_\beta(0)^{x_1-x_\ell} \, 2 \pi \,\int_1^{\infty} \!d\xi \,
K^{O_\infty}(\xi) \, \cP_{-1/2}(\xi)\,. 
\end{equation} 
\item[(ii)] On  $\cC^{\up}_{\rm ainv}$ the expectation functional 
$\cO \mapsto \bra \cO \ket_{\infty,\beta,{\rm bc}}$ is an invariant mean.
\end{itemize}}

{\it Proof.} (i) We decompose $K^{\cO}$ again as 
$K^{\cO} = K^{\cO_{\infty}} +(K^{\cO} - K^{\cO_{\infty}})$. For 
the invariant limiting kernel the manipulations proceed as 
for the translation invariant case and yield Eq.~(\ref{TDTainv_twist})
with the indicated replacement. 
The average of the remainder $K^{\cO} - K^{\cO_{\infty}}$ can 
be bounded by $Q(1;L+x_1) Q(n^{\up}\!\cdot \!n_L;L-x_{\ell})/
\cT_{\beta}(n^{\up}\!\cdot \!n_L;2L)$ with $Q(\xi;x) := 2\pi 
\int_1^{\infty} \! d\xi \overline{\cT}_{\beta}(\xi,\xi';x) p(\xi')$,
for $x \in \N$. The bound vanishes in the limit $L \ra \infty$.

(ii) This is a direct consequence of (\ref{TDainv}).
\hfill $\blacksquare$
 
{\it Remark 1.} For later reference we note again that this reasoning 
remains valid if the lower boundary in the $Q$ integrals was replaced 
with an arbitrarily large constant $\Lambda \gg1$.  
 
{\it Remark 2.} The reason why the expectation functionals do no not 
provide 
an invariant mean for all of $\cC_{\rm ainv}$ is that the ${\rm 
SO}^{\up}\!(2)$
averaging effected by the expectations does not commute with the 
${\rm SO}(1,2)$ action. As a consequence observables in 
$\cC_{\rm ainv}\setminus \cC^{\up}_{\rm ainv}$ will typically signal 
spontaneous symmetry breaking. See (\ref{kappaO2}) and the examples in 
section 4.3.   
Likewise the hyperclustering (\ref{clus}) for ${\rm SO}(1,2)$ invariant 
observables 
trivially generalizes to the class $\cC^{\up}_{\rm ainv}$ but fails 
in general 
for $\cC_{\rm ainv}$: if $\cA,\,\cB \in \cC^{\up}_{\rm ainv}$ have
support as in the premise of (\ref{clus}) the limit of the products equals 
the product of the limits, i.e.~$\overline{(\cA \cB)}_{\infty} = 
\overline{\cA}_{\infty} 
\overline{\cB}_{\infty}$, and to the latter (\ref{clus}) applies. 
A counterexample to hyperclustering in $\cC_{\rm ainv}$ will be given 
in section 4.3.

We can summarize these results by saying that the thermodynamic
limit effectively projects $\cC_{\rm \T\,ainv}$ onto $\cC_{\rm \T\,inv}$ 
and $\cC_{\rm ainv}$ onto $\cC_{\rm inv}$, i.e.~the top row in the diagram 
(\ref{subalgebras}) is projected onto the bottom row. In the first case 
translation invariance emerges but ${\rm SO}(1,2)$ invariance is in 
general still absent, while in the second case, given ${\rm 
SO}^{\up}\!(2)$ invariance as a `seed',  both properties emerge. The 
second result is more interesting because ${\rm SO}(1,2)$ is not 
amenable, so one could not `by hand' switch to invariant states by group 
averaging of noninvariant ones. (Presumably this is still true if one 
adopts a distributional group averaging as in \cite{GuiMar99,GomMar}.)
Rather the thermodynamic limit itself defines a {\it partial invariant 
mean}, that is the subclass of (bounded as well as unbounded) observables 
$\cC^{\up}_{\rm ainv}$ gets averaged to yield a ${\rm SO}(1,2)$ invariant 
result. An invariant mean in the proper sense would do the same for all 
continuous bounded observables $\cC_b^{\up}$, but it cannot exist on 
general grounds.      

%%%%%%%%%%%%%%%%%%%%%%%%%%%%%%%%%%%%%%%%%%%%%%%%%%%%%%%%%%%%%%%%%%%%%%%%%%%%%%%%%%%
\newsubsection{The support of the functional measures} 
 
Here we discuss the results (\ref{1pt_TD}) and (\ref{TDainv})
in more detail. Both properties express a partial symmetry restoration 
and are due to a remarkable concentration (actually rather 
dilution) property of the underlying functional measures. Roughly 
speaking the measures have their support concentrated at configurations 
that are boosted  from the origin by an amount growing at least powerlike 
with the number of sites; the measure of any bounded set of 
configurations goes to zero in the thermodynamic limit. The derivation of 
(\ref{1pt_TD}) given below explicitly makes use of this concentration 
property; the previous derivation of (\ref{TDainv}) did for technical 
reasons not explicitly rely on it. We shall explain later why the 
underlying concentration property is nevertheless visible in the 
derivation. In section 5.2 we shall also describe an alternative proof of 
(\ref{TDainv}) which links it explicitly to the concentration property of 
the 1-spin measures instrumental for (\ref{1pt_TD}). This concentration 
property is due to the large fluctuations present in $D=1$, which in 
compact models are the `enforcers' of the Mermin-Wagner theorem, but which 
are here insufficient to restore the symmetry.

We begin with re-evaluating the thermodynamic limit for asymptotically 
translation invariant observables depending only on a single spin. 
Take some $\cO(n_x) \in \cC_{\rm \T\,ainv}$ (which for $\ell =1$ 
coincides with $\cC_{\rm \,ainv}$ by definition) depending on a single 
spin at site $x$ only. By definition its 
${\rm SO}^{\up}\!(2)$ average has a limit as $n_x \ra \infty$. 
We claim that this limiting value coincides with the thermodynamic 
limits of the  $\cO(n_x)$ expectation. The mechanism behind this is that 
the relevant measures have support `mostly at infinity'. To make 
this precise, recall that in view of (\ref{1pt_per}) and (\ref{1pt_free}) the spin
$n:=n_x$ is for finite $L$ distributed according to the probability 
measures
\ba 
\label{1pt_measures} 
d\mu_{L,\beta,\alpha,{\rm av}}(n;x) \is \frac{\cT_\beta(n^\up\cdot
 n;L+x)\overline{\cT}_{\beta}(n^{\up} \cdot n, \ch\alpha ;L-x)}
{\cT_\beta(\ch \alpha;2L)}\,d\Omega(n)\,,
\nonum  
&& \mbox{for twisted bc} 
\nonum
d\mu_{L,\beta,{\rm free}}(n;x)\is  \cT_\beta(n^\up\cdot n;L+x)
\,d\Omega(n)\,,\nonum
&& \mbox{for free bc}\,. 
\end{eqnarray}
In the first case $\overline{\cT}_{\!\beta}$ is defined as in (\ref{TavgO2}).  
From Eqs.~(\ref{limit}), (\ref{ZTDlimit}) and
(\ref{TlimitO2}) one sees that the densities multiplying $d\Omega$ behave 
for large $L$ as 
\ba
L^{-3/2}\,\left(\cP_{-1/2}(n^\up\cdot n)\right)^2\,,
&\quad & \mbox{for twisted bc}, 
\nonum
\lb_{\beta}(0)^{L+x} \,L^{-3/2}\,\cP_{-1/2}(n^\up\cdot n)\,,
&\quad & \mbox{for free bc}\,; 
\label{mlimit}
\end{eqnarray}
(the approach to this asymptotic form is, however, very nonuniform in 
$n^0$, as can be seen from Eqs.~(\ref{limitder}),(\ref{Tbound})).
In particular the dependence on $\ch \alpha$ drops out in the first case. 
Both expressions in (\ref{1pt_measures}),(\ref{mlimit}) vanish pointwise 
in 
the limit but are not integrable. This implies

{\bf Lemma 4.7.} {\it For any bounded subset $M\subset \H$ and for twisted 
as well as free bc
\be
\lim_{L\to\infty}\int_Md\mu_{L,\rm bc}=0\ , 
\end{equation}  
where $d\mu_{L,\rm bc}$ stands for either of the measures in 
(\ref{1pt_measures})}.

As a consequence these measures do not have a limit as $L\to\infty$; 
they `spread out' over $\H$ (though not evenly); Lemma 4.7 may be interpreted 
as saying that the measure is getting concentrated more and more near infinity.

The measures $d\mu_{L,\beta,{\rm bc}}$ form a sequence of 
bounded, normalized linear functionals (`states') on the space $\cC_b$. 
By the theorem of Banach-Alaoglu \cite{ReedS} there is therefore a 
subsequence convergent to such a functional --  a so-called `mean'; see 
e.g.~\cite{Pat}. Because ${\rm SO}(1,2)$ is not amenable, this mean 
cannot be invariant. We will give below explicit examples of elements of 
$\cC_b$ that show this non-invariance, i.e.~spontaneous symmetry breaking. 
However 1-spin observables invariant under ${\rm SO}^{\up}\!(2)$ still
have a {\it unique} thermodynamic limit, which is independent of $x$ and
$\beta$, see Proposition 4.3(i). In view of Lemma 4.7 this 
expresses the fact that the thermodynamic limit effectively 
projects a one spin observable onto the `boundary at infinity' of the 
hyperbolic plane; for  ${\rm SO}^{\up}\!(2)$ invariant functions we may 
use the one-point compactification of $\H$ so that there is only one such 
boundary point at infinity. 

It is also instructive to estimate the size of the `cup' in the 
hyperboloid whose contribution to the functional integral is negligible. 
We integrate the observable under consideration with the 
pointwise vanishing density in (\ref{mlimit}) over the compact domain 
$\{ n \in \H |\, n^{\up}\! \cdot \!n \leq \Lambda(L)\}$. Demanding that the 
contribution of this domain still vanishes in the limit 
$L \ra \infty$ constrains the permitted growth of $\Lambda(L)$ with $L$.
For twisted bc the relevant integral is 
$\int^{\Lambda(L)} d\xi \cP_{-1/2}(\xi)^2 \sim \int^{\Lambda(L)} 
d \ln \xi (\ln \xi)^2 
\sim (\ln \Lambda(L))^3$, using (\ref{mlimit}) and the asymptotics in 
(\ref{Pasym}). Thus any $\Lambda(L)$ satisfying $\ln \Lambda(L) = o(L^{1/2})$ 
will still give a contribution vanishing in the limit $L \ra \infty$. 
For free bc the relevant integral is 
$\int^{\Lambda(L)} d\xi \cP_{-1/2}(\xi) \sim \Lambda(L) \ln \Lambda(L)$.
Thus any growth $\Lambda(L) = o(L^{3/2}/\ln^2 L)$ is allowed.

To conclude our discussion of 1-point functions let us consider some examples.
A simple example of a bounded ${\rm SO}^{\up}\!(2)$ invariant observable is 
$\cO(n) = \tanh(n^{\up} \cdot n)$. Then (\ref{1pt_TD}) gives 
$1$ for the thermodynamic limit of  its expectation, which in particular 
is ${\rm SO}(1,2)$ invariant. The spin field $n_x^a$ itself is neither bounded 
nor ${\rm SO}^{\up}\!(2)$ invariant. However by the 
${\rm SO}^{\up}\!(2)$ invariance of the 
measures (\ref{1pt_measures}) one has 
\ba 
\bra n_x^a \ket_{L,\beta,{\rm bc}} \is \delta^{a0} \bra n_{-L} \cdot n_x
\ket_{L,\beta,{\rm bc}} \;,
\label{1point}
\end{eqnarray}   
leaving only the ${\rm SO}^{\up}\!(2)$ invariant part 
$\cO(n_x) = n^{\up} \cdot
n_x = n_x^0$ to study. Slightly generalizing the above discussion it follows that the 
$n_x^0$  expectation with both twisted periodic and free bc diverges for
$L\to\infty$: for any constant $\Lambda$, the measure of the (compact) subset of 
$\H$ where $|n_x^0|\le \Lambda$ goes to 0; since the total weight of the measure is 
always 1, the expectation value will eventually be larger than $\Lambda(1-\epsilon)$ 
for any $\epsilon>0$. For the rhs in (\ref{1point}) this is also 
illustrated by the numerical results for the 2-point functions shown in Fig.~2.
In both of these examples the limit is ${\rm SO}(1,2)$ invariant 
and does not signal spontaneous symmetry breaking.  
\medskip

As seen before the computation of the thermodynamic limit for multi-point 
observables can be reduced to that of 1-point functions. Nevertheless 
it is instructive to outline the origin of the concentration property 
also for the multi-point measures. For simplicity we restrict attention 
to twisted bc. The counterpart of the normalized measures 
(\ref{1pt_measures}) for $\ell >1$ are (after integrating out $n_{x_2}, 
\ldots n_{x_{\ell-1}}$)
\ba
\label{measurel1}
&& \nspace d\mu_{L,\beta,\alpha}(n_1,n_{\ell};x_1,x_{\ell}) = 
\\[2mm]
&& \quad \frac{\cT_{\beta}(n^{\up} \cdot n_1;L+ x_1) 
\cT_{\beta}(n_1\cdot n_{\ell}; x_{\ell} - x_1)  
\overline{\cT}_{\beta}(\ch \alpha, n^{\up} \cdot n_{\ell}; L-x_{\ell})}
{\cT_{\beta}(\ch \alpha; 2 L)}  
\, d\Omega(n_1) d\Omega(n_\ell) \,.
\nonumber
\end{eqnarray}
The finite volume expectation of some observable $\cO$ can be written in terms 
of these measures as 
\be 
\bra \cO \ket_{L,\beta,{\rm bc}} = \int \!d\mu_{L,\beta,{\rm bc}}(n_1,n_{\ell})\, 
\frac{K^{\cO}(n_1,n_{\ell})}{\cT_{\beta}(n_1\cdot n_{\ell};x_{\ell} - x_1)}\,.
\label{measurel2}  
\end{equation}
The asymptotics of the density in (\ref{measurel1}) is 
\ba
\label{measurel3}
&& \lb_{\beta}(0)^{x_1-x_\ell}\,  \cP_{-1/2}(n^{\up} \cdot n_1)\, 
\cT_{\beta}(n_1\cdot n_{\ell}; x_{\ell} - x_1) 
\cP_{-1/2}(n^{\up}\cdot n_{\ell})\  L^{-3/2}\,.
\end{eqnarray}
This density vanishes pointwise as $L \ra \infty$ and is integrable wrt one 
but not wrt both variables. As before the limit of the measures 
therefore only exists as a mean.

The concentration property ensued by (\ref{measurel3}) is however more
subtle than for the 1-point measures. This is because invariant 
combinations like $n_1\cdot n_{\ell}$ contribute even for highly boosted 
individual $n_1$ and $n_{\ell}$. Conditions like (asymptotic)  translation invariance 
or (asymptotic) ${\rm SO}(1,2)$ invariance allow one to isolate
the invariant contribution by swapping the order of $K^{\cO}$ and 
$\T^{L + x_1}$ while implying that the commutator does not contribute 
to the invariant part. In order to illustrate the mechanism we set 
\be 
k(n^{\up} \!\cdot \!n_1):= \sup_{n_{\ell}} 
\frac{\overline{K}^{\cO}(n^{\up} \cdot n_1,
n^{\up}\cdot n_{\ell})}{\cT_{\beta}(n_1\cdot n_{\ell};x_{\ell} - x_1)}\,.
\label{k1}
\end{equation}
Clearly $|k(\xi_1)| \leq  \Vert \cO \Vert$. If $\cO$ and hence $K^{\cO}$ is 
${\rm SO}(1,2)$ invariant, $k(\xi_1)$ equals a constant. If $K^{\cO}$ does 
not contain a ${\rm SO}(1,2)$ part the function $k(\xi_1)$ vanishes for 
$\xi_1 \ra \infty$. Then 
\ba 
\label{mu1bound}
&& \int_{n^{\up} \cdot n_1 < \Lambda_1} \!\!\! 
d\mu_{L,\beta,\alpha}(n_1,n_{\ell};x_1,x_{\ell}) 
\frac{K^{\cO}(n_1,n_{\ell})}{\cT_{\beta}(n_1\cdot n_{\ell};x_{\ell} - x_1)}
\nonum
&& \quad \leq \int_1^{\Lambda_1} \!d\xi \,k(\xi) \,\frac{\cT_{\beta}(\xi; L+ x_1)%
\overline{\cT}_{\beta}(\xi, n^{\up}\cdot n_L;L-x_1)}%
{\cT_{\beta}(n^{\up}\cdot n_L;2L)} 
\leq \frac{\cT_{\beta}(1;L+x_1) 
\cT_{\beta}(1;L-x_1)}{ \cT_{\beta}(n^{\up}\cdot n_L;2 L)}
\nonum
&& \quad \times \cP_{-1/2}(n^{\up}\!\cdot \!n_L) 
\int_{1}^{\Lambda_1} \!d\xi\, k(\xi)\,\cP_{-1/2}(\xi)^2 \,
E\Big(\frac{\ln \xi}{\sqrt{L + x_1}} \Big)\,.
\end{eqnarray}  
In the last step we used the ${\rm SO}^{\uparrow}\!(2)$ average of the 
bound in Lemma 2.2(iii) and (\ref{Tbound}). 
The estimates in (\ref{mu1bound}) capture the qualitative features
of the concentration phenomenon. There are two cases to consider:
(i) $K^{\cO}$ does not contain an ${\rm SO}(1,2)$ invariant part,
in which case $k(\xi_1) \ra 0$ as $\xi_1 \ra \infty$. Using the 
first bound in (\ref{mu1bound}) and the argument used for 
1-point functions one sees that its $L\ra \infty$ limit is given
by the $\xi \ra \infty$ limit of $k(\xi)$ and thus vanishes, 
both for finite $\Lambda_1$ and for $\Lambda_1 \ra \infty$.   
(ii)  $K^{\cO}$ does contain an ${\rm SO}(1,2)$ invariant part,     
in which case $\lim_{\xi \ra \infty} k(\xi) \neq 0$. 
In this case it is instructive to estimate the size of the `cup' 
in the $n_1$ hyperboloid that does not contribute significantly to 
the average as $L$ becomes large. To this end we use the second 
bound in (\ref{mu1bound}) 
and note that for fixed $L$ one can take the 
$\Lambda_1 \ra \infty$ limit  at the price that the integral 
scales like $L^{3/2}$ for large $L$. In other words one simple recovers
the normalizibility of the measures in the regime $\ln \Lambda_1 \gg \sqrt{L}$. 
On the other hand for $\ln \Lambda_1 \ll \sqrt{L}$ the integral scales 
like $(\ln \Lambda_1/\sqrt{L})^2$. In particular one can allow    
$\Lambda_1$ to grow with $L$ according to 
\be 
\ln \Lambda_1(L) = o(L^{1/2}) \,,
\label{KLbound}
\end{equation} 
and still have the bound in (\ref{mu1bound}) vanish for $L \ra \infty$. 
(Note that this conclusion only depends on the simple bound 
Lemma 2.2 (iii) and not on (\ref{Tbound}).) The intermediate 
regime can also be analyzed; a typical case is $\ln \Lambda_1 = L^q$ with 
$q>1/2$, for which the integral in (\ref{mu1bound}) 
approaches a finite but nonzero constant as $L \ra \infty$. 
The upshot is that in the original $(n_1,n_{\ell})$ integral  
over $\H \times \H$ only the region $n^{\up} \!\cdot\! n_1
\geq \Lambda_1(L)$, with $\Lambda_1(L)$ as in (\ref{KLbound}), 
contributes significantly to the result for the average as $L$ becomes
large.

For free bc the analysis is similar, except that the change in the rate 
of decay also involves powers of $\lb_{\beta}(0)$. We omit the details 
and simply state that one can likewise allow the cutoff 
$\Lambda_1$ to grow at least powerlike in $L$, without affecting 
the limit formulas.

%%%%%%%%%%%%%%%%%%%%%%%%%%%%%%%%%%%%%%%%%%%%%%%%%%%%%%%%%%%%%%%%%%%%%%%%%%%%%%%%%%
\newsubsection{Examples}

We begin with some examples where a finite thermodynamic 
limit does not necessarily exist, like for the components of the spin field 
or of the Noether current. 

The individual components of the energy observable 
$E^a_{L,\beta,{\rm bc}} := \bra n^a_x \, n^a_{x+1} \ket_{L,\beta,{\rm bc}}$, 
$a =0,1,2$, can be shown to diverge for $L \ra \infty$ by an argument 
similar to the one used in section 4.2. On the other hand the invariant  
combination $(E^0 - 2 E^1)_{L,\beta,{\rm bc}}$ has a finite limit
given by ({\ref{Elimit}). Next consider the Noether current 
$J_x^a = \beta ( n_x \times n_{x+1})^a$, where $n \times n'$ denotes 
the ${\rm SO}(1,2)$ invariant vector product of $n,n' \in \H$.
(Explicitly $(n \times m)^a = \eta^{aa'} \eps_{a'bc} n^b m^c$, with 
$\eps_{abc}$ totally antisymmetric and $\eps_{012} =1$.)   
For the current two-point function one finds 
\ba
&& \bra J_x^0 J_y^0 \ket_{L,\beta,{\rm bc}}  =0\,, \quad  \mbox{for}\;\; x < y\,, 
\nonum
&& \bra J_x^1 J_y^1 \ket_{L,\beta,{\rm bc}}  = 
\bra J_x^2 J_y^2 \ket_{L,\beta,{\rm bc}} 
= - \frac{1}{2}\bra J_x \cdot J_y \ket_{L,\beta,{\rm bc}}  \,, 
\quad  \mbox{for}\;\; x < y\,, 
\label{Jcorr} 
\end{eqnarray}
so that all components have a finite $L \ra \infty$ limit. The first equation
is a special case of the more general result 
\be 
\bra J^0_{x_1} \;\cO(n_{x_2}, \ldots , n_{x_{\ell}}) \ket_{L,\beta, {\rm bc}} =
0\,, 
\quad  \mbox{for}\;\; x_1 < x_2 < \ldots < x_{\ell} \,,           
\label{J0}
\end{equation}
which is obtained by specializing the general formulas (\ref{lpoint_twist}),
(\ref{lpoint_free}) and then using 
\be
\frac{\dd}{\dd \varphi_x} \cT_{\beta}(n_x \cdot n_{x+1};1) 
= - J_x^0 \, \cT_{\beta}(n_x \cdot n_{x+1};1) \;,
\label{TJphi}
\end{equation}
where $n_x\cdot n_{x+1} = \xi_x \xi_{x+1} -  \sqrt{\xi_x^2 -1}
\sqrt{\xi_{x+1}^2 -1} \cos(\varphi_x - \varphi_{x+1})$. Since 
$J_x^0$ is essentially the Noether charge generating infinitesimal 
${\rm SO}^{\up}\!(2)$ rotations (see below) Eq.~(\ref{J0})
expresses the ${\rm SO}^{\up}\!(2)$ invariance of the `ground states' 
$1$ and $\psi_{\up}(n)$, respectively. Conversely, the fact that  
correlators involving $J_x^1,\, J_x^2$ are non-zero is yet another manifestation 
of the ${\rm SO}(1,2)$ symmetry breaking.  

Ward identities expressing the invariance of the measure and of the action 
can be derived along the familiar lines. For example one has 
\be
\bra (J_x^a - J_{x+1}^a) \, n_y^b \ket_{L,\beta,{\rm bc}}
+ \delta_{x,y} \, \bra (t^a n_y)^b \ket_{L,\beta,{\rm bc}} \,,  
\label{Ward1}
\end{equation}
with $(t^a)_c^d = - \eta^{aa'} \eta^{dd'} \eps_{a'd'c}$. Replacing 
$n_y^b$ with a generic (non-invariant) observable $\cO(n_{x_1}, \ldots, n_{x_{\ell}})$ 
a similar identity arises where the correlator with $J_x^a - J_{x+1}^a$ 
produces a sum of contact terms. As is clear from (\ref{Ward1}) these linear
Ward identities will in general not have a non-boring thermodynamic limit.
In particular no conflict, even in spirit, with Coleman's theorem 
\cite{Coleman} arises.
 
Ward identities where the current enters nonlinearly can likewise be 
derived but are hampered by the fact that the `response' are in general 
functions which fail to be translation invariant.
In 2 or more dimensions a useful quadratic Ward identity can be
derived which relates the
components of the longitudinal part of the current-current correlator to
the energies $E^a$; see \cite{PSWard}. In one dimension only the
longitudinal part exists
and only the ${\rm SO}(1,2)$ invariant -- and hence translation invariant 
-- combination of these component Ward identities is useful. It reads
\be 
\bra J_p \cdot J_{-p} \ket_{L,\beta, {\rm bc}} +
2 \beta (E^0 - 2 E^1)_{L,\beta,{\rm bc}} =0 \;,\quad \forall p \neq 0\;,
\label{Ward2}
\end{equation}
where $J^a_p = \sum_x e^{-ip x} J_x^a$, with $p = 2\pi n/(2 L +1)$, 
$n =0, \ldots, 2L$. 

Next we consider some examples of asymptotically translation invariant 
observables. They also serve to highlight the significance of the 
${\rm SO}^{\up}(2)$ averaging in the definition of the algebras in
(\ref{subalgebras}). Recall that $\cC_{\rm ainv} = \{ \cO \in \cC_p\,|\, 
{\rm SO}^{\up}(2) \;\mbox{average lies in} \; \cC^{\up}_{\rm ainv}\}$. 
The point here is that in general $\cO$ and $\rho(A)\cO,\, A\in {\rm SO}(1,2)$, 
will have {\it different} ${\rm SO}^{\up}(2)$ invariant images in 
$\cC^{\up}_{\rm ainv}$. The elements of $\cC_{\rm ainv}\setminus
\cC^{\up}_{\rm ainv}$ will therefore typically signal spontaneous 
symmetry breaking although by section 3.3 they get effectively projected back into 
$\cC^{\up}_{\rm ainv}$. 

An instructive example of such a `symmetry breaking observable' in $\cC_{\rm ainv}$ 
arises as follows: given a spacelike unit vector 
$e= (\sqrt{q^2 -1}, q \sin \gamma, q \cos \gamma)$ we define
\ba
T_e(n) & :=& \tanh (n\cdot e) \in \cC_{\rm ainv}\,,
\nonum
\overline{T}_q(\xi) &:= &\frac{1}{2\pi}\int_{-\pi}^\pi d\varphi\,
\tanh\left(\xi \sqrt{q^2-1} -q \sqrt{\xi^2-1} \,\cos\varphi\right) \in 
\cC^{\up}_{\rm ainv}\,.
\label{Tdef}
\end{eqnarray}
The observable $T_e(n)$ indeed enjoys the property (\ref{ainv_def}): after 
${\rm SO}^{\up}\!(2)$ averaging it has a unique limit 
$\overline{T}_q(\infty)$, which can be obtained by acting with a sequence 
of ${\rm SO}(1,2)$ transformations going to infinity. This limit does not depend 
on $n$ any more, so in a trivial sense it is an invariant function of the 
spins. It does, however, depend on $e$ or rather on the scalar product 
$n^\up\cdot e$ and is therefore not invariant under the action of 
${\rm SO}(1,2)$ on the original observable.
   
Spontaneous symmetry breaking is shown by the following

{\bf Proposition 4.8.} {\it For all bc considered 
\be
\bra T_e(n_x) \ket_{\infty,\beta,{\rm bc}}=\overline{T}_q(\infty)= 
1-\frac{2}{\pi}\arccos \sqrt{1-q^{-2}}\ ;
\label{Tavg}
\end{equation}
this expectation value is manifestly not invariant under ${\rm 
SO}(1,2)$: for a general $A\in {\rm SO}(1,2)$ one has $\bra T_e(n_x) 
\ket_{\infty,\beta,{\rm bc}}\neq 
\bra T_e(An_x) \ket_{\infty,\beta,{\rm bc}}$.}

{\it Proof.} We use the fact that in finite volume expectations we may 
replace $T_e(n_x)$ by it average over ${\rm SO}^{\up}\!(2)$ rotations 
$\overline{T}_q(n^0_x)$. The argument of the $\tanh$, 
i.e.~$\alpha_\xi(\varphi) :=\xi(\sqrt{q^2 -1} - q 
\sqrt{1-\xi^{-2}}\cos\varphi)$, then has its minimum at $\varphi=0$ and 
its maximum at $\varphi=\pm\pi$. Because $e$ 
is spacelike, there is a $\xi_0(q)$ such that for all $\xi>\xi_0(q)$ the 
minimum  $\alpha_\xi(0)$ is negative, the maximum $\alpha_\xi(\pm \pi)$ is 
positive and there are two zeros at $\varphi=\pm \arccos 
[(1 - q^{-2})/(1- \xi^{-2})]^{1/2}$ whose modulus converges to 
$\varphi_0 := \arccos( 1- q^{-2})^{1/2}$. This implies that 
\be 
\lim_{\xi\to\infty} \tanh\alpha_\xi(\varphi)=
\begin{cases}
1, &\text{$|\varphi|>\varphi_0$}\\
-1, &\text{$|\varphi|<\varphi_0$}\\
0, &\text{$|\varphi|=\varphi_0$}.
\end{cases}
\label{philimit}
\end{equation}
By the dominated convergence theorem we can pull the limit $\xi\to\infty$ 
under the integral for the $\varphi$ averaging and obtain
\be
 \lim_{\xi\to\infty}\overline{T}_q(\xi)= 
1-\frac{2}{\pi}\arccos \sqrt{1 -q^{-2}}\,.
\label{Tlimit}
\end{equation}
The result then follows from Eq.~(\ref{1pt_TD}).  \hfill $\blacksquare$

A large class of observables in $\cC_{\rm ainv}$ can now be built by 
algebraic operations. Of course sums and products of $T_e(n)$ at
the same or different sites will lie in $\cC_{\rm ainv}$, but so will be 
algebraic combinations built from elements of $\cC_{\rm inv}$.
The crucial difference to $\cC^{\up}_{\rm ainv}$ is that hyperclustering 
and even ordinary clustering
will now {\it fail} in general. This is because `${\rm SO}^{\up}\!(2)$
averaging' and `taking the $A  \ra \infty$ limit' in (\ref{ainvO}) are 
noncommuting 
operations in general. A simple example is given by the product of 
two tanh-observables (\ref{Tdef}), where 
\be 
1 = \bra T_e(n)^2\ket_{\infty,\beta,{\rm bc}} \neq \bra
T_e(n)\ket_{\infty,\beta,{\rm bc}}^2 = \overline{T}_q(\infty)^2\,,
\end{equation}
from (\ref{philimit}) and (\ref{Tlimit}). 

So we have so far found observables that show hyperclustering and others 
that do not cluster at all. Observables showing ordinary (exponential or 
power-like) clustering presumably also exist in the large space $\cC_b$, 
but it is more difficult to find explicit examples.

%%%%%%%%%%%%%%%%%%%%%%%%%%%%%%%%%%%%%%%%%%%%%%%%%%%%%%%%%%%%%%%%%%%%%%%%%%%%%%%%%%%%%
\newsection{Reconstruction of a Hilbert space and transfer operator}  

The Osterwalder-Schrader type reconstruction allows to reconstruct a
Hilbert space and a transfer matrix from expectation values satisfying
reflection positivity as well as translation invariance. The original
expectation values are recovered as expectations in a genuine, 
i.e.~{\it normalizable} ground state vector. This construction is well
documented in the literature \cite{OS, GJ, S}, but in our case  there are
peculiarities and surprises. For this reason we describe in some
detail how the construction works here.

First there is a rather harmless complication: reflection positivity for 
reflections both in lattice sites and in midpoints between lattice points 
is equivalent to positivity of the transfer operator; as we found in the 
beginning, however, this does not hold in our case. But we still have 
reflection positivity for reflection in lattice points, at least if we 
take the thermodynamic limit with periodic bc, and this is enough for the 
reconstruction of the Hilbert space and a positive two-step transfer 
matrix.

There is a much more serious complication: as stated above, the
reconstruction produces a ground state in the proper sense, whereas we
know that the original transfer matrix $\T$ on $L^2(\H)$ does not have
such a ground state. So it is unavoidable that there is some discrepancy
between the reconstructed quantum mechanics and the one we started from.
This mismatch is also related to the fact that our expectation functional 
in the infinite volume is not given by a measure, but only a mean on the 
configuration space.

In this section we consider periodic bc exclusively and denote the expectation 
functional (the state) $\bra\;.\;\ket_{L,\beta,0}$  in Eq.~(\ref{lpoint_twist}) 
by $\om_L(\,.\,)$. A reconstruction in the usual sense won't work for twisted
or free bc because the $x\geq 0$ and the $x \leq 0$ halves of the chain have 
to enter symmetrically. For the algebra we take $\cC_b$ in order to have 
the usual concept of a state available. For $\cC_b \cap \cC_{\rm \T\, ainv}$ 
we saw before that the thermodynamic limit is explicitly computable 
and translation invariant. For the rest of $\cC_b$ a thermodynamic 
limit exists likewise, though it may be necessary to select 
subsequences and to average over translations in order to have it 
translation invariant. We denote such a weak limiting state by  
$\om_\infty(\,.\,) = w-\lim_{L \ra \infty} \om_L(\,.\,)$.

We denote by $\cC_+$ ($\cC_-$) the subalgebra of bounded observables 
$\cC_b$ depending only on the spins $n_x$ with $x\ge 0$, ($x\le 0$) and 
$\cC_0=\cC_+\cap\cC_-$. Our chain admits a reflection $x \ra -x$ and we
introduce an antilinear time reflection $\vartheta$ acting on on $\cC_b$  
by replacing any function $\cO$ by the same function of the reflected 
arguments and taking the complex conjugate:
\be
\label{theta1}
(\vartheta \cO)(n_{-x_{\ell-1}},\ldots, n_{-x_0})=\cO(n_{-x_0},\ldots,
  n_{-x_{\ell-1}})^\ast\;, \quad x_0 < \ldots < x_{\ell-1}\;,  
\end{equation}
where the asterisk denotes complex conjugation. To interpret this 
formula
correctly note that on the lhs we have written the observable $\vartheta \cO$   
in the customary form as a function of the spins on which it actually
depends, in the order of increasing indices. On the rhs $\cO$ is to be 
read as a function of $\ell$ spins, with the displayed arguments now    
appearing in the order of decreasing indices. For example 
$\cO = n_1 \cdot n_3 + c\, n^{\up} \cdot n_3$ gives 
$\vartheta \cO = n_{-1}\cdot n_{-3} + c^*\, n^{\up} \cdot n_{-3}$. 
We discuss the reconstruction first for a finite and then for an 
infinite chain.

%%%%%%%%%%%%%%%%%%%%%%%%%%%%%%%%%%%%%%%%%%%%%%%%%%%%%%%%%%%%%%%%%%%%%%%%%%%%%%%
\newsubsection{Finite chains}

Recall that we adopt untwisted periodic bc, $n_{-L}=n_L=n^\up$, and 
consider a chain of total length $2L +1$. We begin by assigning to each 
$\cO \in \cC_+$ an element $\cO_{0,L} \in \cC_0$ with the same expectation 
value via 
\ba
\cO_{0,L}(n_0) \!&\!\!:=\!\!&\!
\int \prod_{i=1}^\ell d\Omega(n_i)
\cO(n_1,\dots,n_\ell)\prod_{i=1}^\ell\cT_\beta(n_{i-1}\cdot
n_i;x_i-x_{i-1})\frac{\cT_\beta(n_\ell\cdot n^\up;L-x_\ell)}
{\cT_{\beta}(n_0\cdot n^\up;L)}
\nonum
\!&\!\!=\!\!&\!\int \! d\Omega(n)\, (\T^{x_1} K^{\cO})(n_0,n) 
\frac{\cT_\beta(n^\up \cdot n ;L-x_\ell)}
{\cT_{\beta}(n^\up \cdot n_0;L)}\,,
\label{repres}
\end{eqnarray}
where $x_0=0$ and the first transfer matrix is to be interpreted as the   
identity operator if $x_1=0$. Note the properties 
\ba 
&& |\cO_{0,L}(n)| \le \Vert \cO\Vert\;,\sspace 
(\1)_{0,L}(n) = \1\;,
\nonum
&& (\rho(A) \cO)_{0,L}(n) = \cO_{0,L}(A\inv n)\Big|_{n^{\up} \mapsto
A\inv n^{\up}}\; 
\label{represprop}
\end{eqnarray}
where we denote by $\1$ the unit element of $\cC_+$. Further we set 
\be
\psi^\cO(n) :=\cO_{0,L}(n) \,\frac{\cT_\beta(n\cdot 
n^\up;L)}{\sqrt{\cT_\beta(1;2L)}}\,.
\label{iso}
\end{equation} 
The expressions (\ref{repres}) and (\ref{iso}) are designed such that 
\be
\om_L([\vartheta \cO] \, \cO) = \frac{1}{\cT_{\beta}(1;2L)}
\int \! d \Omega(n) \,|\cO_{0,L}(n)|^2 \,\cT_{\beta}(n \cdot n^{\up};L)^2 = 
\int \! d\Omega(n) \, |\psi^{\cO}(n)|^2\,,
\label{omL}
\end{equation}
holds, as one can verify from (\ref{lpoint_twist}). In particular 
reflection positivity 
\be
\om_L([\vartheta \cO] \cO)\ge 0 \,,\quad \forall \,\cO\in \cC_+\,, 
\label{pos}
\end{equation}
is manifest.

With these preparations at hand the reconstruction of the Hilbert space 
$\cH_L$ for a finite chain works as usual: a positive semidefinite 
scalar product is introduced on $\cC_+$ by    
\be
(\cA,\cB)_L := \om_L([\vartheta \cB]\,\cA)\,;
\label{scalar}
\end{equation}
there will be a nontrivial null space $\cN$ of elements with 
$\om_L([\vartheta \cO]\,\cO)=0$
%, which is also an ideal wrt the multiplication in $\cC_+$
. The Hilbert space $\cH_L$ is then the 
completion of the quotient space $\cC_+/\cN$ with respect to the norm 
induced by $\om_L$. The necessity to divide out $\cN$  becomes clear 
if one notices that for any $\cO\in\cC_+$ one can find a unique
element $\cO_{0,L}\in\cC_0$ such that $\cO-\cO_{0,L}\in\cN$, namely
just the one given in (\ref{repres}). The uniqueness follows from 
(\ref{omL}), which implies $\cC_0\cap \cN= \{0\}$. Note that the OS norm 
for $\cO$ coincides with the $L^2$-norm for $\psi^\cO$. 

The above construction makes it manifest that for a finite chain there is 
a natural isometry between the reconstructed Hilbert space $\cH_L$ and the
original $L^2(\H)$: $\cH_L$ turned out to be the completion of $\cC_0$ with 
respect to the norm induced by (\ref{scalar}), i.e.~$\cH_L = \overline{\cC}_0$. 
Note that although $\cC_0$ is the universal $L$-independent space of bounded 
continuous functions on $\H$, its completion with respect to $(\;,\;)_L$ 
depends on $L$. Of course the $L$-dependence is of a rather trivial nature 
in that by (\ref{iso}) the map 
\be
V_L: \cH_L \rra  L^2(\H)\;, \sspace (V_L\psi)(n) = 
\psi(n) \,\frac{\cT_{\beta}(n\cdot 
n^{\up};L)}{\sqrt{\cT_{\beta}(1;2L)}}\,, 
\end{equation} 
defines an isometry between Hilbert spaces. Alternatively $\cH_L$ could be
regarded as the preimage of $L^2(\H)$ with respect to $V_L$. It is worth 
noting that $\cH_L$ is by itself a commutative $C^\ast$-algebra, so the
reconstruction of the Hilbert space can be considered as an instance of 
the well-known Gel'fand-Na\u{\i}mark-Segal reconstruction, see 
e.g.~\cite{Ruelle, Haag}. 
To sum up, for a finite chain the original Hilbert space $L^2(\H)$ and 
the reconstructed one $\cH_L$ can really be identified.

Unsurprisingly, for a finite chain $\cH_L$ also carries a unitary 
representation $\rho_L$ of ${\rm SO}(2,1)$ (spontaneous symmetry breaking 
can only arise in the thermodynamic limit); it is obtained simply 
by conjugating the representation $\rho$ with $V_L$:
\be
\rho_L=V_L^{-1} \rho V_L\,. 
\end{equation}
Explicitly for $A\in {\rm SO}(2,1)$ and $\psi\in\cC_0$ this gives
\be
(\rho_L(A)\psi)(n)=\psi(A\inv n)\frac{\cT_\beta(A \inv n\cdot n^\up;L)}
{\cT_\beta(n\cdot n^\up;L)}\,.
\label{rhoL}
\end{equation}
For $\cO \in \cC_+$ we define
\be
(\rho_L(A)\cO)(n_{x_0},\ldots,n_{x_{\ell-1}})=\cO(A\inv n_{x_0},\ldots,
A\inv n_{x_{\ell-1}})
\frac{\cT_\beta(A \inv n_{x_{\ell-1}} \cdot n^\up;L-x_{\ell-1})}
{\cT_\beta(n_{x_{\ell-1}}\cdot n^\up;L-x_{\ell-1})}\,,
\label{rhoLO}
\end{equation}
which is compatible with (\ref{rhoL}) and induces it via (\ref{repres})
in that  
$(\rho_L(A)\cO)_{0,L}(n) = (\rho_L(A)\cO_{0,L})(n)$, for all $A \in {\rm SO}(1,2)$.  
This also ensures that $\rho_L$ maps elements $\cO - \cO_{0,L}$ of $\cN$ onto 
other elements of zero norm. The rhs of (\ref{rhoL}), (\ref{rhoLO}) is in 
general no longer a bounded function of $n$ because the 
asymptotics of $\cT_\beta(A\inv n\cdot n^\up;L)$ and $\cT_\beta(n\cdot 
n^\up;L)$ do not match, but it is of course still an element of
$\cH_L$ with the same norm as $\psi$. Likewise by (\ref{rhoL}), (\ref{rhoLO}) 
the completion $\cN_L$ of $\cN$ wrt $\om_L$ is mapped onto itself under
$\rho_L$.

In preparation of the thermodynamic limit let us consider the action of 
$\rho_L$ on the function $\1$ (an approximate ground state for large $L$):
\be
(\rho_L(A)\1)(n)\ = \ \frac{\cT_\beta(A\inv n\cdot n^\up;L)}  
{\cT_\beta(n\cdot n^\up;L)}\,.
\end{equation}     
The scalar product
\ba
(\rho_L(A)\1,\rho_L(B)\1)_L \is \frac{1}{\cT_\beta(1;2L)}\int 
d\Omega(n)\cT_\beta(n\cdot A n^\up;L) \cT_\beta(n\cdot B n^\up;L)
\nonum
\is \frac{\cT_\beta(A \n^\up\cdot B n^\up;2L)}{\cT_\beta(1;2L)}\,,
\label{scalar0}
\end{eqnarray}
then has the finite and nonzero limit $\cP_{-1/2}(A n^{\up} \!\cdot \!B  
n^\up)$, as $L\to\infty$. 

For finite $L$ the state $\omega_L(\,\cdot\,)$ will not be translation 
invariant outside the subalgebra $\cC_{\rm \T\,inv} \cap \cC_+$. As a 
consequence there is no reconstructed transfer matrix for finite $L$. 
Conversely this provides an intrinsic reason to consider the reconstruction 
based on the expectations of the infinite chain.    

%%%%%%%%%%%%%%%%%%%%%%%%%%%%%%%%%%%%%%%%%%%%%%%%%%%%%%%%%%%%%%%%%%%%%%%%%%%%%%%%%
\newsubsection{Thermodynamic limit}

Let us thus turn to the thermodynamic limit $\om_\infty = w-\lim_{L \ra \infty} 
\om_L$.  Reflection positivity remains true in this limit; so one can still 
define a scalar product and a null space $\cN$ as for the finite chain. 
As before a Hilbert space $\cH_{OS}$ can be constructed as the completion 
of $\cC_+/\cN=\cC_0/(\cN\cap\cC_0)$. In order not to clutter the notation 
we continue to use the same symbols for the algebra of observables and 
the spaces $\cN$ etc., however one should keep in mind that the spaces 
$\cC_+$ etc.~for a finite and for the infinite chain cannot be
identified. In particular equation (\ref{iso}) loses its meaning in the limit: 
the left hand side goes to zero pointwise, even though its norm in $\cH_{OS}$ 
in general does not. For observables in $\cC_+ \cap \cC_{\rm \T\,ainv}$ 
the explicit formula (\ref{TDTainv_twist}) can be used to compute 
the inner products $(\;,\;)_{OS}$. Outside this class in general the 
original definition $(\;\,,\;)_{OS} = \lim_{L \ra \infty}(\;\,,\;)_L$ 
has to be used. On the other hand (\ref{repres}) always has a sensible limit:

{\bf Proposition 5.1.}{\it
\vspace{-3mm}

\begin{itemize}
\item[(i)] The limit $\lim_{L\to\infty} \cO_{0,L}(n_0)=:
\cO_{0,\infty}(n_0)$ exists and obeys
\ba
\cO_{0,\infty}(n_0) \!\!&\!\!:=\!\!&\!\!
\int \! d\Omega(n)\, (\T^{x_1} K^{\cO})(n_0,n) 
\frac{\cP_{-1/2}(n^\up \cdot n)}{\cP_{-1/2}(n^\up \cdot n_0)}\,
\lb_{\beta}(0)^{-x_{\ell}}\,.
\label{represTD}
\end{eqnarray} 
\item[(ii)] $\1_{0,\infty} =\1$ and $|\cO_{0,\infty}(n_0)|\leq 
\Vert \cO \Vert$, where $\Vert \cdot\Vert$ denotes the sup norm.
\item[(iii)] If Conjecture 2.5 holds, 
\be
\cO-\cO_{0,\infty}\in\cN\,,
\label{Ninv}
\end{equation}
with respect to $(\,,\,)_{OS}$. 
\end{itemize}}

{\it Proof.} (i) and (ii) are straightforward. (iii), while very 
plausible, requires nevertheless a proof; the one given here 
relies on the validity of Conjecture 2.5. It suffices to show that 
for any $\cA\in\cC_-$ 
\be
\lim_{L\to\infty}\om_L(\cA(\cO-\cO_{0,\infty}))=0\,.
\end{equation}
In order to show this, we write
\ba
\label{Nder1}
&&\om_L(\cA(\cO-\cO_{0,\infty}))\ = \ \nonum
&&\om_L(\cA(\cO-\cO_{0,L}))+
\om_L(\cA(\cO_{0,L}-\cO_{0,\infty}))+
(\om_\infty-\om_L)(\cA(\cO-\cO_{0,\infty}))\, .
\end{eqnarray}
The first term vanishes by construction of $\cO_{0,L}$, the third term 
goes to zero as $L\to\infty$ by definition of $\om_{\infty}$, 
whereas the second term requires a closer look. In view of 
\be
\Big| \om_L(\cA(\cO_{0,L}-\cO_{0,\infty})) \Big| \leq \Vert \cA \Vert \int \! 
d\Omega(n_0)\,\Big| \cO_{0,\infty}(n_0)-\cO_{0,L}(n_0)\Big| \,
\frac{\cT_\beta(n_0\cdot n^\up;L)^2}{\cT_{\beta}(1;2L)} \,, 
\end{equation}
the difference $\vert \cO_{0,\infty}(n_0)-\cO_{0,L}(n_0)\vert$ needs to 
be examined. This however has been done in section 4.2, and the proof that 
the right hand side of (\ref{der5}) vanishes for $L \ra \infty$ carries 
over. This completes the proof of (iii). \hfill $\blacksquare$

The relation (\ref{Ninv}) has several important consequences, which we discuss
consecutively.  

For bounded observables in $\cC_{\rm \T\,ainv}$ a crucial consistency
condition arises from (\ref{Ninv}) and (\ref{TDTainv_twist}). Since an explicit 
formula for the state $\om_{\infty} = \bra \; \ket_{\infty}$ is known for 
these observables it must come out that
\be 
\bra \cA \cO \ket_{\infty} = \bra \cA \cO_{0,\infty} \ket_{\infty}
\sspace \forall \cA \in \cC_-\,,\;\;\; \cC_+ \in \cO\,,
\label{consistency}
\end{equation}
using directly the limiting formulae (\ref{TDTainv_twist}) and 
(\ref{represTD}). This is indeed the case: a computation shows that
both sides of (\ref{consistency}) reduce to 
\ba
&& \lb_{\beta}(0)^{x^{\cA}_< - x^{\cO}_>} \int \! d\Omega(n) 
d\Omega(n') K^{\cA}(n^{\up},n) \cT_{\beta}(n\cdot n';x^{\cA}_> - x^{\cO}_<)
\nonum 
&& \times \int d\Omega(n'') K^{\cO}(n'\cdot n'') \cP_{-1/2}(n^{\up}\cdot n'')
\,.
\label{consistency2}
\end{eqnarray} 
Here we wrote $x^{\cO}_<$ for the leftmost and $x^{\cO}_>$ for the 
rightmost site where an observable $\cO \in \cC$ is supported. 
The consistency condition for $\cO \in \cC_{\rm T\,ainv}$ is therefore
satisfied. On the other hand (\ref{Ninv}) is valid for {\it all} bounded 
observables and the computation leading to (\ref{consistency2}) does not seem to  
leave much room for expressions other than (\ref{TDTainv_twist}) having 
the same property (\ref{consistency}). This suggests that 
(\ref{TDTainv_twist}) is actually valid for all bounded observables 
though our proof is not.

Next let us consider asymptotically invariant observables. 
For them the result (\ref{Oinftylimit}) yields an alternative 
derivation of (\ref{TDainv}). Since it is based on (\ref{Oinftylimit})
this derivation highlights that the origin of the result (\ref{TDainv})
lies in the concentration property of the measures described in 
section 4.3. To this end we write 
$K^{\cO}$ as $K^{\cO_{\infty}} + (K^{\cO} - K^{\cO_{\infty}})$ and 
insert into the definition of $\cO_{0,\infty}(n_0)$. 
Since (\ref{Pinv}) is trivially satisfied for the integral 
operators coming from ${\rm SO}(1,2)$ invariant operators 
the first term gives (\ref{TDTainv_twist}) with $K^{\cO}$ replaced 
by $K^{\cO_{\infty}}$, which is the asserted result. Using 
(\ref{ainv_def}) the modulus of the second term can be bounded by 
\be 
\lb_{\beta}(0)^{-x_{\ell}} \frac{Q(\xi_0,x_1)}{\cP_{-1/2}(\xi_0)}\,,
\label{TDainv_der}
\end{equation}
where $\xi_0 = n^{\up}\! \cdot \!n_0$ and $Q(\xi_0;x_1)$ is defined in 
after Eq.~(\ref{TDainv}). According to (\ref{Oinftylimit}) we have to 
analyze the limit $n_0 \ra \infty$ of this expression. To this end we split the 
region of integration in 
$Q$ into a bounded part $\xi' \in [1,\Lambda]$ and a remainder 
$\xi'\in [\Lambda,\infty[$. For the unbounded part we use 
$\overline{\cT}_{\!\!\beta}(\xi_0,\xi';x_1) \leq \cP_{-1/2}(\xi_0) 
\cP_{-1/2}(\xi') \,\cT_{\beta}(1;x_1)$ to get a $n_0$ independent 
bound $p_1/\Lambda$ on it. In the bounded part we use the 
fact that $\overline{\cT}_{\!\!\beta}(\xi_0,\xi';x_1)$ 
vanishes faster than any power in $\xi_0$, and does so uniformly for 
all $\xi' \in [1,\Lambda]$. For large enough $\xi_0$ the supremum 
$\sup[\overline{\cT}_{\!\!\beta}(\xi_0,\xi';x_1)/\cP_{-1/2}(\xi_0)]$ 
over $\xi' \in [1,\Lambda]$ can be therefore be made smaller than 
$1/\Lambda^2$. The upshot is that (\ref{TDainv_der}) can be made smaller than 
any prescribed quantity. This completes the derivation of 
(\ref{TDainv}) based on (\ref{Oinftylimit}).

A simple consequence of (\ref{Ninv}) is that in contrast to the finite volume 
case $\cC_0$ now also 
intersects the null space $\cN$: for instance all functions going to 
zero for $n \ra \infty$ (i.e.~$n^{\up}\!\cdot\! n \to\infty$) will 
be mapped into the null vector of $\cH_{OS}$, according to the previous 
section. The same is true for all functions that go to zero for $n \ra 
\infty$ after averaging over ${\rm SO}^{\up}\!(2)$. In fact, 
according to the discussion in Section 4.2, this exhausts the intersection 
$\cN\cap\cC_0$. Likewise products of the form $c(n) \psi(n) \in \cC_0$, 
where $c(n) \ra c$ for $n \ra \infty$, differ from $c \psi(n)$ only by 
an element of $\cN$, since their difference goes to zero as $n \to\infty$. 
This means that in linear combinations of vectors constant coefficients 
can always be replaced with coefficients satisfying this decay condition 
without changing the equivalence class mod $\cN$.

Setting $\cA =\1$ in (\ref{Nder1}) and using the results of section 4.3
one infers
\be 
\om_{\infty}(\cO) = \om_{\infty}(\cO_{0,\infty}) = w-\!\!\!\lim_{A\ra \infty} 
\overline{\cO}_{0,\infty}(A n^{\up})\,\quad \mbox{for all} \;\;\;
\cO \in \cC_b\,.
\label{Oinftylimit}
\end{equation}
The weak limit arises because for a 1-point observable only the behavior at 
infinity, defined through some unbounded sequence of $A$'s, is relevant. 
This limit does not necessarily exist, however as the 
$\overline{\cO}_{0,\infty}(A n^{\up})$ form a bounded sequence 
in $\R$ one can always select a convergent subsequence. 
As shown in section 4.2 the limit does exist for all 
$\cO \in \cC_{\rm \T \, ainv}$ without taking subsequences
and is given by (\ref{TDTainv_twist}). For the following discussion 
it is convenient to introduce a somewhat smaller class of observables 
which we call $P$-invariant: 
\begin{equation} 
\cO \in \cC_{\rm P\, inv}\quad \mbox{iff} \quad P K^{\cO} = K^{\cO} P\;.
\label{Pinv}
\end{equation}
One has 
\begin{equation} 
\cC_{\rm \T\,inv} \subset \cC_{\rm P\, inv} \subset \cC_{\rm \T\,ainv}\,.
\end{equation}
The second inclusion is trivial; the first inclusion follows by taking 
the $x \ra \infty$ limit $\cT_{\beta}(1;x)\inv [K^{\cO}, \T^x]=0$.
The condition (\ref{Pinv}) is chosen such that $\cO_{0,\infty}(n_0)$ is 
independent of $n_0$, so that by (\ref{der7}) the value of  $\cO_{0,\infty}$ 
directly coincides with the thermodynamic limit of $\cO \in \cC_P$.  
By a computation similar to the one in (\ref{der7}) one shows from 
(\ref{Oinftylimit}) that for separately 
${\rm SO}^{\up}\!(2)$ invariant $\overline{\cA}, \overline{\cB} 
\in \cC^{\up}_{\rm P\,inv}$ one has the `hyperclustering' relation  
\be 
\om_{\infty}(\overline{\cA}\,\overline{\cB}) =
\om_{\infty}(\overline{\cA})\,\om_{\infty}(\overline{\cB})\,.
\end{equation}  
Since in general $\overline{\cA \cB} \neq \overline{\cA}\, \overline{\cB}$ 
this does not extend to all of $\cC_{\rm P\,inv}$. 

The above properties of $\cC^{\uparrow}_{\rm P\,inv}$ observables render 
them at the same time uninteresting from the viewpoint of the OS reconstruction.
More generally we have

{\bf Proposition 5.2.}~{\it 
Observables $\cO \in \cC_+$ are mapped onto multiples of the 
`canonical' ground state $\psi_0$ in $\cH_{OS}$ if and only if the following 
`hyperclustering relation' holds.  
\be
\om_\infty([\vartheta\cO]\,\cO)=  \om_\infty(\vartheta \cO) \,
\om_\infty(\cO)\;.
\label{hyperOS}
\end{equation}
Sufficient conditions for (\ref{hyperOS}) to hold are: 
(i) $\overline{\cO}_{0,\infty}(A n^{\up})$ in (\ref{Oinftylimit}) has a 
unique (and hence invariant) limit as $A \ra \infty$.
(ii) $\cO \in \cC_{\rm ainv}^{\uparrow} \cup \cC^{\uparrow}_{P {\rm 
inv}}$.} 

{\it Proof.} The relation (\ref{hyperOS}) is equivalent to 
$\cO-\om_\infty(\cO)\in\cN$ being a null vector. This in turn is 
equivalent to $\cO$ and $\om_\infty(\cO)$ giving rise to the same vector 
in $\cH_{OS}$. But the latter is a multiple of the ground state, as
asserted. The condition (i) is sufficient because the Cauchy-Schwarz 
inequality then implies 
$\om_{\infty}([\cO - \om_{\infty}(\cO)] \cA) = 0$ for all 
$\cA \in \cC_+$, which for $\cA = \vartheta[ \cO - \om_{\infty}(\cO)]$
amounts to (\ref{hyperOS}). The fact that observables in 
$\cC_{\rm ainv}^{\uparrow}$ or in $\cC^{\uparrow}_{P {\rm inv}}$
have hyperclustering expectations has been seen before. \hfill $\blacksquare$

%%%%%%%%%%%%%%%%%%%%%%%%%%%%%%%%%%%%%%%%%%%%%%%%%%%%%%%%%%%%%%%%%%%%%%%%%%%%
\newsubsection{The action of ${\rm SO}(1,2)$ on $\cH_{OS}$.}

Next let us consider the action of ${\rm SO}(1,2)$ on the reconstructed
Hilbert space. Both (\ref{rhoL}) and (\ref{rhoLO}) have well defined limits
for $L \ra \infty$ given by 
\ba
\label{rhoOS}
(\rho_{\infty}(A)\psi)(n) \!\is \!\psi(A\inv n)\frac{\cP_{-1/2}(A\inv n\cdot n^\up)}
{\cP_{-1/2}(n\cdot n^\up)}\,,\quad \psi \in \cC_0\,,
\\[2mm]
(\rho_{\infty}(A)\cO)(n_{x_0},\ldots,n_{x_{\ell-1}}) \!\is \!\cO(A\inv n_{x_0},\ldots, 
A\inv n_{x_{\ell-1}})\frac{\cP_{-1/2}(A\inv n_{x_{\ell-1}}\cdot n^\up)}
{\cP_{-1/2}(n_{x_{\ell-1}}\cdot n^\up)}\,,\;\;\, \cO \in \cC_+\,.
\nonumber
\end{eqnarray}
Here $\rho_\infty(A)$ is a well defined bounded linear map from $\cC_+$ 
onto itself because the quotient $\cP_{-1/2}(A\inv n\cdot n^\up)/
\cP_{-1/2}(n\cdot n^\up)$
is a bounded continuous function with a bounded inverse. One readily verifies the 
representation property $\rho_{\infty}(A) (\rho_{\infty}(B) \psi)(n) = 
(\rho_{\infty}(AB)\psi)(n)$. Further the action 
on $\cC_+$ is again compatible with that on $\cC_0$ and induces it 
via (\ref{represTD}), namely: $(\rho_{\infty}(A)\cO)_0(n) =
(\rho_{\infty}(A)\cO_0)(n)$, for all $A \in {\rm SO}(1,2)$. In particular 
this ensures that the null space $\cN$ and its completion are mapped onto 
itself under $\rho_{\infty}$. For clarity's sake let us add the reminder 
that for $\cO \in \cC_+$ the assignment of 
$x_{\ell -1} \geq 0$ as the index of the last argument on which $\cO$ actually 
depends is ambiguous since one may always consider a constant dependence on 
further arguments; see the comment after Eq.~(\ref{kapparho}). 

{\bf Proposition 5.3.}  {\it (i) The representation $\rho_{\infty}$ of 
${\rm SO}(1,2)$ on $\cH_{OS}$ is uniformly bounded and measurable. 
(ii) It does {\it not} act unitarily on all of $\cH_{OS}$.}

{\it Remark 1.}
Uniform boundedness means that 
$\sup_A \Vert \rho_{\infty}(A) \psi \Vert_{OS} < \infty$, measurability of 
the representation means that the functions $A \mapsto
(\psi_1, \rho_{\infty}(A) \psi_2)_{OS}$ and $A \mapsto
(\rho_{\infty}(A) \psi_1, \psi_2)_{OS}$ are measurable wrt the   
Haar measure on ${\rm SO}(1,2)$.
 
{\it Remark 2.} The fact (ii) may be surprising at first sight, upon 
second thought it is not: the inner product $(\;\,,\;)_{OS}$ 
is constructed in terms of the limiting expectation functional 
$w-\lim_{L \ra \infty} \omega_L = \om_{\infty}$, 
and we already know that this functional is not $\rho$ invariant for all $\cO \in
\cC_b$. Of course $\rho_{\infty}$ is different from $\rho$ but it seems 
`unlikely' that the universal ratio $\cP_{-1/2}(A\inv n \!\cdot\! n^{\up})/
\cP_{-1/2}(n \!\cdot \!n^{\up})$ by which they differ could 
`undo' the symmetry breaking for {\it all} of the relevant observables
at the same time. As a consequence the `square root' of a bounded observable 
signaling the $\rho$ symmetry breaking is likely to give rise to a wave function 
in $\cC_0$ on which the unitarity of $\rho_{\infty}$ is violated.  

{\it Proof of Proposition 5.3:} 
(i) By (\ref{nlimit}) in fact $\Vert \rho_{\infty}(A) \psi \Vert_{OS} \leq 
\Vert |\psi|^2 \Vert_{\infty}$ using (\ref{nlimit}). 
Measurability follows from the fact that for each $L$ the functions 
$A \mapsto \om_L( \psi_1^* \,\rho_{\infty}(A) \psi_2)$ and 
$A \mapsto \om_L( (\rho_{\infty}(A) \psi_1^*) \,\psi_2)$ are in 
$L^{\infty}({\rm SO}(1,2))$. 
By construction of the state $\om_{\infty}(\,.\,) = w-\lim_{L \ra \infty} 
\om_L(\,.\,)$ the $L \ra \infty$ limit of the above functions exists 
pointwise for almost all $A \in {\rm SO}(1,2)$ wrt the Haar measure.     
On general grounds the limiting functions are therefore measurable. 
As a warning we should add that for generic $\psi_1,\psi_2$ continuity 
in $A$ may be lost in the limit, as we shall see later. 

(ii) It  suffices to give examples. One class is 
provided by wave functions only depending on the ${\rm SO}^{\up}\!(2)$  
phases. Consider $\psi_l(n) := e^{i l \varphi}$, with $l \in \Z$
and $\varphi = \arctan(n^1/n^2)$. Then 
\be 
0= (\psi_l, \psi_{l'})_{OS} \neq     
(\rho_{\infty}(A)\psi_l,\rho_{\infty}(A)\psi_{l'})_{OS} \,,\quad l \neq l'\,.  
\label{nonunitary2}
\end{equation}
An example for the square root construction mentioned in remark 2 is
\be 
S_e(n) := [T_e(n)]^{1/2}\,,
\label{Sdef} 
\end{equation}
where $T_e(n)$ is the symmetry breaking observable of Eq.~(\ref{Tdef}) and 
the principal branch of the square root is taken. Then 
\be 
\overline{T}_q(\infty) = (S_e, S_e)_{OS} \neq (\rho_{\infty}(A)S_e, 
\rho_{\infty}(A) S_e)_{OS} = \lim_{n^{\up} \cdot n \ra \infty}\!\!  
\overline{T_e(A \inv n) \!\left(\frac{\cP_{-1/2}(n \!\cdot \!A n^\up)}%
{\cP_{-1/2}(n \!\cdot \!n^\up)}\right)^2}\,,
\label{nonunitary1}
\end{equation}
for $A \in {\rm SO}(1,2)/{\rm SO}^{\up}\!(2)$ and with the overbar 
referring to the ${\rm SO}^{\up}\!(2)$ average. 
\hfill $\blacksquare$

Of course one could also extend the action of 
$\rho$ from $L^2(\H)$ to $\cC_0$ and thereby to $\cH_{OS}$ in the obvious 
way. It acts, however, uninterestingly: first of all $\psi_0$ is mapped 
onto itself, likewise all elements of $\cC^{\up}_{\rm ainv}$ are 
mapped onto a multiple of $\psi_0$. Thus $\rho$ acts `unitarily' on 
multiples of $\psi_0$ by not acting at all, and since outside of the 
class $\cC^{\up}_{\rm ainv}$ symmetry breaking is generic, $\rho$ 
cannot be expected to act unitarily on sizeable subspaces of $\cH_{OS}$.      

On which subspaces of $\cH_{OS}$ does $\rho_{\infty}$ act unitarily? 
Let us introduce the following subsets of $\cH_{OS}$: first let 
$\cH^0_{OS}$ be the closed linear subspace generated by the `ground state 
orbit'
\be
\{\psi\in\cH_{OS}\, 
\vert\ 
\psi=\rho_\infty(A)\psi_0,\ A\in 
{\rm SO}(2,1)\}\,,  
\label{vacua}
\end{equation}
and $\cH^{\alpha}_{OS}$ be the closed linear subspace generated by
\be
\{\psi\in\cH_{OS}\, \vert\ \psi=\rho_\infty(A)\psi_{\alpha},\ A\in 
{\rm SO}(2,1)\}\,,  \quad \alpha \in \R\setminus\{0\}\,,
\label{discalpha}
\end{equation}
with $\psi_{\alpha}(n) = \exp(i \alpha \,n^{\up} \!\cdot\! n)$.  
$\cH^0_{OS}$ does not change if we allow the coefficients to be from 
$\cC^{\up}_{\rm ainv}$.  
$\cH^0_{OS}$ and $\cH^{\alpha}_{OS}$ are by construction invariant 
subspaces of $\cH_{OS}$ under the action of the representation 
$\rho_\infty$. It is convenient to introduce the notation 
\be 
\psi_{n_0,\alpha}(n) := \frac{\cP_{-1/2}(n_0\cdot n)}{\cP_{-1/2}(n^{\up} \cdot n)}
\,e^{i \alpha n_0 \cdot n} \,,\quad \n_0 \in \H\,,\;\; \alpha \neq 0\,,
\label{discbasis}
\end{equation} 
for the basis vectors; then $\rho_{\infty}$ acts simply by `rotating' $n_0$, 
i.e.~$\psi_{n_0,\alpha} \ra \psi_{A n_0,\alpha}$, $A \in {\rm SO}(1,2)$. 
Note that this action inherits the properties of the action of ${\rm SO}(1,2)$
on $\H$. As such it is transitive and effective but not free. It is not free 
because $A n = n$ for some fixed $n \in \H$ implies only that $A$ is in 
stability group of $n$. The action is manifestly transitive and also effective 
in that $A n =n$ for all $n \in\H$ implies $A =\1$.   
We define $\cH_{OS}^{\rm p}$ ($p$ being mnemonic for `phase' or 
`polymer') as the closed subspace generated by all 
the vectors $\psi_{n_0,\alpha}$, $\alpha\neq 0$, in (\ref{discbasis}). 

We now describe how $\rho_{\infty}$ acts on $\cH_{OS}^0$ 
and $\cH_{OS}^{\rm p}$:

{\bf Theorem 5.4.} {\it $\cH_{OS}^0$ and $\cH_{OS}^{\rm p}$ are orthogonal 
subspaces of $\cH_{OS}$. 
$\rho_{\infty}$ acts unitarily on both of these subspaces;
the action is {\it continuous} on $\cH_{OS}^0$, but {\it discontinuous}
on $\cH_{OS}^{\rm p}$. 
Furthermore,  on $\cH_{OS}^0$ one has
\be
(\rho_{\infty}(A)\psi_0,\rho_\infty(B)\psi_0)_{OS}= \cP_{-1/2}(A n^{\up}
\cdot B n^{\up})\,,
\label{vacscalar}
\end{equation}
whereas on $\cH_{OS}^{\rm p}$ one has 
\be
(\psi_{n_1,\alpha_1}\,,\,\psi_{n_2,\alpha_2})_{OS} = \delta_{n_1,n_2}
\delta_{\alpha_1,\alpha_2} \,,\quad \forall \, n_1,n_2 \in \H\,,\;\;
\alpha_1,\alpha_2 \in \R\,,\;\alpha_1 \alpha_2 \neq 0\,.
\label{discscalar}
\end{equation}}

{\it Proof.} The derivation of Eqs (\ref{vacscalar}), (\ref{discscalar}) 
as well as the proof of the orthogonality of the two subspaces is 
somewhat technical and is deferred to appendix C.  

From (\ref{vacscalar}) it is easy to see that $\rho_{\infty}$ acts 
continuously on $\cH_{OS}^0$:
By (\ref{vacscalar}) the scalar product of any two elements
in the orbit of $\psi_0$ is a continuous function of the group elements,   
and this continuity trivially extends to finite linear combinations of
elements of this orbit. Denote this linear space by $\cD$. This implies
that for any $\phi\in\cD$ we have $\lim_{A\to 0}
\|\rho_\infty(A)\phi-\phi\|_{OS}=0$.
Now for any element $\psi\in\cH_{OS}^0$ and any $\epsilon>0$ there is a 
$\phi_\epsilon\in\cD$ such that $\|\phi_\epsilon-\psi\|_{OS}<\epsilon$. By 
the triangle inequality therefore $\lim_{A\to 0}
\|\rho_\infty(A)\psi-\psi\|_{OS}\leq \epsilon$, and since $\epsilon$ was
arbitrary, $\lim_{A\to 0}\|\rho_\infty(A)\psi-\psi\|_{OS}=0$ follows. The 
group structure of SO(1,2) yields (strong) continuity of the whole 
representation $\rho_\infty\big|_{\cH^0_{OS}}$.
The discontinuity of the action of $\rho_{\infty}$ on  $\cH_{OS}^{\rm p}$
is obvious from (\ref{discscalar}).
\hfill $\blacksquare$

{\bf Corollary 5.5} {\it $\cH_{OS}$ is nonseparable.}

{\it Proof.} The vectors (\ref{discbasis}) provide an explicit 
nondenumerable orthonormal family. \hfill $\blacksquare$

{\it Remark.} The representation $\rho_{\infty}$ of ${\rm SO}(1,2)$  
acts as a kind of `nondenumerable discrete permutation group',
$\rho_{\infty}(A) \psi_{n_0,\alpha} = \psi_{A n_0, \alpha}$, on the 
orthonormal family (\ref{discbasis}) (see Appendix C). 

The above result should be viewed in the context of an alternative described  
by Segal and Kunze (\cite{SegalKunze}, p.~274) which characterizes 
measurable unitary representations $\pi$ of some 
locally compact group $G$ on a nonseparable Hilbert space $\cH$. Namely 
let $\cH^{s}$ be the subspace of all vectors $\psi_s$ in $\cH$ such that for all 
$\varphi \in L^1(G)$ and all $\psi \in \cH$ one has 
\be 
\int_G \!d\mu(A)\,\varphi(A) (\psi, \pi(A) \psi_s) =0\,.
\label{SK1}
\end{equation}
Then $\cH$ is the direct sum of two invariant subspaces $\cH = \cH^c \oplus 
\cH^s$. The restriction of $\pi$ to $\cH^c$ is continuous while the 
restriction to $\cH^s$ is singular, in the sense that 
$(\psi_s, \pi(A) \psi_s) =0$ for almost all $A \in G$ and all 
$\psi_s \in \cH^s$. If $\cH$ is separable $\cH^s$ is absent, as follows 
from a theorem of von Neumann (see \cite{ReedS}, Theorem VIII.9).      
     
In our case we denote by $\cH_{OS}^{u}$ the maximal 
closed subspace of $\cH_{OS}$ on which $\rho_{\infty}$ is unitary 
and measurable. The above alternative entails that 
$\cH_{OS}^{u}$ decomposes into $\cH_{OS}^{u} = \cH_{OS}^c \oplus 
\cH_{OS}^s$, where the restriction of $\rho_{\infty}$ to 
$\cH_{OS}^c$ and $\cH_{OS}^s$ is continuous and singular, respectively. 
Our results amount to the explicit construction of subspaces
\be 
\cH_{OS}^0 \subset \cH_{OS}^c\,,\sspace 
\cH_{OS}^{\rm p} \subset \cH_{OS}^s\,,
\label{SK2} 
\end{equation} 
together with a formula for the inner products. In particular the singular 
subspace is non-empty (for which the nonseparability of the Hilbert space 
is a necessary but not a sufficient condition). The assumption (\ref{SK1})
is satisfied for $\psi \in \cH^{\rm p}_{OS} \subset \cH^s_{OS}$ because 
(\ref{discscalar}) projects 
onto a 1-dimensional submanifold of the group which has 
zero measure wrt the full Haar measure. The restriction of $\rho_{\infty}$ 
to $\cH^{\rm p} \subset \cH^s$ is indeed singular; in fact by 
(\ref{discscalar}) $(\psi_{n_0,\alpha}, \rho_{\infty}(A) 
\psi_{n_0,\alpha})_{OS}=0$ holds for all $A \neq \1$. The simple explicit 
action $\rho_{\infty}(A) \psi_{n_0,\alpha} = \psi_{A n_0,\alpha}$  
as a permutation group (acting transitively and effectively for fixed $\alpha$) 
is somewhat surprising. The continuous subspace  
$\cH_{OS}^0$ will later be identified as the ground state sector
of the reconstructed transfer operator.  

As outlined in appendix C there are other nondenumerable orthonormal families
in $\cH_{OS}$ which are orthogonal to both $\cH_{OS}^0$ and $\cH_{OS}^{\rm p}$. 
We did not explore the action of $\rho_{\infty}$ on them, but it may well be 
that $\cH_{OS}$ contains other invariant subspaces on which $\rho_{\infty}$ 
acts unitarily. In this case they would likewise be subject to the 
above continuous-discontinuous alternative and render the inclusions
in (\ref{SK2}) proper.

\medskip
We proceed with the construction of a transfer operator $\T_{OS}$ on 
$\cH_{OS}$, which in particular will justify the term `ground state sector'
for $\cH_{OS}^0$ (see Proposition 5.6 below). To this end let $\tau$ be 
the 
map from $\cC_+$ to $\cC_+$ that shifts all variables by 1 unit to the right, 
i.e.~$\tau\cO(n_{x_1}, \ldots, n_{x_{\ell}}) = 
\cO(n_{x_1+1},\ldots n_{x_\ell+1})$. $\tau$ satisfies the relation $\tau 
\vartheta 
\tau = \vartheta$; it maps $\cN$ into itself as can be seen by using 
the Cauchy-Schwarz inequality and translation invariance 
\be
\om_\infty(\vartheta [\tau \cO] \tau\cO)=
\om_\infty([\vartheta \cO]\,[\tau^2\cO])
\le
\om_\infty([\vartheta \cO]\, \,[\tau^4\cO])^{1/2} \;\om_\infty([\vartheta 
\cO]\, O)^{1/2}\,.
\label{TOSdef}
\end{equation}
$\tau$ therefore induces a well-defined operator $\T_{OS}$ on the 
equivalence classes modulo $\cN$, and hence on $\cH_{OS}$. By translation 
invariance $\T_{OS}$ is symmetric. Once known to be bounded it  
extends to a unique selfadjoint operator on $\cH_{OS}$. The boundedness 
follows by iterating the Cauchy-Schwarz inequality (using a classic 
argument of Osterwalder and Schrader) 
\ba
\om_\infty( [\vartheta \cO]\,\tau^2\cO) & \! \le \! & 
\om_\infty( [\vartheta \cO]\, \tau^4\cO)^{1/2}
\om_\infty( [\vartheta \cO]\, \cO)^{1/2}\le\ldots
\nonum 
& \! \le \! &  
\om_\infty( [\vartheta \cO]\, \tau^{2^{n+1}}\cO)^{2^{-n}}
\om_\infty( [\vartheta \cO]\, \cO)^{1-2^{-n}}\,.
\end{eqnarray}
The first factor is bounded by $\Vert \cO\Vert^{2^{-n+1}}$, which goes to 
1 as $n\to\infty$; the second factor goes to 
$\om_\infty([\vartheta\cO]\,\cO)$, which proves that $\Vert \T^2_{OS}\Vert\le 1$ and 
thus also  $\Vert \T_{OS}\Vert\le 1$.

Importantly, the vector $\psi_0$ corresponding to $\cO=\1$ is an 
eigenvector (of norm 1) of the reconstructed transfer operator $\T_{OS}$ 
with eigenvalue $1=\Vert \T_{OS}\Vert_{OS}$, i.e.~$\psi_0$ is a ground state of 
the system. Already the mere existence of at least one normalizable ground state
indicates that the reconstructed quantum mechanics given by
$(\cH_{OS},\T_{OS})$ is very different from the original one given by 
$(L^2(\H),\T)$. This mismatch can be traced back to the purely continuous 
spectrum of the original system, which in turn stems from the noncompactness of
the target space $\H$. A further drastic discrepancy is the nonseparability 
of $\cH_{OS}$. 

Similar surprising features arise already in the much simpler model with
flat target space $\R$, on which $\R$ also acts as an amenable symmetry. 
This example is also instructive because it shows that in the limit of 
an amenable symmetry the symmetry breaking disappears. We therefore 
discuss this example briefly in Appendix B.
 
Returning to the hyperbolic model, we summarise the properties of $\T_{OS}$: 

{\bf Proposition 5.6.}  {\it $\T_{OS}$ is a self-adjoint operator on 
$\cH_{OS}$ with following properties:
\vspace{-3mm}

\begin{itemize}
\item[(i)] $ ||\T_{OS}||=1\,$.
\item[(ii)] $\rho_\infty\circ\T_{OS} = \T_{OS} \circ \rho_{\infty}$.
\item[(iii)] $\T_{OS}\big|_{\cH_{OS}^0} = \1\big|_{\cH_{OS}^0}\,$.
\item[(iv)] $\cH_{OS}^0 = \{ \psi \in \cH_{OS} \,|\, \T_{OS} \psi = \psi \}$.
\item[(v)] $\T_{OS}$ acts on $\cC_+/\cN$, i.e.~on the representatives
(\ref{represTD}) as 
\be
(\T_{OS}^x \psi)(n)=\lb_{\beta}(0)^{-x}\cP_{-1/2}(n\cdot n^{\up})^{-1}\!\!
\int\! d\Omega(n')\cT_\beta(n\cdot n';x)\psi(n')\cP_{-1/2}(n'\cdot n^{\up})\,,
\label{Tinfty}
\end{equation}
up to an element of $\cN$.
\end{itemize}}

{\it Remark.}~(ii) and (iv) show that $\T_{OS}$, in contrast with $\T$,
has at least some point spectrum. Despite the concrete expression in (v), 
it seems difficult to say more about the spectrum of $\T_{OS}$ outside 
the vacuum space $\cH^0_{OS}$.
 
{\it Proof.}~(i) has already been shown; it is a general feature of the 
Osterwalder-Schrader reconstruction.

(ii) Recall that $\T_{OS}$ is defined in terms of
the shift $\tau$ on $\cC_+$. As $\tau$ trivially commutes with the
$\rho_{\infty}$ action (\ref{rhoOS}) of ${\rm SO}(1,2)$
on $\cC_+$ and both $\tau$ and $\rho_{\infty}$ preserve  the nullspace   
$\cN$, the same will be true for $\T_{OS}$ induced on the equivalence
classes. This gives (ii). 

(iii) A simple consequence of (ii) is that $\T_{OS}$ acts like the identity on 
$\cH_{OS}^0$, because $\tau$ acts like the identity on the constants, in 
particular on the unit observable $\cO =\1$, corresponding to the 
`canonical' vacuum $\psi_0$. By (ii) the same must hold for 
all elements of $\cH_{OS}^0$. Equivalently, the eigenspace of $\T_{OS}$ of 
eigenvalue $1$ contains  $\cH_{OS}^0$.

(iv)  Let $\cO\in\cC_+$ be such $\tau\cO-\cO\in\cN$.
Then also $\tau\cO_0 - \cO_0\in\cN$, because
$\tau \cO - \cO = \tau \cO_0 - \cO_0 + \tau(\cO - \cO_0) - (\cO - \cO_0)$,
and the last two terms on the rhs are in $\cN$. By the remark after
(\ref{represTD}) therefore $(\tau\cO_0)_0 - \cO_0\in\cN \cap \cC_0$, and 
it suffices to consider the exact identity $(\tau\cO_0)_0 = \cO_0$. From the
definition of the map (\ref{represTD}) one sees that all solutions
$\psi \in \cC_0$ of $(\tau \psi)_0 = \psi$ are such
that $\psi(n) \cP_{-1/2}(n \cdot n^{\uparrow})$ is an
eigenfunction  of $\cT_{\beta}$ of eigenvalue $\lambda_{\beta}(0)$.
From (\ref{TDlimit2}) one infers that the solutions lie in the closed
subspace of $\cC_0$ spanned by ratios of the form (\ref{discbasis}) with
$\alpha =0$.

(v) Since $(\tau \cO_0)_0 - (\tau \cO)_0 \in \cN$, for all $\cO \in \cC_+$, 
one can use (\ref{represTD}) to compute the action of $\tau$ on the 
representative $\cO_0$ as before. One finds (\ref{Tinfty}), first for $x=1$ 
and then by iteration for all $x \in \N$. 
\hfill $\blacksquare$

In addition to acting unitarily, $\rho_{\infty}$ also acts irreducibly on 
$\cH_{OS}^0$. This follows directly from the definition (\ref{vacua}). 
Alternatively one can use the addition theorem (\ref{Pprop}c) to replace 
the generating set $\rho_{\infty}(A)\psi_0, \, A \in {\rm SO}(1,2)$, by 
the alternative generating set $\cP^l_{-1/2}(\xi)/\cP_{-1/2}(\xi)$, $l 
\in \Z$. These functions are known to span an irreducible and unitary 
representation of ${\rm SO}(1,2)$; in the Bargmann classification it 
corresponds to the limit of the discrete series.

To sum up, we have found that the space $\cH_{OS}$ is nonseparable and that it 
carries a representation $\rho_\infty$ of the symmetry group. This representation 
acts unitarily and discontinuously on a nonseparable proper subspace of $\cH_{OS}$,
and unitarily and continuously on the separable subspace of ground states 
$\cH^0_{OS}$ of the reconstructed transfer operator $\T_{OS}$. This ground state 
sector is irreducible and can be described explicitly as 
\be
\cH_{OS}^0 \simeq \left\{\frac{\cP_{-1/2}(n^\up \!\cdot \!A\inv n)}%  
{\cP_{-1/2}(n^\up \!\cdot \!n)} \Biggr|  A \in {\rm SO}(1,2) \right\} \simeq 
\left\{\frac{\cP_{-1/2}^l(n^{\up} \!\cdot \!n)}%
{\cP_{-1/2}(n^\up \!\cdot \!n)}\Biggr|  l\in\Z \right\}
\subset \cH_{OS}\,,
\label{H0}
\end{equation}
where the symbol `$\simeq$' denotes equality of the span of the lhs and 
rhs for the equivalence classes modulo $\cN$. The inner product on 
$\cH_{OS}^0$ is given by $(\rho_{\infty}(A)\psi_0, \rho_{\infty}(B)\psi_0)_{OS} = 
\cP_{-1/2}(A n^{\up}\!\cdot \!B n^{\up})$. 
\vspace{-5mm}

%%%%%%%%%%%%%%%%%%%%%%%%%%%%%%%%%%%%%%%%%%%%%%%%%%%%%%%%%%%%%%%%%%%%%%%%%%%%%%%%%%%
\newsection{Conclusions and outlook}
We have found that the concept of spontaneous symmetry breaking for a 
nonamenable continuous internal symmetry group differs in some crucial 
ways from the familiar situation of an amenable symmetry group:
\begin{itemize}
\item
Symmetry breaking is unavoidable, even in dimensions 1 and 2, where it is 
forbidden for an amenable continuous symmetry. In one dimension the 
(improper) ground state in the quantum mechanical interpretation is 
infinitely degenerate; in the statistical mechanics interpretation 
invariant states over a `large' algebra cannot be defined by group 
averaging. These features have been worked out in some detail for an 
analytically solvable model, the hyperbolic spin chain with symmetry 
group ${\rm SO}(1,2)$.  
\item
In this 1-dimensional model, however, there is still some vestige of the 
large fluctuations that are responsible for the symmetry restoration in 
the compact and abelian models: the sequence of functional 
measures defined through the thermodynamic limit becomes 
concentrated at configurations `at infinity' of the hyperbolic plane. 
As a consequence a certain subclass of non-invariant observables gets 
averaged to yield an invariant result. The limit of the functional 
measures provides an invariant mean for this subclass of observables, 
while outside this class symmetry breaking is generic. 
\item
While the quantum mechanics described by our model can be simply 
interpreted as motion of a particle on the hyperbolic plane, with 
absolutely continuous spectrum of the  transfer matrix, the
Osterwalder-Schrader reconstruction based on the infinite volume 
expectation values yields some surprises: the reconstructed transfer 
matric has at least some point spectrum, in particular it has normalizable 
ground states, and the full reconstructed Hilbert space is nonseparable.
These features are, however, due to the noncompact nature of the symmetry 
group, not its nonamenability, as can be seen from the `flat' analogue 
discussed in Appendix B.
\item
The Osterwalder-Schrader reconstructed Hilbert space has a nonseparable 
proper subspace on which a unitary representation of ${\rm SO}(1,2)$ acts 
discontinuously as a kind of `nondenumerable discrete permutation group', 
not unlike the way the spatial diffeomorphism group acts on the embedded 
graphs in the framework of \cite{poly1,poly2}. In contrast, the space of 
ground states of the reconstructed transfer operator is separable and a 
nontrivial unitary representation of ${\rm SO}(1,2)$ acts on it 
continuously and irreducibly. These features are specific to the 
case of a nonamenable symmetry and are not present in the `flat' case.

\end{itemize}

In a follow-up paper we study these issues in the $D$-dimensional 
($D\geq 2$) version of the model, i.e.~the nonlinear sigma-model with a 
hyperbolic targetspace; see e.g.~\cite{AmitDav83,CR83,vHol87} for earlier 
investigations. There we use a combination of analytical techniques and 
of numerical simulations \cite{DNS}. We also expect that there will still 
be a marked difference between dimensions $D\le 2$ and $D\ge 3$: whereas 
in the low dimensional case there is, as stated above, dominance of 
highly boosted configurations, we expect that in $D\ge 3$ spontaneous 
symmetry breaking in the usual sense takes place, showing normal, 
approximately Gaussian fluctuations around a fully ordered state, in 
which for instance unbounded observables like $n_0^0$ have finite 
expectation values. Some time after the first version of this paper was 
posted on the web, a paper by Spencer and Zirnbauer \cite{SZ} appeared, 
which showed that indeed in dimensions $D\ge 3$ at low temperature the 
suitable defined spin fluctuations have finite moments.

It would be interesting to elucidate the physical meaning of the 
unavoidable spontaneous symmetry breaking in the context of Anderson 
localization, in which such nonlinear sigma models were studied for 
instance in \cite{Wegner79, HJKP, Hik, Efetov83, Efetov97}.

In order not to blur the discussion with (further) technicalities we 
contrasted here only the simplest compact and noncompact symmetric 
spaces. However the situation would be similar if the sphere 
$S^2 \simeq {\rm SO}(3)/{\rm SO}(2)$ and $\H \simeq {\rm SO}(1,2)/{\rm SO}(2)$ 
were replaced with any other dual pair of compact and noncompact 
riemannian symmetric spaces (see \cite{Campo} for the propagators).
A further generalization would be to consider a 
similar dynamical system where the variables take values in an arbitrary 
riemannian manifold. In particular this would allow one to examine 
the interplay between invariant dynamics and non-invariant states 
for the diffeomorphism group of the target manifold.  
    
Finally we cannot resist mentioning a potential application to quantum
gravity. Supposing that in a suitable topology an appropriate version
of the diffeomorphism group is nonamenable, variants of the above
concepts become applicable. This would suggest a scenario in which 
there is no diffeomorphism invariant ground state, yet a family
of selected observables has invariant expectations in each of
an infinite set of ground states, while outside this family spontaneous 
collapse of diffeomorphism invariance is generic.
\medskip

{\bf Acknowledgments:}  We like to thank A.~Duncan for the enjoyable 
collaboration in \cite{DNS}. M.N. also wishes to thank M. Lashkevich for 
contributing to another aspect of this project, and A. Ashtekar for 
asking about the reconstructed state space. E.S. would like to thank 
S. Ruijsenaars for helpful discussions. This work was supported by the EU 
under contract EUCLID HPRN-CT-2002-00325.

\newpage
%%%%%%%%%%%%%%%%%%%%%%%%%%%%%%%%%%%%%%%%%%%%%%%%%%%%%%%%%%%%%%%%%%%%%%%%%%%%%%%%%%%

\appendix
\newappendix{Harmonic analysis on $\H$} 

Let $a\cdot b = a^0 b^0 - a^1 b^1 - a^2 b^2$ be the bilinear
form of $\R^{1,2}$ and let ${\rm SO}_0(1,2) =:{\rm SO}(1,2)$ be the  
component of its symmetry group connected to the identity. 
Consider the hyperboloid 
$\H = \{ n \in \R^{1,2}\,|\, n\cdot n =1\;,n^0 >0\}$. 
It is isometric to the symmetric space ${\rm SO}(1,2)/{\rm SO}(2)$ and 
can be parameterized either by points $(\Delta,B),\, \Delta >0, B \in \R$, 
in the Poincar\'e upper half plane, or by geodetic polar coordinates 
$(\xi,\varphi),\, \xi\geq 1, -\pi \leq \varphi < \pi$,  via  
\begin{eqnarray}
n^0 \is \frac{1 + \Delta^2 + B^2}{2\Delta} = \xi\,,\sspace 
n^1 = - \frac{B}{\Delta} = \sqrt{\xi^2-1}\sin \varphi \,,
\nonum
n^2 \is \frac{-1 + \Delta^2 + B^2}{2\Delta} =  \sqrt{\xi^2-1}\cos \varphi \,.
\label{spins}
\end{eqnarray} 
The $(\xi,\varphi)$ parameterization is adapted to a preferred ${\rm SO}(2)$ 
subgroup of ${\rm SO}(1,2)$ which leaves $n^{\up} = (1,0,0)$ invariant 
and which we denote by ${\rm SO}^{\up}\!(2)$. We also note the relations 
$\Delta^{-1} = \xi  - \sqrt{\xi^2-1} \cos \varphi$,  
$B = \sqrt{1-\xi^{-2}}\sin \varphi/(\sqrt{1-\xi^{-2}}\cos \varphi -1)$.
For the invariant distance $n\cdot n' \geq 1$ of two points $n,n' \in \H$,
one has 
\be 
n \cdot n' = \frac{ \Delta^2 + {\Delta'}^2 + (B-B')^2}{2\Delta \Delta'} 
= \xi \xi' - (\xi^2 - 1)^{1/2} ({\xi'}^2 -1)^{1/2} \cos(\varphi - \varphi')\,.
\label{ndot}
\end{equation}
Function spaces on $\H$ come naturally equipped with the inner product
\begin{equation} 
(\psi_1,\psi_2) = \int \!d\Omega(n) \,\psi_1(n)^* \psi_2(n)\,,  
\label{inner} 
\end{equation}  
induced by the invariant measure $d\Omega(n) := 2 d^3 n \,\delta(n^2-1)\, 
\theta(n^0)$, which translates into $d\Omega(\Delta,B) = dB d\Delta \Delta^{-2}$ and 
$d \Omega(\xi,\varphi) = d\xi d\varphi$, respectively.  
As indicated we shall freely switch back and forth between the different 
parameterizations. The Schwartz space $\cS(\H)$ is defined as the 
space of smooth functions on $\H$ decaying faster than any power of
$B$ and $\Delta$.  The space of tempered distributions 
$\cS'(\H)$ on it together with $L^2(\H)$ form a Gel'fand space triple
\begin{equation} 
\cS(\H) \subset L^2(\H) \subset \cS'(\H)\,.
\label{Gelfand}
\end{equation}  
The ${\rm SO}(1,2)$ rotations of the `spins' $n$ induce a
unitary representation $\rho$ on $\cS(\H)$ via $\rho(A)\psi(n)    
= \psi(A\inv n)$, $A \in {\rm SO}(1,2)$. On integral operators 
$K$ with kernel $\kappa(n,n')$ it acts as $K \ra \rho(A)^{-1} K \rho(A)$ 
and thus as $\kappa(n,n') \ra \kappa(A n,A n')$ on the kernels. 
Invariant operators
have kernels depending on the inner product $n\cdot n'$ only.
Similarly operators invariant under $\rho$ restricted to the ${\rm SO}^{\up}\!(2)$ 
subgroup have kernels depending on $\xi,\,\xi'$ and the relative 
angle $\varphi - \varphi'$ only. 
In general the representation $\rho$ will not be irreducible. 
Generic functions in $\cS(\H)$ can be expanded into a generalized
Fourier integral whose basis functions form unitary irreps 
of ${\rm SO}(1,2)$. Moreover these basis functions comprise 
$\cS'(\H)$ eigenfunctions of the Laplace-Beltrami operator.   

To make this concrete consider the Killing vectors of $\H$, 
which generate the Lie algebra $sl_2$  
\begin{equation} 
\begin{array}{lll}
{\bf e} = \dd_B\;,\quad & {\bf h} = 2(B \dd_B + \Delta \dd_{\Delta})\;,
\quad & {\bf f} = (\Delta^2 -B^2) \dd_B - 2 B \Delta \dd_{\Delta}\;,
\\[2mm]
[{\bf h},{\bf e}] = -2 {\bf e}\;, &
[{\bf h},{\bf f}] = 2 {\bf f}\;,&
[{\bf f},{\bf e}] = {\bf h}\;,
\label{Ksl2}
\end{array}
\end{equation} 
and are anti-hermitian wrt $(\;,\;)$. Up to a sign the quadratic 
Casimir coincides with the Laplace-Beltrami operator 
\begin{equation} 
- {\bf C} := \frac{1}{4} {\bf h}^2 + \frac{1}{2} ({\bf e}{\bf f} 
+ {\bf f}{\bf e} ) = \Delta^2 (\dd^2_{\Delta} + \dd_B^2)  = 
\frac{\dd}{\dd \xi} (\xi^2 -1) \frac{\dd}{\dd\xi} + 
\frac{1}{\xi^2 -1} \frac{\dd^2}{\dd \varphi^2} \,.
\label{Casimir}
\end{equation}  
If one just blindly lets the differential operators ${\bf e},\,{\bf 
h},\,{\bf f}$ act on the spins (\ref{spins}) (which are not elements of 
$L^2(\H)$) one sees that they act as $3 \times 3$ matrices $t({\bf e}), \, 
t({\bf h}),\, t({\bf f})$ with Casimir ${\bf C} = - 2 \1_3$; the matrices
are however not (anti-)hermitian even though the original differential 
operators (multiplied by $i$) are essentially self-adjoint on 
$\cS(\H) \subset L^2(\H)$.
The exponentiated differential operators therefore extend to the 
unitary action of ${\rm SO}(1,2)$ on $L^2(\H)$
\begin{equation} 
\rho(e^{-s t({\bf x})}) \psi(n) = e^{s {\bf x}} \psi(n) = \psi(e^{s t({\bf x})}n)\,,
\sspace {\bf x} = {\bf e},\, {\bf h},\,{\bf f};\quad s \in \R\,.
\label{rho}
\end{equation} 
A more explicit description of the exponentiated differential
operators is possible on irreducible representations. 

Simultaneous eigenstates of ${\bf C}$ and ${\bf e}$ are given by
\ba
&& \eps_{\om,k}(n) := \eps_{\om,k}(\Delta,B) = 
\Delta^{1/2} K_{i\om}(|k|\Delta) \,e^{ikB} \;,\quad k\neq 0\,.
\nonum
&& \eps_{\om,0}(n) := \eps_{\om,0}(\Delta,B) = \Delta^{i\om +1/2} \,,
\sspace \mbox{with} 
\nonum
&& 
{\bf C} \,\eps_{\om,k} = \Big(\frac{1}{4} + \om^2\Big) \eps_{\om,k}\;,\sspace
{\bf e} \,\eps_{\om,k} = ik \,\eps_{\om,k}\,,\quad \om >0\,,
\label{psis}
\end{eqnarray} 
where $K_{\nu}(x)$ is a modified Bessel function defined e.g.~by 
$K_\nu(\beta)=\int_0^\infty e^{-\beta\  {\rm cosh} t}{\rm cosh} (\nu t) dt$.
The Fourier inversion on $\cS(\H)$ takes the form 
\ba
\psi(n) \is \int_0^{\infty} \! \frac{d\om}{\pi^3}\,
\om \sinh \pi \om\, \int_{\R} dk \,\widehat{\psi}(\om,k)\,\eps_{\om,k}(n) 
\nonum
\widehat{\psi}(\om,k) \is \int \! d\Omega(n) \,\psi(n) \eps_{\om,k}(n)^* \,.
\label{psibasis} 
\end{eqnarray} 

Simultaneous eigenstates of ${\bf C}$ and ${\bf e} 
- {\bf f}$, i.e.~of the ${\rm SO}^{\up}\!(2)$ rotations are given by
\ba
&& \eps_{\om,l}(n) := \eps_{\om,l}(\xi, \varphi) = 
e^{i l \varphi}\,\cP^l_{-1/2 + i \om}(\xi) \,,\sspace l \in \Z\,,\;\;\om >0\,,
\nonum
&& {\bf C} \,\eps_{\om,l} = \Big(\frac{1}{4} + \om^2\Big) \eps_{\om,l}\;,\sspace
({\bf e} - {\bf f})\,\eps_{\om,l} = il \,\eps_{\om,l}\,,
\label{psils}
\end{eqnarray} 
where $\cP_s^l(\xi)$ are Legendre functions, defined e.g.~by 
\be
\cP_s^l(\xi) = \frac{\Gamma(s+ l + 1)}{2\pi \Gamma(s+1)} 
\int_0^{2\pi} \! du \, e^{il u} [\xi + \sqrt{\xi^2 -1} \cos u]^s\,,
\sspace \xi \geq 1\,.
\label{Pdef}
\end{equation}
We further note the following properties
\begin{subeqnarray}
&& \cP_s^l(\xi) = \cP_{-s-1}^l(\xi) = \frac{\Gamma(s+1+l)}{\Gamma(s+1-l)}
\cP_s^{-l}(\xi)\,,  
\\
&& \int_1^{\infty} \! d \xi \,\cP^l_{-1/2+ i\om}(\xi) \cP^{-l}_{-1/2 + i\om'}(\xi)  
= \frac{(-)^l}{\om \tanh \pi \om}\,\delta(\om - \om')\,, 
\\
&& \cP_s\Big(\xi \xi' - (\xi^2 -1)^{1/2} ({\xi'}^2 -1)^{1/2} \cos\varphi \Big) 
=\sum_{l \in \Z} (-)^l e^{i l \varphi} \, \cP_s^{-l}(\xi) \, \cP_s^l(\xi') \;,
\label{Pprop}
\end{subeqnarray} 
as well as the asymptotics for $\xi \ra \infty$
\ba 
\cP^l_{-1/2 + i\om}(\xi) &\sim& \frac{\Gamma(i\om)}{\sqrt{\pi} 
\Gamma(\frac{1}{2}+i\om -l )}  
(2 \xi)^{-1/2 + i\om} + c.c.\;,  \quad \om > 0\,,
\nonum
\cP^l_{-1/2}(\xi) & \sim & \frac{2}{\sqrt{\pi} \Gamma(\frac{1}{2} -l)}\, 
\frac{\ln \xi}{\sqrt{2 \xi}}\,.
\label{Pasym}
\end{eqnarray} 
The Fourier inversion in the basis (\ref{psilbasis}) takes the form 
\ba
\psi(n) \is \sum_{l \in \Z} (-)^l \int_0^{\infty} \! \frac{d\om}{2\pi}\,
\om \tanh \pi \om\,  
\widehat{\psi}(\om,l)\,\eps_{\om,-l}(n)\,, 
\nonum
\widehat{\psi}(\om,l) \is \int \! d\Omega(n) \,\psi(n) \eps_{\om,l}(n) \,.
\label{psilbasis} 
\end{eqnarray} 
In group theoretical terms the expansions (\ref{psibasis}), 
(\ref{psilbasis})
correspond to the decomposition of the unitary representation $\rho$ on 
$L^2(\H)$ into a direct integral of unitary irreducible representations, namely 
those of the type 0 principal series in the Bargmann classification, see
e.g.~\cite{VilKlim}.  In terms of the representation spaces
\be
L^2(\H) = \int^{\oplus} \!\!d\mu(\om) \,\cC_{\om}(\H) \;,
\label{regdecom}
\end{equation}
with the spectral weight $d\mu(\om) = \frac{d\om}{2\pi} \,\om \tanh\om$.   
We shall frequently encounter ${\rm SO}^{\up}\!(2)$ invariant functions 
$\psi = \psi(\xi)$, 
for which (\ref{psilbasis}) reduces to the Mehler-Fock transform
\ba
\psi(\xi) \is \int_0^{\infty} \! \frac{d \om}{2\pi}\, \om \tanh (\pi\om)\,
\cP_{-1/2+ i\om}(\xi) \,\widehat{\psi}(\om)\;,
\nonum
\widehat{\psi}(\om) \is 2\pi \int_1^{\infty} \!d\xi\,
\cP_{-1/2 + i\om}(\xi) \,\psi(\xi) = \widehat{\psi}(\om,0)\,.
\label{MehlerFock}
\end{eqnarray}   
It holds in the classical sense provided 
\begin{equation}
\int_1^\infty 
d\xi|\psi(\xi)|^2<\infty \quad 
\Longleftrightarrow \quad 
\int_0^\infty|\widehat{\psi}(\om)|^2\om \tanh\pi\om<\infty\,,
\label{MehlerFock2}
\end{equation}
see e.g.~\cite{Dym}.  It is possible, however, to interprete the 
Mehler-Fock transform in the distributional sense and therefore give it a 
wider range of applicability. 

The Fourier decomposition of a kernel $\kappa(n,n')$ defining an integral
operator $K$ makes some of its properties manifest. Subject to suitable 
regularity conditions the generic form of the expansion wrt the basis 
(\ref{psilbasis}) is 
\be
\kappa(n,n') = \sum_{l_1, l_2\in \Z}  (-)^{l_1 + l_2} 
\int_0^{\infty} \! \frac{d\om_1}{2\pi}  \frac{d\om_2}{2\pi} \,
\om_1 {\rm th} \pi \om_1 \,\om_2 {\rm th} \pi \om_2 \, 
\widehat{\kappa}_{l_1.l_2}(\om_1,\om_2) \, \eps_{\om_1,-l_1}(n) \, 
\eps_{\om_2, -l_2}(n')\,,
\label{KFT}
\end{equation} 
Depending on the properties of the spectral weight 
$\widehat{\kappa}_{l_1,l_2}(\om_1,\om_2) = (\eps_{\om_1,l_1}, 
\, K \eps_{\om_2,l_2})$ the corresponding integral operator 
$K$ will enjoy certain bonus properties:
\begin{subeqnarray} 
\label{Kprop}
\widehat{\kappa}_{l_1,l_2}(\om_1, \om_2) &=& 
\frac{2\pi \widehat{\kappa}_{l_1,l_2}(\om_1)} 
{\om_1 \tanh \pi \om_1} \delta(\om_1 - \om_2) \;\bspace \,\mbox{translation inv.}
\\
\widehat{\kappa}_{l_1,l_2}(\om_1, \om_2) &=& \delta_{l_1 + l_2 ,0}\;
\widehat{\kappa}_{l_1}(\om_1,\om_2)  \sspace \bspace \;
{\rm SO}^{\up}\!(2)\;\;\mbox{inv.}
\\
\widehat{\kappa}_{l_1,l_2}(\om_1, \om_2) &=& \delta_{l_1 + l_2,0} \;
\frac{2\pi \widehat{\kappa}(\om_1)} {\om_1 \tanh \pi \om_1} \delta(\om_1 -
\om_2) 
\sspace {\rm SO}(1,2)\;\;\mbox{inv.}
\end{subeqnarray} 
For $K$ itself these properties amount to a vanishing commutator with 
$\T,\, \rho|_{{\rm SO}^{\up}\!(2)}$ and $\rho$, respectively. 
The fact that the spectral weights (\ref{Kprop}c) lead to ${\rm SO}(1,2)$ invariant
operators follows from (\ref{Pprop}c); the kernels $\kappa(n,n')$ of these operators 
depend on the invariant distance $n\cdot n'$ only. 

As an example of a  spectral decomposition consider the transfer matrix itself 
where the weights are just the eigenvalues (\ref{lb}). Using e.g.~\cite{Grad}, 
p.804 and the completeness relation for the Legendre functions one verifies 
\be
\frac{\beta}{2\pi} \exp\{ \beta( 1- n\cdot n')\} 
= \int_0^{\infty} \! \frac{d\om}{2\pi} \, \om \tanh \pi \om \,
\cP_{-1/2 + i \om}(n\cdot n') \, \lb_{\beta}(\om) \,.
\label{VUsinglet}
\end{equation} 
In representation theoretical terms this expresses the exponential of a 
singlet wrt the non-unitary vector irrep as a superposition of singlets 
wrt the unitary irrep (\ref{regdecom}). We note that the inverse 
Mehler-Fock transform (\ref{MehlerFock}) gives 
\begin{equation}
\lambda_\beta(\om)=\beta \int_1^\infty d\xi 
e^{\beta(1-\xi)}\cP_{-1/2+i\om}(\xi).
\label{lbP}
\end{equation} 
Clearly the integral kernels $\kappa(n\cdot n')$ that give rise to well-defined
operators on $L^2(\H)$ must have suitable regularity and decay properties.
The asymptotics in (\ref{Pasym}) suggests that the kernels 
$\kappa(\xi)$ should also decay at least like $\xi^{-1/2}$. Some decay stronger 
than $\xi^{-1/2}$ is also necessary in order for $\kappa$ to be the integral 
kernel of a densely defined {\it operator} from $L^2(\H)$ to $L^2(\H)$. 
A sufficient condition seems to be more difficult to obtain, 
but in any case kernels like $n\cdot n'$ do {\it not} correspond to densely 
defined operators on $L^2(\H)$ (they give rise only to densely defined quadratic forms). 

The integrands of the Legendre functions (\ref{Pdef}) likewise provide 
eigenfunctions of the Laplace-Beltrami operator (\ref{Casimir}). Explicitly
\be
E_{\om,u}(n) := E_{\om,u}(\xi,\varphi) =
[\xi - \sqrt{\xi^2 -1} \cos(u - \varphi)]^{-1/2 - i \om}\;,
\label{Edef}
\end{equation}
are bounded complex solutions for all $|\varphi-u {\rm 
(mod}2\pi)|>\epsilon>0$), decaying like $\xi^{-1/2 + i \om}$ for $\xi \ra 
\infty$. The upper bound will diverge as $\epsilon\to 0$ because for 
$\phi=u$ one has $|E_{\om,u}(\xi,u)| \sim \sqrt{2 \xi}$, for $\xi 
\ra \infty$. The Legendre functions (\ref{Pdef}) are recovered as the Fourier 
modes of (\ref{Edef}) and vice versa. The orthogonality 
and completeness relations take the form
\ba
&& (E_{\om,\th},\,E_{\om',\th'}) = \frac{(2\pi)^2}{\om \tanh \pi \om}
\delta(\om-\om') \delta(\th - \th')\,,
\nonum
&& \frac{1}{(2\pi)^2} \int_0^{\infty} \! d\om \,\om \tanh \om\,
\int_0^{2\pi} \! d\th\, E_{\om,\th}(n)^* E_{\om,\th}(n') = \delta(n,n')\,.
\label{Ebasis}
\end{eqnarray}
The main virtue of these solutions is their simple transformation law under 
${\rm SO}(1,2)$, see e.g.~\cite{BV}. For a boost $A\inv = A(\th,\alpha)\inv$ mapping 
$\xi = n^0$ into $\xi \ch \th -\sh \th \cos(\varphi -\alpha)$ one has 
\be
E_{\om,u}(A\inv n) = [\ch \th + \cos(\varphi -\alpha) \sh \th]^{-1/2 - i \om}  
E_{\om,u'}(n)\,,
\label{Etrans}
\end{equation}
for some angle $u' = u'(\th,\alpha)$. This is also a convenient starting 
point to show that the Fourier decomposition (\ref{psilbasis}) 
indeed has the representation theoretical significance (\ref{regdecom}),
see e.g.~\cite{VilKlim}.  

%%%%%%%%%%%%%%%%%%%%%%%%%%%%%%%%%%%%%%%%%%%%%%%%%%%%%%%%%%%%%%%%%%%%%
\newappendix{Flat noncompact spin chain} 

In order to elucidate the relation between the original Hilbert space and
the one obtained by Osterwalder-Schrader reconstruction it is useful to 
consider the simplest noncompact spin chain where the target space is 
$\R$. The symmetry group in this case is also $\R$, which in contrast to 
SO(1,2) is amenable. Some of the unusual aspects of this model have 
been 
analyzed already in \cite{LMS}. Of course all results generalize trivially to 
target spaces $\R^n,\,n>1$. 

We consider the Hilbert space $L^2(\R)$ and take as the one-step 
transfer matrix simply the heat kernel $\exp [\beta^{-1}\Delta](u,v)$, so that
\ba
(\T^x\psi)(u) \is \int_{-\infty}^\infty \!dv \,\cT_{\beta}(u-v;x) \,\psi(v)\,,
\quad x \in \N\,,
\nonum
\cT_{\beta}(u;x) \is \sqrt{\frac{\beta}{2\pi x}} \exp
\left[-\frac{\beta}{2 x}\,u^2 \right].
\label{flattrans}
\end{eqnarray}
The transfer operator trivially commutes with the action of $\R$ on the wave
functions, i.e.~$\T \circ \rho = \rho \circ \T$, with $\rho(a) \psi(u) = \psi(u-a)$. 
It is well known that the spectrum of $\T$ is continuous and covers the 
interval $[0,1]$; the generalized eigenfunctions are imaginary 
exponentials. As in the hyperbolic case, a gauge fix is necessary; we 
simply fix the leftmost `spin' $u_{-L}$ to 0, which is analogous to fixing 
$n_{-L}=n^\up$. For the purpose of the Osterwalder-Schrader 
reconstruction we choose again in addition the bc $u_L=0$, i.e. we choose 
0 Dirichlet conditions. As observable algebra we take $\cC=\cC_b$, the algebra 
of continuous bounded functions of finitely many variables $u_{x_1},\ldots u_{x_\ell}$, 
and we introduce the subalgebras $\cC_+\, ,\cC_-$ and $\cC_0=\cC_+\cap 
\cC_-$ as in Section 5.
 
For a finite chain the reconstruction of the Hilbert space $\cH_L$ 
proceeds as in Section 5; we define for each $\cO\in\cC_+$
\begin{equation}
\cO_{0,L}(u_0)= \int\prod_{i=1}^\ell du_i\,
\cO(u_1,\dots,u_\ell)\prod_{i=1}^\ell\cT_\beta(u_i-u_{i-1};x_i-x_{i-1})
\frac{\cT_\beta(u_\ell;L-x_\ell)}{\cT_\beta(u_0;L)}\,,
\label{repflat}
\end{equation}
and
\begin{equation}
\psi^\cO(u)=\cO_{0,L}(u)\frac{\cT_\beta(u;L)}{\sqrt{\cT_\beta(0;2L)}}\,.
\label{isoflat}
\end{equation}
The reconstructed Hilbert space $\cH_L$ is the completion of $\cC_0$ 
wrt $\omega_L$ in (\ref{om1}). It can be 
identified with the original $L^2(\R)$ by the isometry
\be
V_L: \cH_L \rra  L^2(\R)\;, \sspace (V_L\psi)(n) =
\psi(u) \,\frac{\cT_{\beta}(u;L)}{\sqrt{\cT_{\beta}(0;2L)}}\,.
\end{equation}
Equivalently $\cH_L$ can be viewed as the preimage of $L^2(\R)$ wrt $V_L$.

The thermodynamic limit can be readily understood here. 
The ratio $\cT_{\beta}(u;L)/\cT(0;L)$ approaches a constant for 
$L \ra \infty$, signaling a unique ground state.  On elements $\cO_{0,\infty}\in\cC_0$ 
the expectation functionals becomes an invariant mean, 
which exists in this case. There is a subspace $\cH_{AP}$ of almost 
periodic functions on which this mean is unique, see \cite{Pat}; this 
subspace consists of the completion (in the Hilbert space norm defined by 
the mean) of the space of trigonometric polynomials. A brief account of 
the theory of almost periodic functions on $\R$, which is due to H. Bohr, 
can be found in \cite{AG}. For $\psi\in\cH_{AP}$ the mean is
\be
\om(\psi)=\lim_{L\to\infty} \frac{1}{\sqrt{2\pi L}}
\int \!du \,\exp\Big(\!-\frac{u^2}{2L}\Big)\psi(u)
=: \lim_{L \ra \infty} \om_L(\psi)\,.
\label{om1}
\end{equation}
A better known expression of the invariant mean on $\cH_{AP}$ is
\begin{equation}
\om(\psi)=\lim_{L\to\infty} \frac{1}{2L}\int_{-L}^L du\, \psi(u)\,,
\end{equation}
see for instance \cite{AG}. By the uniqueness these two expressions have 
to be the same for an almost periodic $\psi$  and it is straightforward to 
verify this equivalence for the dense subspace of trigonometric polynomials.
The scalar product induced by this invariant mean can be written as 
\begin{equation}
(\psi',\psi)_{OS}=\lim_{L\to\infty} \frac{1}{2L}\int_{-L}^L du\, 
\psi(u)^\ast \psi(u)\,,
\end{equation}
and the unitarity of $\rho$ on $\cH_{AP}$ is manifest.

It might be surprising that the Hilbert space obtained by the OS 
reconstruction from $\cC_0$ is nonseparable; but it is well known that 
already the space $\cH_{AP}$ is nonseparable \cite{AG}: there is an 
uncountable set of mutually orthonormal functions, namely the set 
\begin{equation}
\{\psi_\alpha(u)=e^{i\alpha u}\;\vert \;\alpha\in \R\}\,. 
\label{flateigen}
\end{equation}
One can introduce a shift automorphism $\tau$ like the one used in 
the hyperbolic case. From this one obtains a 
reconstructed transfer operator $\T_{OS}$ acting on $\cH_{OS}$; in this 
case it is nonnegative and has again norm 1. $\T_{OS}$ acts on $\cC_0$ 
simply by Eq.~(\ref{flattrans}). This shows that the functions 
(\ref{flateigen}) are eigenvectors (in the proper sense) of $\T_{OS}$ 
with eigenvalue $\exp(-\frac{1}{2\beta} \alpha^2)$. 

The relation between the original system $(L^2(\R),\T)$ and the  
reconstructed one $(\cH_{OS},\T_{OS})$ turns out to be simply that the 
spectrum as a set remains the same, namely the interval $[0,1]$. 
However there is now pure point spectrum on every point of the 
spectral interval and the generalized eigenfunctions become normalizable 
eigenstates. With respect to the representation of symmetry group $\R$ 
the original $L^2(\R)$ 
is a direct integral of the one-dimensional irreducible representations
on the imaginary exponentials (\ref{flateigen}), whereas $\cH_{AP}$ 
is (and hence $\cH_{OS}$ contains) a direct sum over the continuous 
parameters $\alpha$:
\begin{equation}
\cH_{OS} \supset \cH_{AP}=\bigoplus_{\alpha\in\R}\cH_\alpha\,,
\end{equation}
where $\cH_\alpha$ is the one-dimensional Hilbert space spanned by 
$e^{i\alpha u}$.

Let us end this appendix with the remark that the space $\cH_{AP}$, 
huge as it is, is still only a small subspace of the full space $\cH_{OS}$. 
It turns out that there are uncountably many more functions orthogonal to 
the exponentials discussed so far, for instance the functions 
$p_\alpha(u)=|u|^{i\alpha}$. Using distributional Fourier transformation 
one can show that
\be
(p_\alpha, \psi_{\alpha'})_{OS}=0\;, \quad \forall  
\alpha\neq 0\,,\;\alpha'\in\R.
\end{equation}
Presumably these functions belong to the continuous spectrum overlaying 
the point spectrum we have found.

%%%%%%%%%%%%%%%%%%%%%%%%%%%%%%%%%%%%%%%%%%%%%%%%%%%%%%%%%%%%%%%%%%%%%%%%%%%%%%%%
\newappendix{Inner products on $\cH_{OS}^0$ and $\cH_{OS}^{\rm p}$}

Here we derive the formulas (\ref{vacscalar}) and (\ref{discscalar}) for the 
inner products on  $\cH_{OS}^0$ and $\cH_{OS}^{\rm p}$. We begin with 
(\ref{vacscalar}), i.e.~$(\rho_{\infty}(A) \psi_0, \rho_{\infty}(B) \psi_0)_{OS} = 
\cP_{-1/2}(A n^{\up} \! \cdot \! B n^{\up})$. By (\ref{1pt_TD}) this 
is equivalent to 
\ba
\lim_{n^{\up} \cdot n\to\infty} f_{A,B}(n^{\up} \cdot n) 
\is \cP_{-1/2}(A n^{\up} \!\cdot \!B n^{\up})\,,
\quad \mbox{with}
\nonum 
f_{A,B}(n^{\up}\cdot n) &:=& \overline{
\left(\frac{\cP_{-1/2}(n \!\cdot \!A n^\up)}{\cP_{-1/2}(n \!\cdot \!n^\up)}\right) 
\left(\frac{\cP_{-1/2}(n\!\cdot \!B n^\up)}{\cP_{-1/2}(n \!\cdot \!n^\up)}\right) 
}\,,
\label{nlimit}
\end{eqnarray}
where the bar as before denotes the average over ${\rm SO}^{\up}\!(2)$.
Writing  $\xi=n\cdot n^\up$ and momentarily $n \cdot An^\up= 
\xi \xi_A - \sqrt{\xi^2 -1} \sqrt{\xi_A^2 -1} \cos(\varphi - \varphi_A)$,
and similarly for $n \cdot Bn^\up$, the ${\rm SO}^{\up}\!(2)$ average  
evaluates by means of (\ref{Pprop}c) to 
\be
f_{A,B}(\xi)= \sum_{l\in\Z} e^{-i l (\varphi_A - \varphi_B)} 
\cP_{-1/2}^l(\xi_A)\cP_{-1/2}^{-l}(\xi_B) 
\bigg( \frac{\cP_{-1/2}^{l}(\xi) \cP_{-1/2}^{-l}(\xi)}{\cP_{-1/2}(\xi)^2}\bigg)\,.
\end{equation}
The series converges uniformly in $\xi$: using the Cauchy-Schwarz
inequality, the geometric-arithmetic mean inequality  and the bound  
$|\cP_{-1/2}^l(\xi)\cP_{-1/2}^{-l}(\xi)|\leq \cP_{-1/2}(\xi)^2$
the rhs is bounded by 1 and likewise the tail of the sum can be bounded 
uniformly in $\xi$.
Taking now the limit $\xi\to\infty$ under the sum, which is permitted 
because of the uniform convergence of the series, one obtains 
\be
\lim_{\xi\to\infty}f_{A,B}(\xi)=
\sum_{l\in\Z}e^{il (\varphi_A - \varphi_B)} (-)^l 
\cP_{-1/2}^l(\xi_A)\cP_{-1/2}^{-l}(\xi_B)
=\cP_{-1/2}(A n^{\up} \!\cdot \!B n^{\up})\,,
\end{equation}
using (\ref{Pasym}) and (\ref{Pprop}c). This gives (\ref{vacscalar}); note
that the result coincides with the one obtained from the `correlated' limit in 
(\ref{scalar0}).  

The derivation of (\ref{discscalar}) we break up in several steps. 
Recall the notation $\psi_\alpha(n)=\exp(i\alpha n^{\up} \cdot n)$, 
$\alpha\in\R\setminus\{0\}$. We first show that these functions form an 
orthonormal 
system 
\be
(\psi_\alpha, \psi_\alpha)_{OS} = 1\,,\sspace
(\psi_\alpha, \psi_{\alpha'})_{OS} =0\,,\quad {\rm for}\  \alpha\neq\alpha'\,.
\label{nonsep}
\end{equation}
The normalization is clear. For the orthogonality consider for $\alpha\neq 
0$
\be
I_{\alpha}(L):= \int_{1}^{\infty} \! d\xi \,e^{i\alpha\xi}\,
\frac{\cT_{\beta}(\xi;L)^2}{\cT_{\beta}(1;2L)}\,.
\label{nonsep1}
\end{equation}
To analyze this expression we integrate by parts and obtain
\be
I_\alpha(L)=-\frac{\cT_\beta(1;L)^2}{i\alpha\cT_\beta(1;2L)} 
\left\{e^{i\alpha}+ \int_{1}^{\infty} \! d\xi e^{i\alpha\xi}\ 
\frac{\partial}{\partial\xi}\left(\frac{\cT_\beta(\xi;L)}
{\cT_\beta(1;L)}\right)^2\right\}\,.
\label{nonsep2}
\end{equation}
The first term is $O(L^{-3/2})$ by (\ref{ZTDlimit}); the modulus of 
the second term can be bounded, using the monotonicity of 
$\cT_\beta(\xi;L)$ by 
\be 
- \frac{\cT_\beta(1;L)^2}{\alpha\cT_\beta(1;2L)}\int_{1}^{\infty} \! d\xi
\frac{\partial}{\partial\xi}\left(\frac{\cT_\beta(\xi;L)}
{\cT_\beta(1;L)}\right)^2=\frac{\cT_\beta(1;L)^2}{\alpha\cT_\beta(1;2L)}\,,
\label{nonsep3}
\end{equation}
which is also $O(L^{-3/2})$. Together, $\lim_{L\to\infty}I_\alpha(L)= 0$ and 
(\ref{nonsep}) is proven.

We remark that this construction readily generalizes to all wave functions 
oscillating `sufficiently fast' as $\xi \ra \infty$. Consider 
\be 
\psi_p(\n) = \exp\Big\{ i \int^{\xi}_1 du \,p(u)\Big\}\,\quad 
\mbox{with} \quad \lim_{\xi \ra \infty} \frac{(\ln \xi)^2}{\xi p(\xi)} =0\,.
\label{psip}
\end{equation}
Then every pair of wave functions $\psi_{p_1}(n), \,\psi_{p_2}(n)$, 
where the difference $p_1(\xi) - p_2(\xi)$ is strictly monotonous for suffiently 
large $\xi$ and obeys the decay condition in (\ref{psip}) is orthogonal: 
$(\psi_{p_1}, \psi_{p_2})_{OS} =0$, using Lemma 
2.2 (iii) to get bounds 
uniform in $L$ for the $\xi \ra \infty$ limits. 
For example $\exp\{i \alpha (\ln \xi)^4\}$, $\alpha \in \R$, provides another
nondenumerable orthonormal family, each  member of which is orthogonal to 
each of the plain exponentials in (\ref{nonsep}). Here we shall only pursue 
the plain exponentials $\psi_{\alpha},\,\alpha \in \R$, further.

Repeating the above computations with the transformed exponentials 
$\rho_{\infty}(A) \psi_{\alpha}$ one readily shows that they remain 
orthogonal if they were initially. For the computation of the norms the 
phases are irrelevant, so 
they remain unity if $(\rho_{\infty}(A)\psi_0,\rho_\infty(A)\psi_0)_{OS}=
(\psi_0,\psi_0)_{OS} =1$. This however is a special case of (\ref{vacscalar}).
Thus 
\be 
(\rho_{\infty}(A) \psi_{\alpha}, \rho_{\infty}(A) \psi_{\alpha'})_{OS} = 
(\psi_{\alpha}, \psi_{\alpha'})_{OS} \,,\quad \alpha,\alpha' \in \R\,.
\label{unitarityp}
\end{equation}
In a last step we show
\be 
(\rho_{\infty}(A) \psi_{\alpha}\,,\,\psi_{\alpha'})_{OS} = 0 \,,\sspace 
\forall\, A \in {\rm SO}(1,2)\,,\;\; \alpha\,, \alpha' \in \R\,,\;\;
\alpha \alpha' \neq 0\,.
\label{discorth}
\end{equation}
By definition one has 
\ba
\label{discorth1}
(\rho_{\infty}(A) \psi_{\alpha}\,,\,\psi_{\alpha'})_{OS} &=&
\lim_{L \ra \infty} 2\pi \int_1^{\infty} \! d\xi \,e^{i \alpha' \xi} 
J_{\alpha}(\xi,\xi_A) \,\frac{\cT_{\beta}(\xi,L)^2}{\cT_{\beta}(1,2L)}\;,
\\[2mm]
J_{\alpha}(\xi,\xi_A) &:=& \int_0^{2\pi} \! \frac{d \varphi}{2\pi}\, 
e^{-i \alpha A n^{\up} \cdot n}\, 
\frac{\cP_{-1/2}(A n^{\up} \cdot n)}{\cP_{-1/2}(\xi)}\,,
\nonumber
\end{eqnarray}
where we view $A n^{\up} \!\cdot \!n =  \xi \xi_A - (\xi^2 -1)^{1/2} (\xi_A^2 -1)^{1/2} 
\cos(\varphi -\varphi_A)$ as a function of $\xi,\xi_A$ and $\varphi - \varphi_A$.   
As anticipated by the notation $J_{\alpha}(\xi,\xi_A)$ is independent of 
$\varphi_A$. Clearly $J_{\alpha}(\xi,1) = e^{-i \alpha \xi}$ and 
$|J_{\alpha}(\xi,\xi_A)| \leq \cP_{-1/2}(\xi_A)$ by the addition theorem 
(\ref{Pprop}c). We take now $\xi_A > 1$ and by (\ref{unitarityp}) we may also assume 
that $\alpha \neq 0$ and wlog $\alpha >0$ (while $\alpha' \in \R$ may be zero). 
By the argument familiar from section 4.2 only the behavior of 
$J_{\alpha}(\xi,\xi_A)$ for large $\xi$ will be relevant for the inner product 
(\ref{discorth1}). We claim that 
\ba
&& J_{\alpha}(\xi,\xi_A) \sim \frac{1}{\sqrt{\alpha \xi}} \Big( Q_+(\xi_A) 
e^{- i \alpha p_+(\xi_A) \xi} + Q_-(\xi_A) e^{- i \alpha p_-(\xi_A) \xi} \Big)
\quad \mbox{as}\;\;\xi \ra\infty\,,
\nonum
&& \mbox{with}\quad p_{\pm}(\xi_A) = \xi_A \pm \sqrt{\xi_A^2 -1} \,,
\label{Jasym}
\end{eqnarray}
and some complex constants $Q_{\pm}(\xi_A)$ nowhere zero for $\xi_A > 1$.  
Note that $1 < \xi_A < p_+(\xi_A)$ and $0 < p_-(\xi_A) < 1$. 

We first show that the rhs of (\ref{Jasym}) is the leading term in an 
asymptotic expansion of $J_{\alpha}(\xi,\xi_A)$ for large $\xi$.    
The point to observe is that from (\ref{Pasym}) we have 
\be 
\cP_{-1/2}\Big(\xi \xi_A - (\xi^2 -1)^{1/2} (\xi_A^2 -1)^{1/2} \cos\varphi \Big)
\cP_{-1/2}(\xi)^{-1} \sim \frac{1}{[\xi_A - \sqrt{\xi_A^2 -1} \cos \varphi]^{1/2}}\,,
\label{Pratio}
\end{equation}
with additive corrections of $O(1/\ln \xi)$. Asymptotically the integral becomes
\be
J_{\alpha}(\xi,\xi_A) \sim e^{-i\alpha \xi \xi_A} 
\int_0^{2\pi} \frac{d\varphi}{2\pi} \frac{e^{i \alpha \xi \sqrt{\xi_A^2 -1} \cos \varphi}}%
{[\xi_A - \sqrt{\xi_A^2 -1} \cos \varphi ]^{1/2}}\,.
\label{Jasym2}
\end{equation}
For large $\xi$ this integral can now be evaluated by the method of stationary phase 
(see e.g.~\cite{Olver}) with the result (\ref{Jasym}). The constants $Q_{\pm}(\xi_A)$ 
come out as 
\be 
Q_{\pm}(\xi_A) = 2^{\pm 1/2} e^{\pm i \pi/4} \Big[2\pi \sqrt{\xi_A^2-1} 
\Big(\xi_A \pm \sqrt{\xi_A^2 - 1} \Big)\Big]^{-1/2} \,.
\quad
\label{Jasym3}
\end{equation}
Subleading terms in the asymptotic expansion of $J_{\alpha}(\xi,\xi_A)$ could 
be worked out similarly, but are not needed. The properties relevant in 
the following
are that $|J_{\alpha}(\xi,\xi_A)|$ vanishes for $\xi \ra \infty$, and that 
the phases are linear in $\xi$ with the given frequencies. To make sure 
that these are properties of $J_{\alpha}(\xi,\xi_A)$ and not just of its 
asymptotic expansion, we verified them numerically.

With (\ref{Jasym}) at our disposal, the rest of the derivation of (\ref{discorth}) 
is straightforward. Substituting (\ref{Jasym}) into (\ref{discorth1}) one shows for 
generic $\xi_A$ the vanishing of the $L \ra \infty$ limit along the lines of (\ref{nonsep1}) --
(\ref{nonsep3}). If both $\alpha$ and $\alpha'$ are nonzero one of the $\xi$-dependent 
phases might cancel for the special boost parameter 
$\xi_A = \frac{1}{2}( \frac{\alpha}{\alpha'} + \frac{\alpha'}{\alpha})$.    
The modulus of this term in the asymptotics of $J_{\alpha}(\xi,\xi_A)$ then is proportional 
to $(\alpha \xi)^{-1/2}$ which is an element of $\cC^{\up}_{\rm ainv}$, 
and the $L \ra \infty$ limit vanishes on account of (\ref{1pt_TD}). 
This establishes (\ref{discorth}). The result (\ref{discscalar}) then follows 
by combining (\ref{nonsep}), (\ref{unitarityp}), and (\ref{discorth}).

%%%%%%%%%%%%%%%%%%%%%%%%%%%%%%%%%%%%%%%%%%%%%%%%%%%%%%%%%%%%%%%%%%%
\newpage 


\begin{thebibliography}{10}
\small

\bibitem{Ruelle} D. Ruelle, {\it Statistical Mechanics}, W. A. Benjamin,
Reading, Mass. 1969.

\bibitem{Sewell} G. Sewell, Quantum mechanics and its emergent macrophysics,
Princeton UP, 2002. 

\bibitem{NT} H. Narnhofer and W. Thirring, Spontaneously broken     
symmetries, Ann.~Inst.~Henri Poincar\'e, {\bf 70} (1999) 1.

\bibitem{Pat} A. Paterson, {\it Amenability}, American Mathematical
Society, Providence, R.I.1988.

\bibitem{qernst} M. Niedermaier, Dimensionally reduced gravity theories
are asymptotically safe, \NP{B673} (2003) 131; M. Niedermaier and H.
Samtleben, An algebraic bootstrap for dimensionally reduced gravity,
\NP{B579} (2000)

\bibitem{Korch} L. Faddeev and G. Korchemsky, High energy QCD as a
completely integrable system, \PL{B342} (1995) 311; S. Derkachov, G. 
Korchemsky, and A. Manashov, Noncompact Heisenberg spin chains from high 
energy QCD, \NP{B617} (2001) 375; \NP{B661} (2003) 533.

\bibitem{Wegner79} F. Wegner, The mobility edge problem: continuous
symmetry and a conjecture, Z.~Phys.~{\bf B35} (1979) 207.

\bibitem{HJKP} A. Houghten, A. Jevicki, R. Kenway, and A. Pruisken,
Noncompact sigma-models and the existence of a mobility edge
in disordered electronic systems near two dimensions,
\PRL{45} (1980) 394.

\bibitem{Hik} S. Hikami, Anderson localization in a nonlinear sigma-model 
representation, \PR{B 24} (1981) 2671.


\bibitem{Efetov83} K.~B.~Efetov, Supersymmetry and theory of
disordered metals, Adv. Phys. {\bf 32} (83) 53.

\bibitem{Efetov97} K.~B.~Efetov, Supersymmetry in Disorder and Chaos,
Cambridge University Press, Cambridge, U.K. 1997.

\bibitem{MW} D. Mermin and H. Wagner, Absence of ferromagnetism or
anti-ferromagnetism in one or two-dimensional isotropic Heisenberg models,
Phys.~Rev.~Lett. {\bf 17} (1966) 1133.

\bibitem{DS} R. L. Dobrushin and S. B. Shlosman, Absence of 
breakdown of continuous symmetry in two-dimensional models of 
statistical physics, \CMP 42 (1975) 31.

\bibitem{DNS} T. Duncan, M. Niedermaier, and E. Seiler, Vacuum orbit and
spontaneous symmetry breaking in hyperbolic sigma-models, hep-th/0405143.

\bibitem{Georgii} H. O. Georgii, Gibbs measures and phase transitions, de
Gruyter, Berlin and New York 1988. 

\bibitem{greenleaf} F. P. Greenleaf, Amenable actions of locally compact 
groups, J. Funct. Anal. {\bf 4} (1969) 295.

\bibitem{eymard} P.~Eymard, Moy\'{e}nnes invariantes et
repr\'{e}sentations unitaires, Lecture Notes in Mathematics {\bf 300},
Springer-Verlag, Berlin-New York 1972.

\bibitem{AmitDav83} D. Amit and A. Davies, Symmetry breaking in the
non-compact sigma model, \NP{B225} (1983) 221.

\bibitem{vHol87} J.~W. van Holten, Quantum noncompact sigma models,
J. Math. Phys. {\bf 28}~(1987)~1420.

\bibitem{BO} D. Buchholz and I. Ojima, Spontaneous collapse of
supersymmetry, \NP{B498} (1997) 228.

\bibitem{LMS} J. L\"offelholz, G. Morchio and F. Strocchi, Spectral
stochastic processes arising in quantum mechanical models with a non-$L^2$
ground state, Lett. Math. Phys. {\bf 35} (1995) 251.

\bibitem{poly1} A. Ashtekar, J. Lewandowski, and H. Sahlmann,
Polymer and Fock representations for a scalar field, Class. Quant.
Grav. {\bf 20} (2003) L1.

\bibitem{poly2} A. Ashtekar, S. Fairhurst, and J. Willis, Quantum
gravity, shadow states, and quantum mechanics, Class. Quant. Grav. 
{\bf 20} (2003) 1031.

\bibitem{GroschSt88} C. Grosche and F. Steiner, The path integral on the
pseudosphere, Ann. Phys. {\bf 182} (1988) 120.

\bibitem{Schaefer} J. Schaefer, Covariant path integral on hyperbolic
surfaces, J. Math. Phys. {\bf 38} (1997) 11.

\bibitem{Campo} R. Camporesi, Harmonic analysis and propagators on
homgeneous spaces, Phys. Repts. {\bf 196} (1990) 1.

\bibitem{Anker} J.P. Anker and P. Ostellari, The heat kernel on noncompact
symmetric spaces, in: Lie groups and symmetric spaces, pp. 27-46, 
Amer. Math. Soc. Transl. Ser.2, 210, AMS. Providence, RI 2003.

\bibitem{Grad} I. Gradshteyn and I. Ryzhik, Table of integrals and
products, Academic Press, New York and London 1980.

\bibitem{Olver} F.~Olver, {\it Introduction to asymptotics
and special functions}, Academic Press, New York and London 1978.

\bibitem{ReedS} M. Reed and B. Simon, {\it Methods of Modern Mathematical
Physics}, vol. 1, Academic Press, New York and London 1972.

\bibitem{dixmier} J. Dixmier, $C^\ast$ algebras, North Holland, Amsterdam
1977.

\bibitem{SeilerY03} E. Seiler and K. Yildirim, Critical behavior in a
quasi D-dimensional spin model, J. Statist. Phys. {\bf 112} (2003) 457;
[hep-lat/0209166].

\bibitem{Ziegler} K.~Ziegler, Divergencies in a Vector Model with 
Hyperbolic Symmetry on a Chain, Z. Phys. B {\bf 43} (1981) 275.

\bibitem{GuiMar99} D. Giulini and D. Marolf, A uniqueness theorem for
constraint quantization, Class. Quant. Grav.~{\bf 16} (1999) 2489;
[gr-qc/9902045].

\bibitem{GomMar} A. Gomberoff and D. Marolf, On group averaging for  
${\rm SO}(n,1)$, Int. J. Mod. Phys. {\bf D8} (1999); [gr-gc/9902069].

\bibitem{Coleman} S.~Coleman, There are no Goldstone bosons in two 
dimensions,
\CMP {\bf 31}~(1973)~259.

\bibitem{PSWard} A. Patrascioiu and E. Seiler, Continuum limit of 2D spin
models  with continuous symmetry and conformal field theory, \PR {\bf E
57}~(1998)~111; Does conformal quantum field theory describe the
continuum limits of 2D spin models with continuous symmetry? \PL {\bf B
417}~(1998)~123.

\bibitem{OS} K. Osterwalder and R. Schrader, Axioms for Euclidean Green's
functions, \CMP 31, (1973) 83; Axioms for Euclidean Green's functions 2,
\CMP 42, (1975) 281.

\bibitem{GJ} J. Glimm and A. Jaffe, Quantum Physics, Springer-Verlag, New
York etc. 1987.

\bibitem{S} E. Seiler, Gauge Theories as a Problem of Constructive Quantum
Field Theory and Statistical Mechanics, Lecture Notes in Physics vol. 159,
Springer-Verlag Berlin etc. 1982.

\bibitem{Haag} R. Haag, Local Quantum Physics, Springer-Verlag Berlin etc.
1992.

\bibitem{CR83} Y. Cohen and E. Rabinovici, A study of the non-compact
non-linear sigma-model: A search for dynamical realizations of
non-compact symmetries, Phys.~Lett. {\bf B124} (1983) 371.

\bibitem{VilKlim} N. Vilenkin and A. Klimyk, {\it Representations of Lie
groups  and special functions}, Kluwer, Dordrecht 1993.

\bibitem{Dym} H. Dym and H. P. McKean, {\it Fourier Series and Integrals},
Academic Press, New York and London 1972.

\bibitem{BV} N. Balazs and A. Voros, Chaos on the pseudosphere,
Phys. Repts. {\bf 143} (1986) 109.

\bibitem{SegalKunze} I. Segal and R. Kunze, Integrals and operators,
Springer-Verlag, Berlin -- New York 1978.  

\bibitem{SZ} T. Spencer and M.~R.~Zirnbauer, Spontaneous symmetry breaking
of a hyperbolic sigma model in three dimensions,
\CMP {\bf 252} (2004) 167 [arXiv:math-phys/0410032].

\bibitem{AG} N. I. Akhiezer and I. M. Glazman, Theory of linear operators
in Hilbert space, Dover, New York 1993.


\end{thebibliography}
\end{document}